\documentclass[reprint,amsmath,amsfonts,amssymb,aps,prx,preprintnumbers,superscriptaddress]{revtex4-2}

\usepackage[utf8]{inputenc}
\usepackage[T1]{fontenc}
\usepackage{lmodern}
\usepackage{graphicx}
\usepackage{dcolumn}
\usepackage{bm}
\usepackage{textcomp}
\usepackage{ulem}
\usepackage{ifpdf}
\usepackage[squaren,Gray]{SIunits}
\usepackage{color}
\usepackage[export]{adjustbox}
\definecolor{red}{rgb}{1,0,0}
\definecolor{blue}{rgb}{0,0,1}
\definecolor{darkred}{rgb}{0.6,0,0}
\definecolor{darkblue}{rgb}{0,0,.6}
\definecolor{darkgreen}{rgb}{0,0.5,0}
\definecolor{green}{rgb}{0.1,0.5,0.1}
\definecolor{brown}{rgb}{0.4,0.25,0.12}
\definecolor{grey}{rgb}{0.2,0.2,0.2}

\usepackage{natbib}
\ifpdf
\usepackage{epstopdf}
\usepackage[pdftex,unicode,pdfstartview={FitH},pdfborder={0 0 0}]{hyperref}
\usepackage{hypcap}
\else
\usepackage[hypertex]{hyperref}
\fi
\hypersetup{
    bookmarksnumbered = true,
    colorlinks = true, linkcolor = darkblue,
    citecolor = darkblue, filecolor = darkblue,
    menucolor = darkblue, urlcolor = darkblue
}

\newcolumntype{R}{>{$\displaystyle}r<{$}}
\newcolumntype{C}{>{$\displaystyle}c<{$}}

\begin{document}
    
\title{Stacking-dependent exciton multiplicity in WSe$_2$ bilayers}

\author{Zhijie Li}
\affiliation{Fakult\"at f\"ur Physik, Munich Quantum Center, and Center for NanoScience (CeNS), Ludwig-Maximilians-Universit\"at M\"unchen, Geschwister-Scholl-Platz 1, 80539 M\"unchen, Germany}
\author{Jonathan F\"orste}
\affiliation{Fakult\"at f\"ur Physik, Munich Quantum Center, and Center for NanoScience (CeNS), Ludwig-Maximilians-Universit\"at M\"unchen, Geschwister-Scholl-Platz 1, 80539 M\"unchen, Germany}
\author{Kenji Watanabe}
\affiliation{Research Center for Functional Materials, National Institute for Materials Science, 1-1 Namiki, Tsukuba 305-0044, Japan}
\author{Takashi Taniguchi}
\affiliation{International Center for Materials Nanoarchitectonics, National Institute for Materials Science, 1-1 Namiki, Tsukuba 305-0044, Japan}
\author{Bernhard Urbaszek}
\affiliation{Université de Toulouse, INSA-CNRS-UPS, LPCNO, 135 Avenue Rangueil, 31077, Toulouse, France}
\author{Anvar~S.~Baimuratov}
\affiliation{Fakult\"at f\"ur Physik, Munich Quantum Center, and Center for NanoScience (CeNS), Ludwig-Maximilians-Universit\"at M\"unchen, Geschwister-Scholl-Platz 1, 80539 M\"unchen, Germany}
\author{Iann C. Gerber}
\affiliation{Université de Toulouse, INSA-CNRS-UPS, LPCNO, 135 Avenue Rangueil, 31077, Toulouse, France}
\author{Alexander H{\"o}gele}
\affiliation{Fakult\"at f\"ur Physik, Munich Quantum Center, and Center for NanoScience (CeNS), Ludwig-Maximilians-Universit\"at M\"unchen, Geschwister-Scholl-Platz 1, 80539 M\"unchen, Germany}
\affiliation{Munich Center for Quantum Science and Technology (MCQST), Schellingtra\ss{}e 4, 80799 M\"unchen, Germany}
\author{Ismail Bilgin}
\affiliation{Fakult\"at f\"ur Physik, Munich Quantum Center, and Center for NanoScience (CeNS), Ludwig-Maximilians-Universit\"at M\"unchen, Geschwister-Scholl-Platz 1, 80539 M\"unchen, Germany}

\date{\today}
    
\begin{abstract}
Twisted layers of atomically thin two-dimensional materials realize a broad range of novel quantum materials with engineered optical and transport phenomena arising from spin and valley degrees of freedom and strong electron correlations in hybridized interlayer bands. Here, we report experimental and theoretical studies of WSe$_2$ homobilayers obtained in two stable configurations of 2H ($60^\circ$ twist) and 3R ($0^\circ$ twist) stackings by controlled chemical vapor synthesis of high-quality large-area crystals. Using optical absorption and photoluminescence spectroscopy at cryogenic temperatures, we uncover marked differences in the optical characteristics of 2H and 3R bilayer WSe$_2$ which we explain on the basis of beyond-DFT theoretical calculations. Our results highlight the role of layer stacking for the spectral multiplicity of momentum-direct intralayer exciton transitions in absorption, and relate the multiplicity of phonon sidebands in the photoluminescence to momentum-indirect excitons with different spin valley and layer character. Our comprehensive study generalizes to other layered homobilayer and heterobilayer semiconductor systems and highlights the role of crystal symmetry and stacking for interlayer hybrid states.
\end{abstract}
    
\maketitle
    
\section{Introduction}
The optical properties of transition metal dichalcogenide (TMD) semiconductors are governed by excitons in different spin, valley and layer configurations \cite{wang2018colloquium,wilson2021excitons}. Among possible realizations of TMD systems, heterobilayers and homobilayers stand out as hosts of excitons with layer-indirect character. Initial studies of homobilayers were limited to natural 2H layer stacking with antiparallel alignment or $60^\circ$ twist angle \cite{Zhao2013a,Arora2017,Molas2017}, extracted by exfoliation from native crystals and structurally different from 3R stacking with parallel alignment or $0^\circ$ twist. Recently, variations of the twist angle in WSe$_2$ bilayers (BLs) away from 2H and 3R stackings revealed novel phenomena ranging from correlated electronic phases~\cite{Wang2020} to moir\'e exciton physics \cite{Brem2020,Andersen2021} with angle-controlled exciton valley coherence and dynamics~\cite{Gong2013,Scuri2020,Andersen2021}, Coulomb correlations in effectively flat moir\'e exciton bands~\cite{Merkl2020} or optical nonlinearities~\cite{Lin2021}. 

\begin{figure*}[t!]
\centering
\includegraphics[scale=0.98]{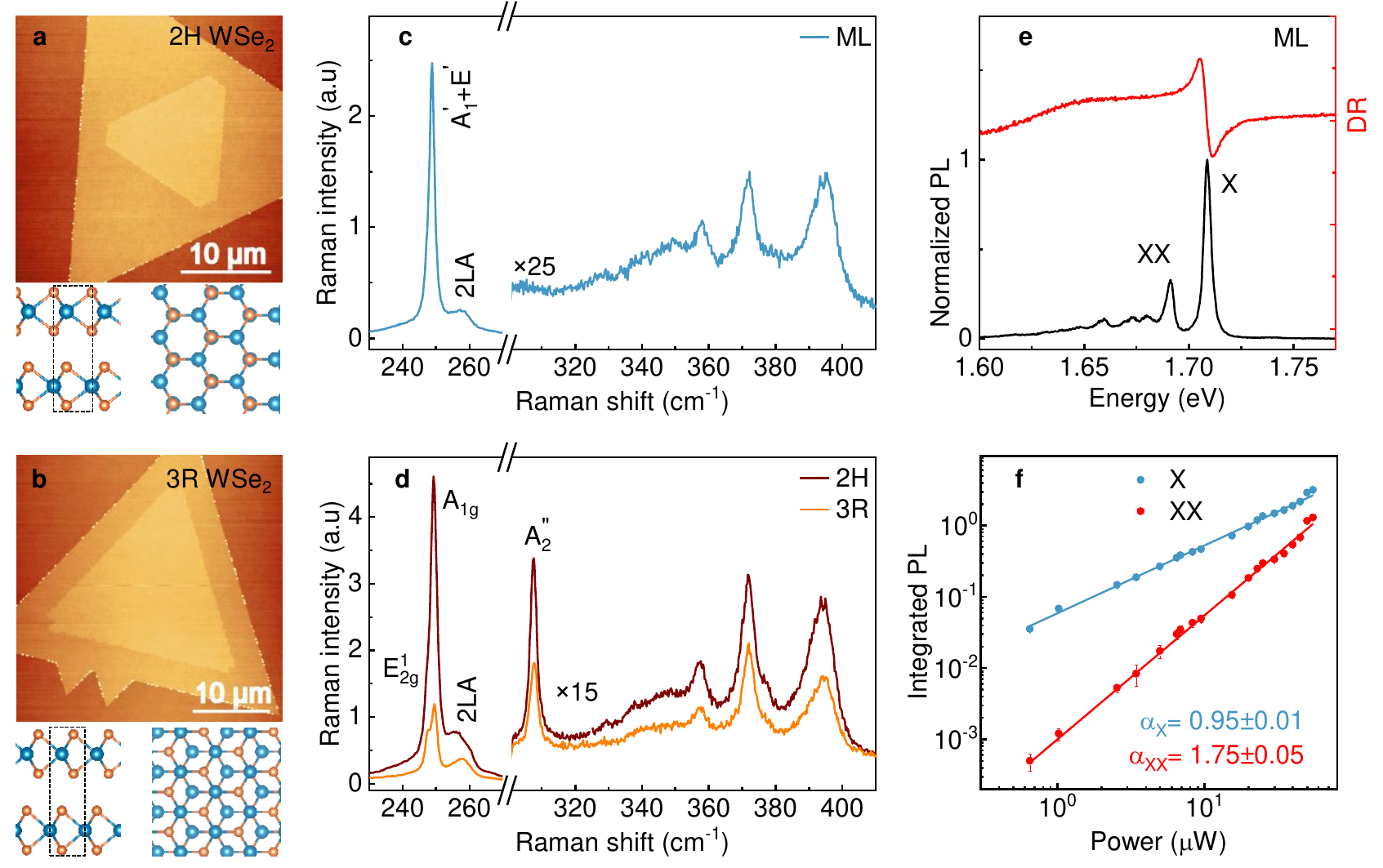}
\caption{\textbf{Characteristics of high-quality CVD-grown WSe$_2$ homobilayers.} \textbf{a} and \textbf{b}, Atomic force micrographs of 2H and 3R homobilayer WSe$_2$ on SiO$_2$/Si, respectively, with side and top view illustrations of the respective stacking orientations shown below (blue and orange spheres represent W and Se atoms, respectively, dashed rectangles indicate the unit cell). \textbf{c} and \textbf{d}, Room-temperature Raman spectra of ML and BL WSe$_2$ in 2H and 3R stacking (the spectra were recorded with excitation at $532$~nm and normalized to the peak intensity of the Si phonon mode at $520$~cm$^{-1}$; note that the intensity of higher-order Raman modes associated with multi-phonon processes were scaled for increased visibility). The in-plane and out-of-plane phonon modes, $E'$ and $A'_1$, are degenerate in the ML but split by $2$~cm$^{-2}$ both in 2H and 3R homobilayer WSe$_2$. \textbf{e}, Normalized photoluminescence (black) and differential reflectivity (red) spectra at $3.2$~K, recorded in the ML region of hBN-encapsulated BL WSe$_2$ crystals. The ML exciton (X) and biexciton (XX) peaks exhibit characteristic linear and superlinear scaling of the integrated PL intensity as a function of the excitation power, as evident from data and respective power-law fits $I_{PL} \propto P^\alpha$ shown by solid lines in \textbf{f}.}
\label{fig1}
\end{figure*}

A particularly attractive feature of interlayer excitons in related homobilayer systems is provided by the permanent dipole moment of layer-indirect excitons \cite{Calman2018,Deilmann2018,Gerber2019} which promotes exciton-exciton dipolar interactions, allows employing electric fields to tune the optical properties via the Stark effect~\cite{WangMak2018,Leisgang2020,Scuri2020,Peimyoo2021,Lorchat2021,Altaiary2021arxiv,Huang2021arxiv} or implementing exciton trapping and routing~\cite{Liu2021}. The electrostatic dipole moment depends on the degree of exciton layer delocalization, which in turn is sensitive to the interlayer coupling and thus to the stacking order, as was shown recently by optical absorption for 2H and 3R MoS$_2$ BLs for transitions involving momentum direct $KK$ excitons \cite{Hsu2019,Paradisanos2020}. In contrast to absorption probing zero-momentum exciton transitions, the photoluminescence (PL) of BLs is dominated by momentum-indirect excitons~\cite{Mak2010,Splendiani2010} via luminescence phonon sidebands~\cite{Lindlau2018,Forste2020,Funk2021} which exhibit shift linear energy shifts in perpendicular electric fields according to the Stark effect \cite{WangMak2018,Scuri2020,Altaiary2021arxiv,Huang2021arxiv}. The most recent observation of two different slopes in the energy dispersion of phonon sideband emission from 2H WSe$_2$ BL in electric field ~\cite{Altaiary2021arxiv,Huang2021arxiv} indicates the presence of two degrees of electron-hole layer separation, with respective dipole moments attributed to excitons in distinct $QK$ and $Q\Gamma$ reservoirs~\cite{Altaiary2021arxiv,Huang2021arxiv}. Experiments employing strain tuning \cite{Aslan2020} and magnetic fields \cite{Forste2020}, on the other hand, suggest that the PL sidebands stem exclusively from $QK$ excitons, indicating shortcomings in the present understanding of the underlying lowest-energy exciton reservoirs with finite center-of-mass momenta.

To provide comprehensive insight into the nature of exciton reservoirs in BL WSe$_2$ with different spin, momentum and layer character, we performed experimental and theoretical studies of 2H and 3R stackings with contrasting spectroscopic responses. To this end we synthesized WSe$_2$ crystals by chemical vapor deposition (CVD), yielding 2H and 3R as two stable limits of relative layer orientation. As opposed to BL crystals of MoS$_2$~\cite{Liu2014,Hsu2019,Paradisanos2020} and WS$_2$~\cite{Schneider2019,Du2019} synthesized by CVD, studies of WSe$_2$ BLs in 3R stacking have remained elusive due to the challenge of perfect layer alignment. The only realization of nominal zero-angle twist so far has been obtained from exfoliation stacking~\cite{Scuri2020}, which can only approximate the ideal 3R layer order inherent to CVD growth. Moreover, the spectral features of BLs aligned near zero twist can be compromised by marginal-angle reconstruction ~\cite{Carr2018,Weston2020,Enaldiev2020,Andersen2021} with interfacial ferroelectricity effects~\cite{Magorrian2021} as observed recently for reconstructed homobilayers of hexagonal boron nitride (hBN)~\cite{Yasuda2021,Vizner2021,Woods2021} and TMDs~\cite{Ferreira2021,Weston2021}.

In the following, we present a comprehensive study of excitons in 2H and 3R WSe$_2$ BLs, performed with cryogenic optical spectroscopy on  CVD-synthesized high-quality crystals. With a set of complementary spectroscopy techniques, we identify contrasting responses of 2H and 3R BL configurations and observe rich exciton multiplicity in the 3R case. We relate the differences between the optical spectra of 2H and 3R BL stackings to the nature of excitons in different spin, valley and layer configurations. Our interpretation is based on first-principles calculations of the band structure, optical absorption and exciton $g$-factors. Theory analysis shows that in contrast to 2H stacking, the top and bottom layers in 3R stacking differ in their band structure and exhibit different optical bandgaps. This insight resolves the recent puzzle of two distinct intralayer $KK$ exciton transitions observed in absorption on nearly aligned WSe$_2$ BLs~\cite{Scuri2020}. Moreover, our theoretical analysis highlights the multiplicity and degree of layer delocalization FOR interlayer excitons in both 2H in 3R stackings, thus providing an intuitive explanation for different electrostatic dipole moments of $QK$ excitons without the requirement of involving $Q\Gamma$ states~\cite{Altaiary2021arxiv,Huang2021arxiv}.

\section{Results and Discussion}
WSe$_2$ BL crystals in 2H and 3R stacking were obtained from CVD synthesis. Atomic force micrographs in Fig.~\ref{fig1}a and b show respective WSe$_2$ crystals with ML and BL regions in 2H and 3R configurations with $60$° and $0$° rotational alignment of the two WSe$_2$ monolayers (MLs), as can be deduced directly from the edges of CVD flakes with the same edge termination. The corresponding atomic registries are illustrated in the bottom panels of Fig.~\ref{fig1}a and b, showing side (left) and top (right) views and the BL unit cell by dashed boxes. 

Initial characterization of the two distinct crystal configurations was performed with Raman spectroscopy at ambient conditions. A typical Raman spectrum from a ML region is shown in Fig.~\ref{fig1}c. 
It features degenerate in-plane $E'$ and out-of-plane $A_1'$ first-order Raman modes around $249$~cm$^{-1}$, the double resonance 2LA mode at $257$~cm$^{-1}$, and a series of multi-phonon modes at $358$, $372$, and $395$~cm$^{-1}$, all of which are consistent with ML features \cite{Li2013,Zeng2013,Sahin2013,Zhao2013,Luo2013,Terrones2014}. The degeneracy of the $E'$ and $A_1'$ modes, characteristic of MLs \cite{Luo2013,Terrones2014}, is lifted in the Raman spectra of both 2H and 3R BLs shown in Fig.~\ref{fig1}d. We determined by peak decomposition that in both BL stackings the respective $E^{1}_{2g}$ and $A_{1g}$ modes exhibit a $\sim 2$~cm~$^{-1}$ frequency splitting due to interactions between the layers. We also observe a red-shift of the $E'$/$E^{1}_{2g}$  and a blue shift of the $A_1'$/$A_{1g}$ modes for both stackings~\cite{Puretzky2015}. The peak at $308$~cm$^{-1}$, labelled as $A''_2$, is absent in the ML limit but very pronounced in the Raman spectra of both 2H and 3R stackings as a hallmark of BL regions~\cite{Li2013,Zhao2013,Luo2013,Terrones2014}. The spectra of multi-phonon modes at frequencies above $350$~cm~$^{-1}$ are similar in ML and BLs.

Having identified 2H and 3R BL crystals by AFM and Raman spectroscopy, we performed cryogenic micro-PL and differential reflection (DR) spectroscopy. To this end, the CVD-grown crystals were encapsulated in hexagonal boron nitride (hBN) by standard lift-off and exfoliation and transferred onto a Si/SiO$_2$ substrate. Typical PL and DR spectra of ML regions, recorded at $3.2$~K, are shown in Fig.~\ref{fig1}e. Both spectra feature the resonance of the fundamental exciton (X) around $1.71$~eV with a spectrally narrow full-width at half-maximum (FWHM) linewidth of $4$~meV, exceeding the transform-limited linewdith of $1$~meV in best hBN-encapsulated MLs obtained from native crystals \cite{Cadiz2017,Ajayi2017,Wierzbowski2017,Shree2019} yet slightly narrower than $5.2$~meV reported for CVD-grown MoS$_2$ ML in hBN~\cite{Shree2019}. The second pronounced peak in the PL spectrum at $1.69$~eV corresponds to the biexciton (XX) emission with a characteristic $20$~meV red-shift \cite{You2015,Barbone2018,Steinhoff2018,Li2018,Ye2018} and superlinear dependence on the excitation power shown in Fig.~\ref{fig1}f.   All main features in PL and DR as well as the absence of trion signatures indicate vanishing residual doping and the overall high quality of our CVD-grown crystals.

\begin{figure}[t!]
\centering
\includegraphics[scale=0.98]{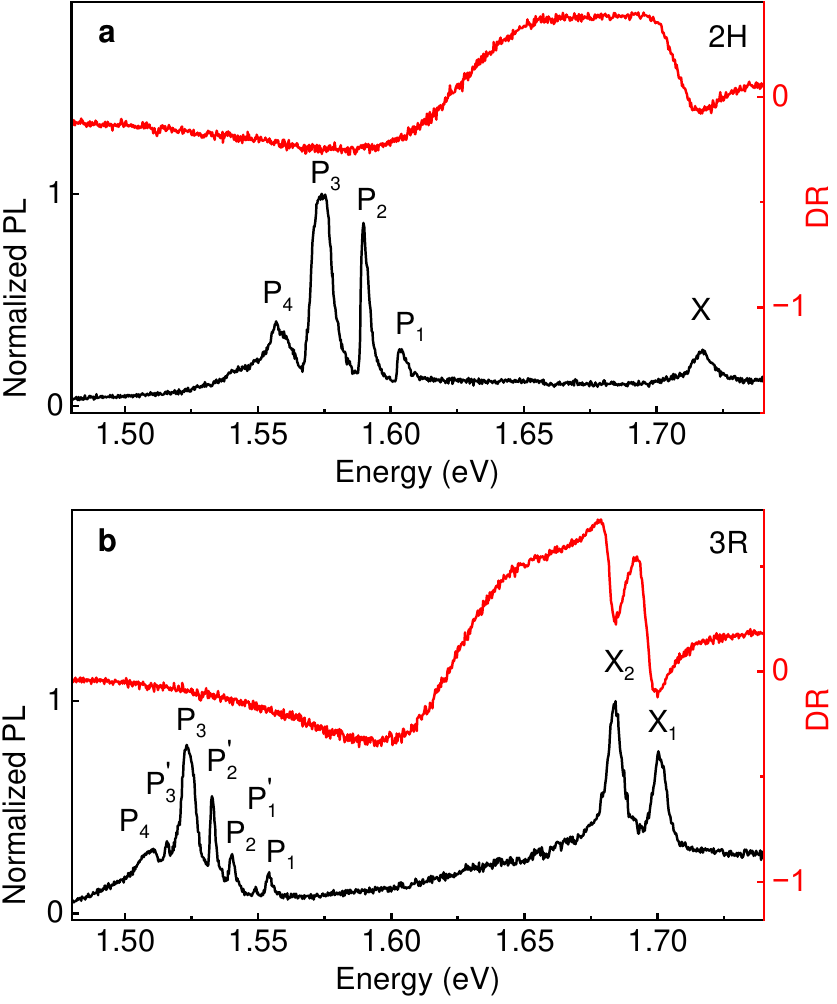}
\caption{\textbf{Cryogenic spectroscopy of 2H and 3R WSe$_2$ homobilayers.} \textbf{a} and \textbf{b}, Normalized photoluminescence (PL, black) and differential reflectivity (DR, red) spectra of 2H and 3R WSe$_2$ at $3.2$~K. In contrast to 2H with only one resonance of the fundamental exciton X, the spectra of 3R exhibit two exciton peaks X$_1$ and X$_2$ with a splitting of $17$~meV both in PL and DR. Moreover, the series of phonon sideband PL peaks P$_1$ to P$_4$ in 2H red-shifts by $50$~meV in 3R and features additional peaks P$'_1$ to P$'_3$. The contrast change in DR between the phonon sideband groups and the main exciton resonances are presumably due to phonon-assisted excitation of momentum-indirect states.}
\label{fig2}
\end{figure}

\begin{figure*}[t!]
\centering
\includegraphics[scale=0.99]{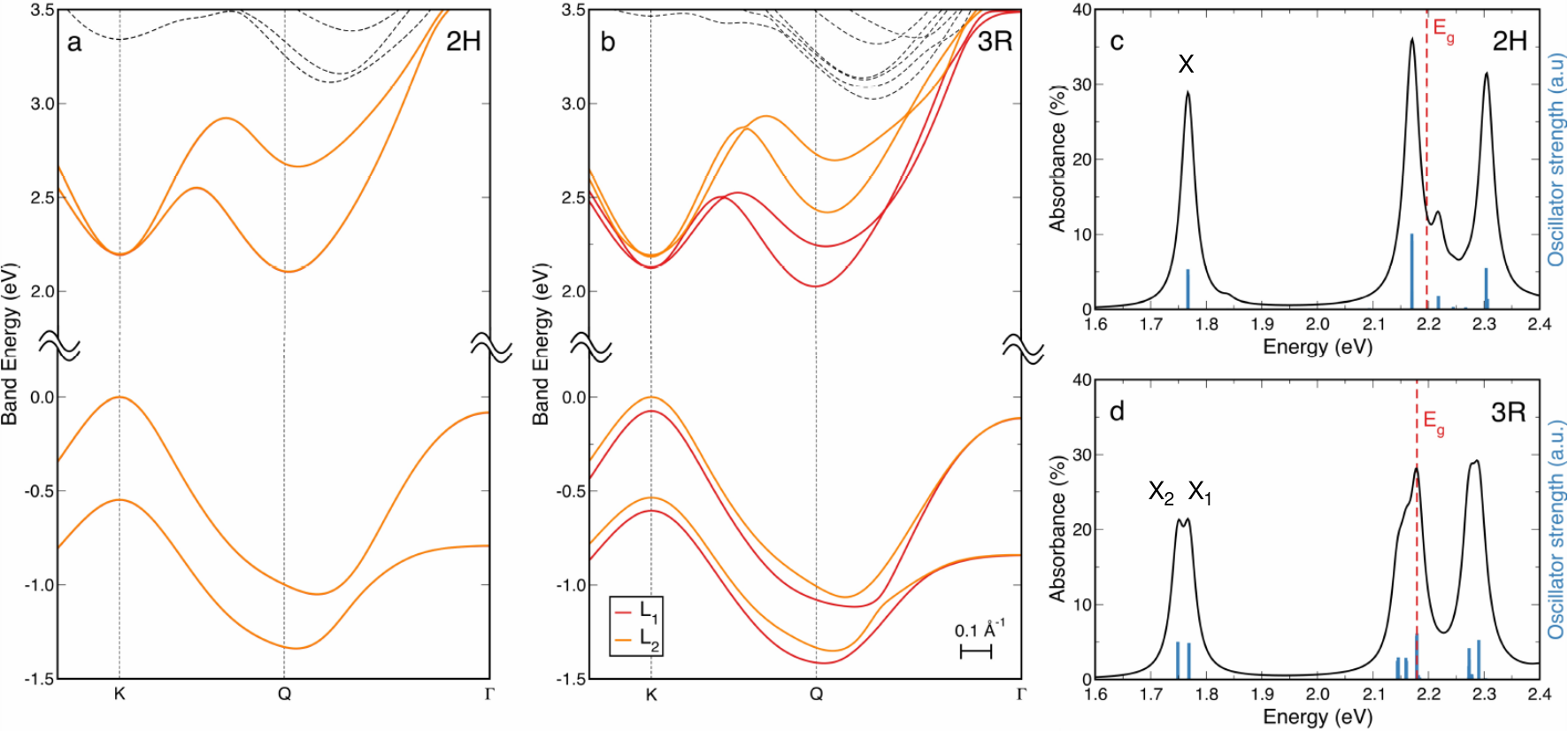}
\caption{\textbf{Theoretical calculations for 2H and 3R WSe$_2$ homobilayers.} \textbf{a} and \textbf{b}, Band structure for 2H and 3R WSe$_2$ homobilayer, respectively. Red and orange colors represent the bottom and top layer, L$_1$ and L$_2$. \textbf{c} and \textbf{d}, Absorption spectra (black) and oscillator strength (blue) for 2H and 3R stacking, respectively, calculated with the GW+BSE method. The dashed red lines indicate the energy gap $E_g$.}
\label{fig3}
\end{figure*}

The cryogenic PL and DR of BL regions with 2H and 3R stacking are shown in Fig.~\ref{fig2}a and b with pronounced differences in the spectral characteristics. The 2H spectra in Fig.~\ref{fig2}a agree with the results of previous studies on exfoliated BLs from bulk WSe$_2$ \cite{Arora2015,Lindlau2017BL,Scuri2020,Aslan2020,Forste2020,Funk2021}, with the neutral exciton resonance X at around $1.72$~eV associated with momentum direct transitions between conduction and valence band states at $K$, as well as a series of PL peaks between $1.55$ and $1.61$~eV we label as P$_1$ to P$_4$ with increasing red-shift. Consistent with the indirect band gap of WSe$_2$ BL between the conduction band minima at the six inequivalent $Q$-points of the first Brillouin zone and the valence band maximum at $K$ \cite{Wickramaratne2014,Terrones2014}, the lowest energy exciton states form as $QK$ and $Q'K$ with finite center-of-mass momentum and characteristic phonon sideband luminescence into the peaks P$_1$ through P$_4$ \cite{Forste2020,Funk2021}. As on ML regions, the high-quality of our CVD-synthesized crystals is reflected by spectrally narrow features of BLs in both stackings.

The characteristic features of 2H WSe$_2$ BL are contrasted by the spectra from 3R stacking. First, the peak multiplicity increases both in PL and DR spectra of Fig.~\ref{fig2}b. We observe two distinct exciton resonances X$_1$ and X$_2$ around $1.70$ and $1.68$~eV with an energy splitting of $17$~meV and spot-to-spot variations between $16$ and $20$~meV, which is very different from the single transition observed throughout the 2H stacked sample. As a second important observation, we find in the region of phonon sidebands between $1.50$ and $1.55$~eV three additional peaks emerging in between P$_1$ through P$_4$, which we label as P$'_1$, P$'_2$ and P$'_3$. Moreover, we observe a red-shift of $16$~meV from the exciton resonance X in 2H to X$_1$ in 3R, as well as a larger red-shift to the first peak P$_1$, increasing from $\sim 100$~meV in 2H to $\sim 150$~meV in 3R as a consequence of stronger interlayer hybridization in the latter case.

To understand the increased multiplicity of exciton transitions as well as their energy shifts and splittings in 3R stacking, we carried out band structure calculations including excitonic effects via the GW+BSE approach. Our calculations, performed for WSe$_2$ BLs in vacuum without hBN-encapsulation, yield the band structure shown in Fig.~\ref{fig3}a and b for 2H and 3R stacking, respectively. Due to the different atomic registries for top and bottom layers in 3R stacking, the band structure for these two layers is distinct, with different transition energies at the $K$-point. This is in stark contrast to 2H stacking, where both layers, and their associate intralayer states, are energetically degenerate.

\begin{figure*}[t!]
\centering
\includegraphics[scale=0.98]{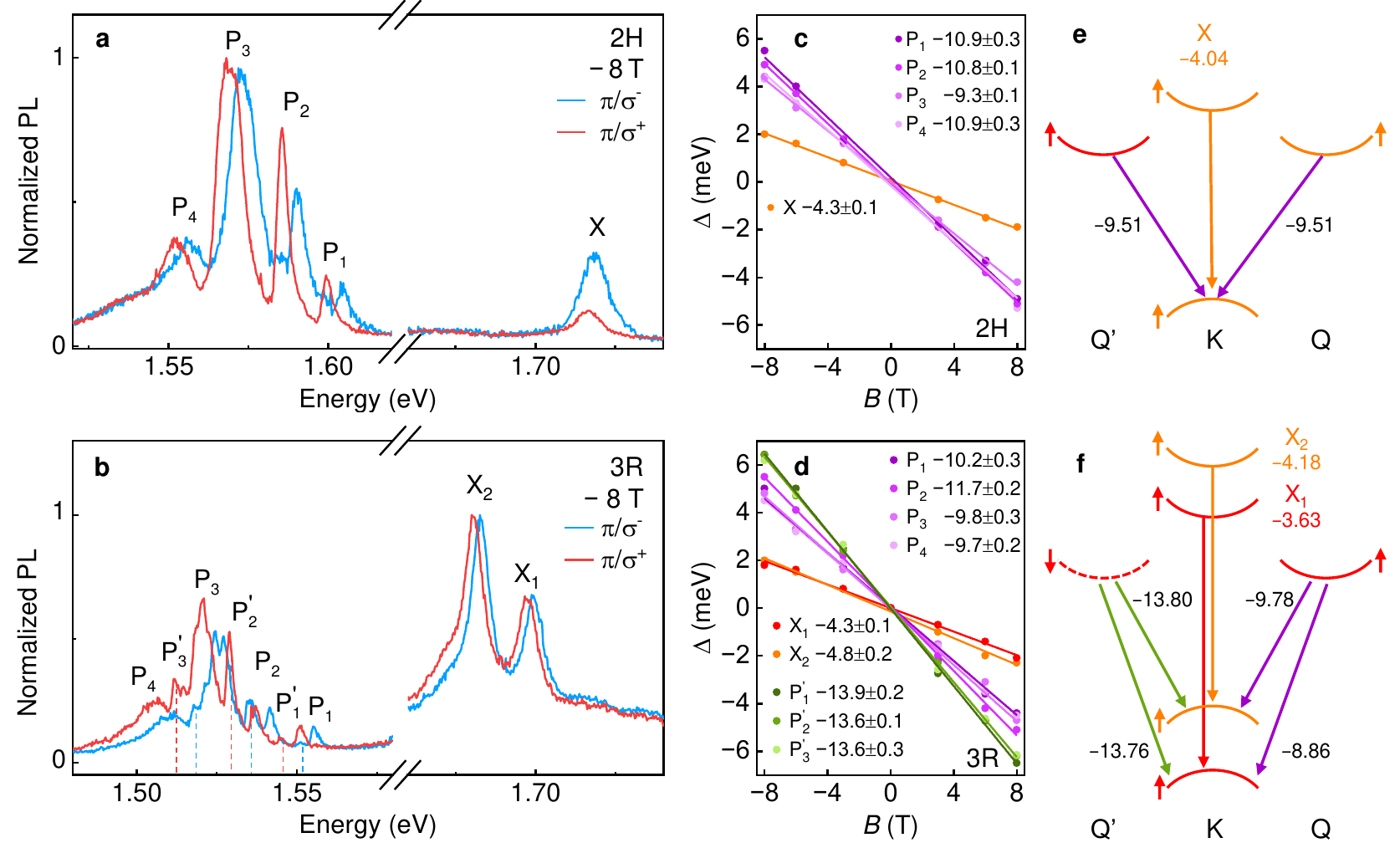}
\caption{\textbf{Magneto-optics of 2H and 3R WSe$_2$ homobilayers in experiment and theory.} \textbf{a} and \textbf{b}, Photoluminescence spectra of 2H and 3R WSe$_2$ in a magnetic field of $-8$~T recorded with $\sigma^+$ (red) $\sigma^-$ (blue) polarized detection under linearly polarized excitation ($\pi$). Magneto-induced splittings of $P'$ peaks in 3R WSe$_2$ are indicated in \textbf{b} by dashed lines. \textbf{c} and \textbf{d}, Valley Zeeman splitting $\Delta$ of all peaks identified in 2H and 3R WSe$_2$ magneto-luminescence as a function of magnetic field. The respective exciton g-factors were extracted from linear fits (solid lines) to the data and grouped by colors. \textbf{e} and \textbf{f}, Schematic illustration of momentum-direct and phonon-assisted transitions in 2H and 3R, tagged with theoretical exciton $g$-factor values (solid and dashed lines denote spin-up and spin-down bands; orange and red colors indicate the bands of the first and second layer degenerate in 2H but split in 3R stacking).}
\label{fig4}
\end{figure*}

For direct comparison of our theory with PL and DR experiments, we calculated the absorbance of momentum-direct exciton transitions for both stackings. For the 2H stacking we found only one transition at about $1.75$~eV, which corresponds to the A-exciton or the X transition in the spectra of Fig.~\ref{fig2}a. Note that the feature at $1.85$~eV with very low oscillator strength is the interlayer $KK$ exciton with a much lower oscillator strength here than in BL MoSe$_2$ and MoS$_2$ due to the larger spin orbit coupling in WSe$_2$ \cite{Gong2013,Paradisanos2020,slobodeniuk2019fine,horng2018observation}. As in the experiment, we find in our calculated absorption for the 3R stacking two transitions around $1.75$~eV which are separated by about $20$~meV. From our calculations we infer that the two transitions correspond to A-excitons in the top and bottom layer. This insight provides a microscopic explanation for the splitting we observe for the 3R stacked sample in Fig.~\ref{fig2}a. Note that a similar lifting of the intralayer exciton degeneracy was anticipated for 3R MoS$_2$ BLs but not clearly discernible in the experimental data \cite{Paradisanos2020}. The interlayer exciton feature is completely absent in 3R stacking as hybridization of electronic states at the $K$-point is forbidden by symmetry \cite{Gong2013}. Lower-energy  momentum-indirect transitions that dominate BL emission \cite{shree2021guide} are not captured by our absorbance calculation due to the vanishingly small oscillator strength.

To complement our theory analysis of excitons in 2H and 3R WSe$_2$ BLs, we determined the exciton $g$-factors from magneto-luminescence measurements and correlated them with theoretically calculated values in different spin, valley and layer configurations \cite{Forste2020,Forg2021}. Out-of-plane magnetic fields break the time-reversal symmetry and lift the valley degeneracy between $\sigma^+$ and $\sigma^-$ polarized PL peaks with energies $E^+$ and $E^-$ to induce a valley Zeeman splitting $\Delta =E^+-E^-=g \mu_\mathrm{B}B$ proportional to the exciton $g$-factor, the Bohr magneton $\mu_\mathrm{B}$, and the magnetic field $B$ \cite{Aivazian2015,Srivastava2015,Wang2015,Koperski2018}.  Polarization-resolved magneto-luminescence spectra of 2H and 3R WSe$_2$ BLs, recorded at $B = -8$~T with $\sigma^+$ and $\sigma^-$ circular detection under linearly polarized ($\pi$) excitation, are shown in Fig.~\ref{fig4}a and b. The exciton $g$-factors follow from linear fits to the data shown by solid lines in Fig.~\ref{fig4}c and d and grouped in color according to the $g$-factor values.  

For 2H (Fig.~\ref{fig4}c), we found for the fundamental exciton peak X the characteristic $g$-factor of $-4.3$, as well as $g$-factors of momentum-indirect $QK$ excitons (i.\,e. Coulomb-correlated states of the conduction band electron in the $Q$-valley and the valence band vacancy in the $K$-valley) in the range from $-9.3$ to $-10.9$ (peaks P$_1$ through P$_4$), consistent with previous findings for exfoliated WSe$_2$ BLs in 2H stacking \cite{Forste2020}. In 3R stacking, our high-quality sample with spectrally narrow PL emission allows us to quantify two distinct $g$-factors of $-4.3$ and $-4.8$ for the momentum-direct $KK$ exciton doublet X$_1$ and X$_2$ (Fig.~\ref{fig4}d). For the emission peaks at lower energy between $1.50$ and $1.55$ eV, we determined $g$-factors of about $-11$ as in 2H (peaks P$_1$ through P$_4$) and an additional group of peaks (peaks P$'_1$ through P$'_3$) with a $g$-factor close to $-14$. According to our understanding of the band structure and exciton $g$-factors in 2H WSe$_2$ BL, the peaks P$_1$ through P$_4$ stem from phonon-assisted recombination of $QK$ excitons, without contribution of $Q\Gamma$ states with small $g$-factors between $0$ and $4$ \cite{Forste2020}.

\begin{table}[t!]
\caption{\textbf{Theoretical exciton $g$-factors in 2H and 3R WSe$_2$ bilayers.} We distinguish between intra- and interlayer excitons and show for momentum-direct $KK$ transitions only the $g$-factors of intralayer excitons. For momentum-indirect $QK$ and $Q'K$ transitions, we select spin-valley configurations of intra- and interlayer excitons with lowest energies. The transition column ($L_\text{cb} \rightarrow L_\text{vb}$) indicates the layers with conduction and valence band electrons involved in phonon-assisted recombination.}
    \begin{ruledtabular}
        \begin{tabular}{RRRRRR}
            \multicolumn{3}{c}{Exciton} & & \multicolumn{2}{c}{$g$-factors}\\
            \cline{1-3} \cline{5-6}
            \text{Intra} & \text{Inter} & \text{Spin}          & \text{Transition}   & \text{2H} & \text{3R} \\
            \hline          
            KK           &              & \uparrow \uparrow    & L_1 \rightarrow L_1 & -4.04     & -3.63     \\
            KK           &              & \uparrow \uparrow    & L_2 \rightarrow L_2 &           & -4.18     \\
            \hline
            QK           &              & \uparrow \uparrow    & L_1 \rightarrow L_1 & -9.51     & -8.86     \\
                         & QK           & \uparrow \uparrow    & L_1 \rightarrow L_2 &           & -9.78     \\
                         & Q'K          & \uparrow \uparrow    & L_1 \rightarrow L_2 & -9.51     &           \\
            \hline
            Q'K          &              & \downarrow \uparrow  & L_1 \rightarrow L_1 & -13.71    & -13.76    \\
                         & Q'K          & \downarrow \uparrow  & L_1 \rightarrow L_2 &           & -13.80    \\ 
                         & QK           & \downarrow \uparrow  & L_1 \rightarrow L_2 & -13.71    &              
        \end{tabular}   
    \end{ruledtabular}
    \label{tab_gfactor}
\end{table}

To interpret the origin of additional peaks in 3R stacking with distinct magneto-luminescence features as well as their difference to the 2H configuration, we calculated the exciton $g$-factors from first principles~\cite{Wozniak2020,Forste2020,Deilmann2020,Xuan2020,Forg2021}. The results for both stackings are presented in Table~\ref{tab_gfactor} and pictorially also in Fig.~\ref{fig4}e and f. In the case of 2H, the $g$-factor of X corresponds to spin-like $(\uparrow \uparrow )$ $KK$ transition, whereas momentum-indirect exciton peaks (P$_1$ to P$_4$) relate to spin-like intralayer $QK$ and interlayer $Q'K$ transitions with $g = -9.51$. For transitions in 3R, our results also identify spin-like $KK$ excitons X$_1$ and  X$_2$ with two different $g$-factors of $-4.18$ and $-3.63$. This difference in the theoretical values arises from different $g$-factors of the conduction bands at $K$ in the top and bottom layers $L_1$ and $L_2$. 

Momentum-indirect excitons with peaks P$_1$ to P$_4$ in 3R with similar $g$-factors to the corresponding peaks in 2H stem from spin-like intralayer and interlayer $QK$ transitions with $g=-8.86$ and $g=-9.78$, respectively. Additional 3R peaks with relatively large $g$-factors (P$'_1$ to P$'_3$), on the other hand, originate from spin-unlike intralayer and interlayer $Q'K$ transitions with $g$-factors of $-13.80$ and $-13.76$. We note that in 2H, spin-unlike intralayer $Q'K$ and interlayer $QK$ transitions with $g = -13.71$ from Table~\ref{tab_gfactor} are not observed within the signal-to-noise ratio of our experiment. These states are degenerate with the two respective spin-like interlayer $Q'K$ and intralayer $QK$ transitions, yet their exciton population seems to be vanishingly small, presumably due to spin-conserving relaxation \cite{Scuri2020}. However, once the degeneracy of the states is removed by the lack of inversion symmetry in 3R stacking, our analysis shows that spin-unlike $Q'K$ excitons light up in the PL of peaks P$'_1$ to P$'_3$.

\section{Conclusions}
Our extensive experimental and theoretical studies of BL WSe$_2$ identify stacking-dependent optical response of 2H and 3R crystals. We find the marked differences to arise from distinct multiplicity of intralayer and interlayer excitons in the two different stable realizations of BL stackings. Using high-quality samples of both stacking configurations by CVD crystal synthesis and hBN-encapsulation, we obtain spectrally narrow resonances which allow us to identify in great detail the characteristics of momentum-direct excitons in absorption and momentum-indirect excitons in emission of 2H and 3R BLs. For both stackings we find that the lowest-energy PL is dominated by momentum-indirect $QK$ and $Q'K$ excitons with intralayer and interlayer character. The different degrees of layer delocalization give rise to two different electrostatic dipoles, providing an explanation for the recent observations of two distinct slopes in the first-order Stark effect~\cite{Altaiary2021arxiv,Huang2021arxiv}. Moreover, our results demonstrate that exciton state multiplicity, transition energies and oscillator strengths sensitively depend on the BL stacking order, and highlight how layer-degenerate excitons in inversion symmetric 2H BLs split into exciton doublets in 3R stacking with broken inversion symmetry and different atomic registries for the top and bottom layers. Our findings are relevant for the understanding of other TMD homobilayer systems as well as semiconductor van der Waals heterobilayers with varying degrees of symmetry, twist and interlayer hybridization. 

\section{Methods}
\noindent \textbf{Synthesis of homobilayer WSe$_2$:} Homobilayer WSe$_2$ crystals were synthesized on thermally oxidized silicon substrates (with a SiO$_2$ thickness of $285$~nm) using WO$_2$ and Se powders (99.99\%, Sigma-Aldrich) as precursors and the vapor phase chalcogenization method to obtain high-quality crystalline samples \citep{Bilgin2015,Bilgin2018} with addition of NaCl \citep{Li2015} for higher yield. A three-zone furnace CVD system (Carbolite Gero) equipped with a 1-inch quartz tube was used for the growth. An alumina boat containing mixed powder of WO$_2$ ($40$~mg) and NaCl ($5$~mg) was placed at the center of the first zone and a SiO$_2$/Si substrate was located face-down above the WO$_2$ powder. Another crucible boat containing Se was placed $25$~cm upstream from the center of the first zone. After the tube was evacuated to $\sim10$~mTorr several times to remove air and moisture, the reaction chamber pressure was increased to ambient pressure through $500$~sccm argon gas flow. Then, the furnace was heated with a ramping rate of $50^{\circ}$C/min to the target growth temperature and kept there for $5$~min with $140/10$~sccm Ar/H$_2$ gas flow before cooling down. In general, the yield of 3R stacked crystals was highest at $960^{\circ}$C, while 2H stacking was obtained at higher temperatures.\\

\noindent \textbf{Sample fabrication:} A PDMS/PC stamp was used to sequentially pick up the exfoliated hBN layers (NIMS) and CVD-grown homobilayer WSe$_2$ crystals using the dry transfer method \cite{Pizzocchero2016,Purdie2018}. Poly-(Bisphenol A-carbonate) pellets (Sigma Aldrich) were dissolved in chloroform with a weight ratio of $8$. The mixture was stirred at $500$~rpm using a magneton bar at room temperature overnight. The well-dissolved PC film was mounted on a PDMS block on a glass slide. First, the top hBN layer with a thickness of $163$~nm for 2H ($161$~nm for 3R) was picked up with the stamp, followed by 2H or 3R WSe$_2$ BL and the bottom hBN layer with a thickness of $67$~nm for 2H ($66$~nm for 3R). The pick-up temperatures for the hBN flakes and WSe$_2$ BLs were around $50^{\circ}$C and $140^{\circ}$C, respectively. The entire stack was released at a temperature of $180^{\circ}$C onto a Si/SiO$_2$ target substrate, then soaked in chloroform solution for $20$~min to remove PC residues, cleaned by acetone and isopropanol and annealed at $200^{\circ}$C under ultrahigh vacuum for $15$~hours.\\

\noindent \textbf{Optical spectroscopy:}
Raman spectra were recorded at room temperature under ambient conditions using a Raman system (T64000, Horiba) with $40~\mu$W laser excitation at $532$~nm and a spectral resolution of $0.3$~cm$^{-1}$ with a $2400$~grooves/mm grating. Low-temperature PL and DR measurements were performed with a micro-spectroscopy set-up assembled around a closed-cycle magneto-cryostat (attocube systems, attoDRY1000) equipped with nanopositioners (attocube systems, ANP101 series), a low-temperature apochromatic micro-objective (attocube systems, LT-APO/VISIR/0.82) and a bi-directional solenoid for magnetic fields of up to $9$~T in Faraday configuration. A halogen lamp (HL 2000, Ocean Optics) or pulsed supercontinuum laser (NKT, SuperK Extreme) was used for reflectivity measurements. Differential reflectivity spectra were obtained by normalizing the reflected light intensity from the hBN encapsulated homobilayer region to that from a bare hBN region. For PL measurements, the sample was excited at powers ranging from $10$ to $50~\mu$W either by a continuous-wave laser diode at $670$~nm or pulsed supercontinuum laser with a $10$~nm bandwidth around $670$~nm. The spectra were detected with a CCD cooled by liquid nitrogen (Roper Scientific, Spec 10:100BR/LN) dispersed with a monochromator (Roper Scientific, Acton SP2500).\\

\noindent \textbf{Theoretical calculations:}
The atomic structures, the quasi-particle band structures and optical spectra have been obtained using the VASP package \cite{Kresse:1993a,Kresse:1996a}. Core electrons have been treated by the plane-augmented wave scheme \cite{Blochl1994,Mostofi2008}. A lattice parameter value of $3.32~\AA$ has been set for all calculation runs. A grid of $15 \times 15 \times 1$ k-points has been used, in conjunction with a vacuum height of $21.9~\AA$, for all the calculation cells. The geometry's optimization process has been performed at the PBE-D3 level~\cite{Grimme2010} in order to include van der Waals interaction between layers. All the atoms were allowed to relax with a force convergence criterion below $0.005$~eV/\AA. Heyd-Scuseria-Ernzerhof (HSE) hybrid functional~\cite{Heyd2004,Heyd2005,Paier2006} has been used as approximation of the exchange-correlation electronic term, including SOC, to determine eigenvalues and wave functions as input for g-factor calculations
and the full-frequency-dependent $GW$ calculations~\cite{Shishkin2006} performed at the $G_0W_0$ level. An energy cutoff of $400$~eV and a gaussian smearing of $0.05$~eV width have been chosen for partial occupancies, when a tight electronic minimization tolerance of $10^{-8}$~eV was set to determine with a good precision the corresponding derivative of the orbitals with respect to $k$ needed in quasi-particle band structure calculations. The total number of states included in the $GW$ procedure is set to $1280$, in conjunction with an energy cutoff of $100$~eV for the response function, after a careful check of the direct band gap convergence (smaller than $0.1$~eV as a function of k-points sampling). Band structures have been obtained after a Wannier interpolation procedure performed by the WANNIER90 program~\cite{Mostofi2008}. All optical excitonic transitions have been calculated by solving the Bethe-Salpeter equation~\cite{Hanke1979,Rohlfing1998}, using the twelve highest valence bands and the sixteen lowest conduction bands to obtain eigenvalues and oscillator strengths on all systems. From these calculations, we report the absorbance values by using the imaginary part of the complex dielectric function constructed with a broadening of $12$~meV.

\section{Acknowledgements}
We thank P.~Altpeter and C.~Obermayer for assistance in the clean room and the group of T. Weitz for providing access to their Raman spectrometer. The research was funded by the European Research Council (ERC) under the Grant Agreement Number~772195 as well as the Deutsche Forschungsgemeinschaft (DFG, German Research Foundation) within the Priority Programme SPP~2244 2DMP (Project Number 443405595) and the Germany's Excellence Strategy Munich Center for Quantum Science and Technology (MCQST) (EXC-2111-390814868). Z.\,L. acknowledges funding by the China Scholarship Council (CSC) through the Grant Number 201808140196. I.\,B. gratefully acknowledges financial support from the Alexander von Humboldt Foundation. A.\,H. acknowledges support from the Center for NanoScience (CeNS) and the LMUinnovativ project Functional Nanosystems (FuNS). B.\,U. acknowledges funding from ANR IXTASE. I.\,C.\,G. thanks the CALMIP initiative for the generous allocation of computational time, through Project Number p0812, as well as GENCI-CINES and GENCI-IDRIS for Grant Number A010096649. K.\,W. and T.\,T. acknowledge support from the Elemental Strategy Initiative conducted by the MEXT, Japan, Grant Number JPMXP0112101001, JSPS KAKENHI Grant Numbers JP20H00354 and the CREST (JPMJCR15F3), JST.\\

Note: During the preparation of our manuscript we became aware of a related work on stacking-dependent optical properties in bilayer WSe$_2$~\cite{Mccreary2021}.


\begin{thebibliography}{89}%
\makeatletter
\providecommand \@ifxundefined [1]{%
 \@ifx{#1\undefined}
}%
\providecommand \@ifnum [1]{%
 \ifnum #1\expandafter \@firstoftwo
 \else \expandafter \@secondoftwo
 \fi
}%
\providecommand \@ifx [1]{%
 \ifx #1\expandafter \@firstoftwo
 \else \expandafter \@secondoftwo
 \fi
}%
\providecommand \natexlab [1]{#1}%
\providecommand \enquote  [1]{``#1''}%
\providecommand \bibnamefont  [1]{#1}%
\providecommand \bibfnamefont [1]{#1}%
\providecommand \citenamefont [1]{#1}%
\providecommand \href@noop [0]{\@secondoftwo}%
\providecommand \href [0]{\begingroup \@sanitize@url \@href}%
\providecommand \@href[1]{\@@startlink{#1}\@@href}%
\providecommand \@@href[1]{\endgroup#1\@@endlink}%
\providecommand \@sanitize@url [0]{\catcode `\\12\catcode `\$12\catcode
  `\&12\catcode `\#12\catcode `\^12\catcode `\_12\catcode `\%12\relax}%
\providecommand \@@startlink[1]{}%
\providecommand \@@endlink[0]{}%
\providecommand \url  [0]{\begingroup\@sanitize@url \@url }%
\providecommand \@url [1]{\endgroup\@href {#1}{\urlprefix }}%
\providecommand \urlprefix  [0]{URL }%
\providecommand \Eprint [0]{\href }%
\providecommand \doibase [0]{https://doi.org/}%
\providecommand \selectlanguage [0]{\@gobble}%
\providecommand \bibinfo  [0]{\@secondoftwo}%
\providecommand \bibfield  [0]{\@secondoftwo}%
\providecommand \translation [1]{[#1]}%
\providecommand \BibitemOpen [0]{}%
\providecommand \bibitemStop [0]{}%
\providecommand \bibitemNoStop [0]{.\EOS\space}%
\providecommand \EOS [0]{\spacefactor3000\relax}%
\providecommand \BibitemShut  [1]{\csname bibitem#1\endcsname}%
\let\auto@bib@innerbib\@empty
\bibitem [{\citenamefont {Wang}\ \emph
  {et~al.}(2018{\natexlab{a}})\citenamefont {Wang}, \citenamefont {Chernikov},
  \citenamefont {Glazov}, \citenamefont {Heinz}, \citenamefont {Marie},
  \citenamefont {Amand},\ and\ \citenamefont {Urbaszek}}]{wang2018colloquium}%
  \BibitemOpen
  \bibfield  {author} {\bibinfo {author} {\bibfnamefont {G.}~\bibnamefont
  {Wang}}, \bibinfo {author} {\bibfnamefont {A.}~\bibnamefont {Chernikov}},
  \bibinfo {author} {\bibfnamefont {M.~M.}\ \bibnamefont {Glazov}}, \bibinfo
  {author} {\bibfnamefont {T.~F.}\ \bibnamefont {Heinz}}, \bibinfo {author}
  {\bibfnamefont {X.}~\bibnamefont {Marie}}, \bibinfo {author} {\bibfnamefont
  {T.}~\bibnamefont {Amand}},\ and\ \bibinfo {author} {\bibfnamefont
  {B.}~\bibnamefont {Urbaszek}},\ }\bibfield  {title} {\bibinfo {title}
  {Excitons in atomically thin transition metal dichalcogenides},\ }\href
  {https://doi.org/10.1103/RevModPhys.90.021001} {\bibfield  {journal}
  {\bibinfo  {journal} {Rev. Mod. Phys.}\ }\textbf {\bibinfo {volume} {90}},\
  \bibinfo {pages} {3721} (\bibinfo {year} {2018}{\natexlab{a}})}\BibitemShut
  {NoStop}%
\bibitem [{\citenamefont {Wilson}\ \emph {et~al.}(2021)\citenamefont {Wilson},
  \citenamefont {Yao}, \citenamefont {Shan},\ and\ \citenamefont
  {Xu}}]{wilson2021excitons}%
  \BibitemOpen
  \bibfield  {author} {\bibinfo {author} {\bibfnamefont {N.~P.}\ \bibnamefont
  {Wilson}}, \bibinfo {author} {\bibfnamefont {W.}~\bibnamefont {Yao}},
  \bibinfo {author} {\bibfnamefont {J.}~\bibnamefont {Shan}},\ and\ \bibinfo
  {author} {\bibfnamefont {X.}~\bibnamefont {Xu}},\ }\bibfield  {title}
  {\bibinfo {title} {{Excitons and emergent quantum phenomena in stacked 2D
  semiconductors}},\ }\href {https://doi.org/10.1038/s41586-021-03979-1}
  {\bibfield  {journal} {\bibinfo  {journal} {Nature}\ }\textbf {\bibinfo
  {volume} {599}},\ \bibinfo {pages} {383} (\bibinfo {year}
  {2021})}\BibitemShut {NoStop}%
\bibitem [{\citenamefont {Zhao}\ \emph
  {et~al.}(2013{\natexlab{a}})\citenamefont {Zhao}, \citenamefont {Ribeiro},
  \citenamefont {Toh}, \citenamefont {Carvalho}, \citenamefont {Kloc},
  \citenamefont {Castro~Neto},\ and\ \citenamefont {Eda}}]{Zhao2013a}%
  \BibitemOpen
  \bibfield  {author} {\bibinfo {author} {\bibfnamefont {W.}~\bibnamefont
  {Zhao}}, \bibinfo {author} {\bibfnamefont {R.~M.}\ \bibnamefont {Ribeiro}},
  \bibinfo {author} {\bibfnamefont {M.}~\bibnamefont {Toh}}, \bibinfo {author}
  {\bibfnamefont {A.}~\bibnamefont {Carvalho}}, \bibinfo {author}
  {\bibfnamefont {C.}~\bibnamefont {Kloc}}, \bibinfo {author} {\bibfnamefont
  {A.~H.}\ \bibnamefont {Castro~Neto}},\ and\ \bibinfo {author} {\bibfnamefont
  {G.}~\bibnamefont {Eda}},\ }\bibfield  {title} {\bibinfo {title} {{Origin of
  Indirect Optical Transitions in Few-Layer MoS$_2$, WS$_2$, and WSe$_2$}},\
  }\href {https://doi.org/10.1021/nl403270k} {\bibfield  {journal} {\bibinfo
  {journal} {Nano Lett.}\ }\textbf {\bibinfo {volume} {13}},\ \bibinfo {pages}
  {5627} (\bibinfo {year} {2013}{\natexlab{a}})}\BibitemShut {NoStop}%
\bibitem [{\citenamefont {Arora}\ \emph {et~al.}(2017)\citenamefont {Arora},
  \citenamefont {Dr{\ifmmode\ddot{u}\else\"{u}\fi}ppel}, \citenamefont
  {Schmidt}, \citenamefont {Deilmann}, \citenamefont {Schneider}, \citenamefont
  {Molas}, \citenamefont {Marauhn}, \citenamefont {Michaelis~de Vasconcellos},
  \citenamefont {Potemski}, \citenamefont {Rohlfing},\ and\ \citenamefont
  {Bratschitsch}}]{Arora2017}%
  \BibitemOpen
  \bibfield  {author} {\bibinfo {author} {\bibfnamefont {A.}~\bibnamefont
  {Arora}}, \bibinfo {author} {\bibfnamefont {M.}~\bibnamefont
  {Dr{\ifmmode\ddot{u}\else\"{u}\fi}ppel}}, \bibinfo {author} {\bibfnamefont
  {R.}~\bibnamefont {Schmidt}}, \bibinfo {author} {\bibfnamefont
  {T.}~\bibnamefont {Deilmann}}, \bibinfo {author} {\bibfnamefont
  {R.}~\bibnamefont {Schneider}}, \bibinfo {author} {\bibfnamefont {M.~R.}\
  \bibnamefont {Molas}}, \bibinfo {author} {\bibfnamefont {P.}~\bibnamefont
  {Marauhn}}, \bibinfo {author} {\bibfnamefont {S.}~\bibnamefont {Michaelis~de
  Vasconcellos}}, \bibinfo {author} {\bibfnamefont {M.}~\bibnamefont
  {Potemski}}, \bibinfo {author} {\bibfnamefont {M.}~\bibnamefont {Rohlfing}},\
  and\ \bibinfo {author} {\bibfnamefont {R.}~\bibnamefont {Bratschitsch}},\
  }\bibfield  {title} {\bibinfo {title} {{Interlayer excitons in a bulk van der
  Waals semiconductor}},\ }\href {https://doi.org/10.1038/s41467-017-00691-5}
  {\bibfield  {journal} {\bibinfo  {journal} {Nat. Commun.}\ }\textbf {\bibinfo
  {volume} {8}},\ \bibinfo {pages} {1} (\bibinfo {year} {2017})}\BibitemShut
  {NoStop}%
\bibitem [{\citenamefont {Molas}\ \emph {et~al.}(2017)\citenamefont {Molas},
  \citenamefont {Nogajewski}, \citenamefont {Slobodeniuk}, \citenamefont
  {Binder}, \citenamefont {Bartos},\ and\ \citenamefont
  {Potemski}}]{Molas2017}%
  \BibitemOpen
  \bibfield  {author} {\bibinfo {author} {\bibfnamefont {M.~R.}\ \bibnamefont
  {Molas}}, \bibinfo {author} {\bibfnamefont {K.}~\bibnamefont {Nogajewski}},
  \bibinfo {author} {\bibfnamefont {A.~O.}\ \bibnamefont {Slobodeniuk}},
  \bibinfo {author} {\bibfnamefont {J.}~\bibnamefont {Binder}}, \bibinfo
  {author} {\bibfnamefont {M.}~\bibnamefont {Bartos}},\ and\ \bibinfo {author}
  {\bibfnamefont {M.}~\bibnamefont {Potemski}},\ }\bibfield  {title} {\bibinfo
  {title} {The optical response of monolayer{,} few-layer and bulk tungsten
  disulfide},\ }\href {https://doi.org/10.1039/C7NR04672C} {\bibfield
  {journal} {\bibinfo  {journal} {Nanoscale}\ }\textbf {\bibinfo {volume}
  {9}},\ \bibinfo {pages} {13128} (\bibinfo {year} {2017})}\BibitemShut
  {NoStop}%
\bibitem [{\citenamefont {Wang}\ \emph {et~al.}(2020)\citenamefont {Wang},
  \citenamefont {Shih}, \citenamefont {Ghiotto}, \citenamefont {Xian},
  \citenamefont {Rhodes}, \citenamefont {Tan}, \citenamefont {Claassen},
  \citenamefont {Kennes}, \citenamefont {Bai}, \citenamefont {Kim},
  \citenamefont {Watanabe}, \citenamefont {Taniguchi}, \citenamefont {Zhu},
  \citenamefont {Hone}, \citenamefont {Rubio}, \citenamefont {Pasupathy},\ and\
  \citenamefont {Dean}}]{Wang2020}%
  \BibitemOpen
  \bibfield  {author} {\bibinfo {author} {\bibfnamefont {L.}~\bibnamefont
  {Wang}}, \bibinfo {author} {\bibfnamefont {E.-M.}\ \bibnamefont {Shih}},
  \bibinfo {author} {\bibfnamefont {A.}~\bibnamefont {Ghiotto}}, \bibinfo
  {author} {\bibfnamefont {L.}~\bibnamefont {Xian}}, \bibinfo {author}
  {\bibfnamefont {D.~A.}\ \bibnamefont {Rhodes}}, \bibinfo {author}
  {\bibfnamefont {C.}~\bibnamefont {Tan}}, \bibinfo {author} {\bibfnamefont
  {M.}~\bibnamefont {Claassen}}, \bibinfo {author} {\bibfnamefont {D.~M.}\
  \bibnamefont {Kennes}}, \bibinfo {author} {\bibfnamefont {Y.}~\bibnamefont
  {Bai}}, \bibinfo {author} {\bibfnamefont {B.}~\bibnamefont {Kim}}, \bibinfo
  {author} {\bibfnamefont {K.}~\bibnamefont {Watanabe}}, \bibinfo {author}
  {\bibfnamefont {T.}~\bibnamefont {Taniguchi}}, \bibinfo {author}
  {\bibfnamefont {X.}~\bibnamefont {Zhu}}, \bibinfo {author} {\bibfnamefont
  {J.}~\bibnamefont {Hone}}, \bibinfo {author} {\bibfnamefont {A.}~\bibnamefont
  {Rubio}}, \bibinfo {author} {\bibfnamefont {A.~N.}\ \bibnamefont
  {Pasupathy}},\ and\ \bibinfo {author} {\bibfnamefont {C.~R.}\ \bibnamefont
  {Dean}},\ }\bibfield  {title} {\bibinfo {title} {{Correlated electronic
  phases in twisted bilayer transition metal dichalcogenides}},\ }\href
  {https://doi.org/10.1038/s41563-020-0708-6} {\bibfield  {journal} {\bibinfo
  {journal} {Nat. Mater.}\ }\textbf {\bibinfo {volume} {19}},\ \bibinfo {pages}
  {861} (\bibinfo {year} {2020})}\BibitemShut {NoStop}%
\bibitem [{\citenamefont {Brem}\ \emph {et~al.}(2020)\citenamefont {Brem},
  \citenamefont {Lin}, \citenamefont {Gillen}, \citenamefont {Bauer},
  \citenamefont {Maultzsch}, \citenamefont {Lupton},\ and\ \citenamefont
  {Malic}}]{Brem2020}%
  \BibitemOpen
  \bibfield  {author} {\bibinfo {author} {\bibfnamefont {S.}~\bibnamefont
  {Brem}}, \bibinfo {author} {\bibfnamefont {K.-Q.}\ \bibnamefont {Lin}},
  \bibinfo {author} {\bibfnamefont {R.}~\bibnamefont {Gillen}}, \bibinfo
  {author} {\bibfnamefont {J.~M.}\ \bibnamefont {Bauer}}, \bibinfo {author}
  {\bibfnamefont {J.}~\bibnamefont {Maultzsch}}, \bibinfo {author}
  {\bibfnamefont {J.~M.}\ \bibnamefont {Lupton}},\ and\ \bibinfo {author}
  {\bibfnamefont {E.}~\bibnamefont {Malic}},\ }\bibfield  {title} {\bibinfo
  {title} {{Hybridized intervalley moir{\ifmmode\acute{e}\else\'{e}\fi}
  excitons and flat bands in twisted WSe$_2$ bilayers}},\ }\href
  {https://doi.org/10.1039/D0NR02160A} {\bibfield  {journal} {\bibinfo
  {journal} {Nanoscale}\ }\textbf {\bibinfo {volume} {12}},\ \bibinfo {pages}
  {11088} (\bibinfo {year} {2020})}\BibitemShut {NoStop}%
\bibitem [{\citenamefont {Andersen}\ \emph {et~al.}(2021)\citenamefont
  {Andersen}, \citenamefont {Scuri}, \citenamefont {Sushko}, \citenamefont
  {De~Greve}, \citenamefont {Sung}, \citenamefont {Zhou}, \citenamefont {Wild},
  \citenamefont {Gelly}, \citenamefont {Heo}, \citenamefont
  {B{\ifmmode\acute{e}\else\'{e}\fi}rub{\ifmmode\acute{e}\else\'{e}\fi}},
  \citenamefont {Joe}, \citenamefont {Jauregui}, \citenamefont {Watanabe},
  \citenamefont {Taniguchi}, \citenamefont {Kim}, \citenamefont {Park},\ and\
  \citenamefont {Lukin}}]{Andersen2021}%
  \BibitemOpen
  \bibfield  {author} {\bibinfo {author} {\bibfnamefont {T.~I.}\ \bibnamefont
  {Andersen}}, \bibinfo {author} {\bibfnamefont {G.}~\bibnamefont {Scuri}},
  \bibinfo {author} {\bibfnamefont {A.}~\bibnamefont {Sushko}}, \bibinfo
  {author} {\bibfnamefont {K.}~\bibnamefont {De~Greve}}, \bibinfo {author}
  {\bibfnamefont {J.}~\bibnamefont {Sung}}, \bibinfo {author} {\bibfnamefont
  {Y.}~\bibnamefont {Zhou}}, \bibinfo {author} {\bibfnamefont {D.~S.}\
  \bibnamefont {Wild}}, \bibinfo {author} {\bibfnamefont {R.~J.}\ \bibnamefont
  {Gelly}}, \bibinfo {author} {\bibfnamefont {H.}~\bibnamefont {Heo}}, \bibinfo
  {author} {\bibfnamefont {D.}~\bibnamefont
  {B{\ifmmode\acute{e}\else\'{e}\fi}rub{\ifmmode\acute{e}\else\'{e}\fi}}},
  \bibinfo {author} {\bibfnamefont {A.~Y.}\ \bibnamefont {Joe}}, \bibinfo
  {author} {\bibfnamefont {L.~A.}\ \bibnamefont {Jauregui}}, \bibinfo {author}
  {\bibfnamefont {K.}~\bibnamefont {Watanabe}}, \bibinfo {author}
  {\bibfnamefont {T.}~\bibnamefont {Taniguchi}}, \bibinfo {author}
  {\bibfnamefont {P.}~\bibnamefont {Kim}}, \bibinfo {author} {\bibfnamefont
  {H.}~\bibnamefont {Park}},\ and\ \bibinfo {author} {\bibfnamefont {M.~D.}\
  \bibnamefont {Lukin}},\ }\bibfield  {title} {\bibinfo {title} {{Excitons in a
  reconstructed moir{\ifmmode\acute{e}\else\'{e}\fi} potential in twisted
  WSe$_2$/WSe$_2$ homobilayers}},\ }\href
  {https://doi.org/10.1038/s41563-020-00873-5} {\bibfield  {journal} {\bibinfo
  {journal} {Nat. Mater.}\ }\textbf {\bibinfo {volume} {20}},\ \bibinfo {pages}
  {480} (\bibinfo {year} {2021})}\BibitemShut {NoStop}%
\bibitem [{\citenamefont {Gong}\ \emph {et~al.}(2013)\citenamefont {Gong},
  \citenamefont {Liu}, \citenamefont {Yu}, \citenamefont {Xiao}, \citenamefont
  {Cui}, \citenamefont {Xu},\ and\ \citenamefont {Yao}}]{Gong2013}%
  \BibitemOpen
  \bibfield  {author} {\bibinfo {author} {\bibfnamefont {Z.}~\bibnamefont
  {Gong}}, \bibinfo {author} {\bibfnamefont {G.-B.}\ \bibnamefont {Liu}},
  \bibinfo {author} {\bibfnamefont {H.}~\bibnamefont {Yu}}, \bibinfo {author}
  {\bibfnamefont {D.}~\bibnamefont {Xiao}}, \bibinfo {author} {\bibfnamefont
  {X.}~\bibnamefont {Cui}}, \bibinfo {author} {\bibfnamefont {X.}~\bibnamefont
  {Xu}},\ and\ \bibinfo {author} {\bibfnamefont {W.}~\bibnamefont {Yao}},\
  }\bibfield  {title} {\bibinfo {title} {Magnetoelectric effects and
  valley-controlled spin quantum gates in transition metal dichalcogenide
  bilayers},\ }\href {https://www.nature.com/articles/ncomms3053} {\bibfield
  {journal} {\bibinfo  {journal} {Nat. Commun.}\ }\textbf {\bibinfo {volume}
  {4}},\ \bibinfo {pages} {1} (\bibinfo {year} {2013})}\BibitemShut {NoStop}%
\bibitem [{\citenamefont {Scuri}\ \emph {et~al.}(2020)\citenamefont {Scuri},
  \citenamefont {Andersen}, \citenamefont {Zhou}, \citenamefont {Wild},
  \citenamefont {Sung}, \citenamefont {Gelly}, \citenamefont
  {B{\ifmmode\acute{e}\else\'{e}\fi}rub{\ifmmode\acute{e}\else\'{e}\fi}},
  \citenamefont {Heo}, \citenamefont {Shao}, \citenamefont {Joe}, \citenamefont
  {Mier~Valdivia}, \citenamefont {Taniguchi}, \citenamefont {Watanabe},
  \citenamefont {Lon{\ifmmode\check{c}\else\v{c}\fi}ar}, \citenamefont {Kim},
  \citenamefont {Lukin},\ and\ \citenamefont {Park}}]{Scuri2020}%
  \BibitemOpen
  \bibfield  {author} {\bibinfo {author} {\bibfnamefont {G.}~\bibnamefont
  {Scuri}}, \bibinfo {author} {\bibfnamefont {T.~I.}\ \bibnamefont {Andersen}},
  \bibinfo {author} {\bibfnamefont {Y.}~\bibnamefont {Zhou}}, \bibinfo {author}
  {\bibfnamefont {D.~S.}\ \bibnamefont {Wild}}, \bibinfo {author}
  {\bibfnamefont {J.}~\bibnamefont {Sung}}, \bibinfo {author} {\bibfnamefont
  {R.~J.}\ \bibnamefont {Gelly}}, \bibinfo {author} {\bibfnamefont
  {D.}~\bibnamefont
  {B{\ifmmode\acute{e}\else\'{e}\fi}rub{\ifmmode\acute{e}\else\'{e}\fi}}},
  \bibinfo {author} {\bibfnamefont {H.}~\bibnamefont {Heo}}, \bibinfo {author}
  {\bibfnamefont {L.}~\bibnamefont {Shao}}, \bibinfo {author} {\bibfnamefont
  {A.~Y.}\ \bibnamefont {Joe}}, \bibinfo {author} {\bibfnamefont {A.~M.}\
  \bibnamefont {Mier~Valdivia}}, \bibinfo {author} {\bibfnamefont
  {T.}~\bibnamefont {Taniguchi}}, \bibinfo {author} {\bibfnamefont
  {K.}~\bibnamefont {Watanabe}}, \bibinfo {author} {\bibfnamefont
  {M.}~\bibnamefont {Lon{\ifmmode\check{c}\else\v{c}\fi}ar}}, \bibinfo {author}
  {\bibfnamefont {P.}~\bibnamefont {Kim}}, \bibinfo {author} {\bibfnamefont
  {M.~D.}\ \bibnamefont {Lukin}},\ and\ \bibinfo {author} {\bibfnamefont
  {H.}~\bibnamefont {Park}},\ }\bibfield  {title} {\bibinfo {title}
  {{Electrically Tunable Valley Dynamics in Twisted
  ${\mathrm{WSe}}_{2}/{\mathrm{WSe}}_{2}$ Bilayers}},\ }\href
  {https://doi.org/10.1103/PhysRevLett.124.217403} {\bibfield  {journal}
  {\bibinfo  {journal} {Phys. Rev. Lett.}\ }\textbf {\bibinfo {volume} {124}},\
  \bibinfo {pages} {217403} (\bibinfo {year} {2020})}\BibitemShut {NoStop}%
\bibitem [{\citenamefont {Merkl}\ \emph {et~al.}(2020)\citenamefont {Merkl},
  \citenamefont {Mooshammer}, \citenamefont {Brem}, \citenamefont {Girnghuber},
  \citenamefont {Lin}, \citenamefont {Weigl}, \citenamefont {Liebich},
  \citenamefont {Yong}, \citenamefont {Gillen}, \citenamefont {Maultzsch},
  \citenamefont {Lupton}, \citenamefont {Malic},\ and\ \citenamefont
  {Huber}}]{Merkl2020}%
  \BibitemOpen
  \bibfield  {author} {\bibinfo {author} {\bibfnamefont {P.}~\bibnamefont
  {Merkl}}, \bibinfo {author} {\bibfnamefont {F.}~\bibnamefont {Mooshammer}},
  \bibinfo {author} {\bibfnamefont {S.}~\bibnamefont {Brem}}, \bibinfo {author}
  {\bibfnamefont {A.}~\bibnamefont {Girnghuber}}, \bibinfo {author}
  {\bibfnamefont {K.-Q.}\ \bibnamefont {Lin}}, \bibinfo {author} {\bibfnamefont
  {L.}~\bibnamefont {Weigl}}, \bibinfo {author} {\bibfnamefont
  {M.}~\bibnamefont {Liebich}}, \bibinfo {author} {\bibfnamefont {C.-K.}\
  \bibnamefont {Yong}}, \bibinfo {author} {\bibfnamefont {R.}~\bibnamefont
  {Gillen}}, \bibinfo {author} {\bibfnamefont {J.}~\bibnamefont {Maultzsch}},
  \bibinfo {author} {\bibfnamefont {J.~M.}\ \bibnamefont {Lupton}}, \bibinfo
  {author} {\bibfnamefont {E.}~\bibnamefont {Malic}},\ and\ \bibinfo {author}
  {\bibfnamefont {R.}~\bibnamefont {Huber}},\ }\bibfield  {title} {\bibinfo
  {title} {{Twist-tailoring Coulomb correlations in van der Waals
  homobilayers}},\ }\href {https://doi.org/10.1038/s41467-020-16069-z}
  {\bibfield  {journal} {\bibinfo  {journal} {Nat. Commun.}\ }\textbf {\bibinfo
  {volume} {11}},\ \bibinfo {pages} {1} (\bibinfo {year} {2020})}\BibitemShut
  {NoStop}%
\bibitem [{\citenamefont {Lin}\ \emph {et~al.}(2021)\citenamefont {Lin},
  \citenamefont {Faria~Junior}, \citenamefont {Bauer}, \citenamefont {Peng},
  \citenamefont {Monserrat}, \citenamefont {Gmitra}, \citenamefont {Fabian},
  \citenamefont {Bange},\ and\ \citenamefont {Lupton}}]{Lin2021}%
  \BibitemOpen
  \bibfield  {author} {\bibinfo {author} {\bibfnamefont {K.-Q.}\ \bibnamefont
  {Lin}}, \bibinfo {author} {\bibfnamefont {P.~E.}\ \bibnamefont
  {Faria~Junior}}, \bibinfo {author} {\bibfnamefont {J.~M.}\ \bibnamefont
  {Bauer}}, \bibinfo {author} {\bibfnamefont {B.}~\bibnamefont {Peng}},
  \bibinfo {author} {\bibfnamefont {B.}~\bibnamefont {Monserrat}}, \bibinfo
  {author} {\bibfnamefont {M.}~\bibnamefont {Gmitra}}, \bibinfo {author}
  {\bibfnamefont {J.}~\bibnamefont {Fabian}}, \bibinfo {author} {\bibfnamefont
  {S.}~\bibnamefont {Bange}},\ and\ \bibinfo {author} {\bibfnamefont {J.~M.}\
  \bibnamefont {Lupton}},\ }\bibfield  {title} {\bibinfo {title} {{Twist-angle
  engineering of excitonic quantum interference and optical nonlinearities in
  stacked 2D semiconductors}},\ }\href
  {https://doi.org/10.1038/s41467-021-21547-z} {\bibfield  {journal} {\bibinfo
  {journal} {Nat. Commun.}\ }\textbf {\bibinfo {volume} {12}},\ \bibinfo
  {pages} {1} (\bibinfo {year} {2021})}\BibitemShut {NoStop}%
\bibitem [{\citenamefont {Calman}\ \emph {et~al.}(2018)\citenamefont {Calman},
  \citenamefont {Fogler}, \citenamefont {Butov}, \citenamefont {Hu},
  \citenamefont {Mishchenko},\ and\ \citenamefont {Geim}}]{Calman2018}%
  \BibitemOpen
  \bibfield  {author} {\bibinfo {author} {\bibfnamefont {E.~V.}\ \bibnamefont
  {Calman}}, \bibinfo {author} {\bibfnamefont {M.~M.}\ \bibnamefont {Fogler}},
  \bibinfo {author} {\bibfnamefont {L.~V.}\ \bibnamefont {Butov}}, \bibinfo
  {author} {\bibfnamefont {S.}~\bibnamefont {Hu}}, \bibinfo {author}
  {\bibfnamefont {A.}~\bibnamefont {Mishchenko}},\ and\ \bibinfo {author}
  {\bibfnamefont {A.~K.}\ \bibnamefont {Geim}},\ }\bibfield  {title} {\bibinfo
  {title} {{Indirect excitons in van der Waals heterostructures at room
  temperature}},\ }\href {https://doi.org/10.1038/s41467-018-04293-7}
  {\bibfield  {journal} {\bibinfo  {journal} {Nat. Commun.}\ }\textbf {\bibinfo
  {volume} {9}},\ \bibinfo {pages} {1} (\bibinfo {year} {2018})}\BibitemShut
  {NoStop}%
\bibitem [{\citenamefont {Deilmann}\ and\ \citenamefont
  {Thygesen}(2018)}]{Deilmann2018}%
  \BibitemOpen
  \bibfield  {author} {\bibinfo {author} {\bibfnamefont {T.}~\bibnamefont
  {Deilmann}}\ and\ \bibinfo {author} {\bibfnamefont {K.~S.}\ \bibnamefont
  {Thygesen}},\ }\bibfield  {title} {\bibinfo {title} {Interlayer excitons with
  large optical amplitudes in layered van der waals materials},\ }\href
  {https://doi.org/10.1021/acs.nanolett.8b00438} {\bibfield  {journal}
  {\bibinfo  {journal} {Nano Lett.}\ }\textbf {\bibinfo {volume} {18}},\
  \bibinfo {pages} {2984} (\bibinfo {year} {2018})}\BibitemShut {NoStop}%
\bibitem [{\citenamefont {Gerber}\ \emph {et~al.}(2019)\citenamefont {Gerber},
  \citenamefont {Courtade}, \citenamefont {Shree}, \citenamefont {Robert},
  \citenamefont {Taniguchi}, \citenamefont {Watanabe}, \citenamefont
  {Balocchi}, \citenamefont {Renucci}, \citenamefont {Lagarde}, \citenamefont
  {Marie},\ and\ \citenamefont {Urbaszek}}]{Gerber2019}%
  \BibitemOpen
  \bibfield  {author} {\bibinfo {author} {\bibfnamefont {I.~C.}\ \bibnamefont
  {Gerber}}, \bibinfo {author} {\bibfnamefont {E.}~\bibnamefont {Courtade}},
  \bibinfo {author} {\bibfnamefont {S.}~\bibnamefont {Shree}}, \bibinfo
  {author} {\bibfnamefont {C.}~\bibnamefont {Robert}}, \bibinfo {author}
  {\bibfnamefont {T.}~\bibnamefont {Taniguchi}}, \bibinfo {author}
  {\bibfnamefont {K.}~\bibnamefont {Watanabe}}, \bibinfo {author}
  {\bibfnamefont {A.}~\bibnamefont {Balocchi}}, \bibinfo {author}
  {\bibfnamefont {P.}~\bibnamefont {Renucci}}, \bibinfo {author} {\bibfnamefont
  {D.}~\bibnamefont {Lagarde}}, \bibinfo {author} {\bibfnamefont
  {X.}~\bibnamefont {Marie}},\ and\ \bibinfo {author} {\bibfnamefont
  {B.}~\bibnamefont {Urbaszek}},\ }\bibfield  {title} {\bibinfo {title}
  {Interlayer excitons in bilayer {MoS}$_{2}$ with strong oscillator strength
  up to room temperature},\ }\href {https://doi.org/10.1103/PhysRevB.99.035443}
  {\bibfield  {journal} {\bibinfo  {journal} {Phys. Rev. B}\ }\textbf {\bibinfo
  {volume} {99}},\ \bibinfo {pages} {035443} (\bibinfo {year}
  {2019})}\BibitemShut {NoStop}%
\bibitem [{\citenamefont {Wang}\ \emph
  {et~al.}(2018{\natexlab{b}})\citenamefont {Wang}, \citenamefont {Chiu},
  \citenamefont {Honz}, \citenamefont {Mak},\ and\ \citenamefont
  {Shan}}]{WangMak2018}%
  \BibitemOpen
  \bibfield  {author} {\bibinfo {author} {\bibfnamefont {Z.}~\bibnamefont
  {Wang}}, \bibinfo {author} {\bibfnamefont {Y.-H.}\ \bibnamefont {Chiu}},
  \bibinfo {author} {\bibfnamefont {K.}~\bibnamefont {Honz}}, \bibinfo {author}
  {\bibfnamefont {K.~F.}\ \bibnamefont {Mak}},\ and\ \bibinfo {author}
  {\bibfnamefont {J.}~\bibnamefont {Shan}},\ }\bibfield  {title} {\bibinfo
  {title} {{Electrical Tuning of Interlayer Exciton Gases in WSe$_2$
  Bilayers}},\ }\href {https://doi.org/10.1021/acs.nanolett.7b03667} {\bibfield
   {journal} {\bibinfo  {journal} {Nano Lett.}\ }\textbf {\bibinfo {volume}
  {18}},\ \bibinfo {pages} {137} (\bibinfo {year}
  {2018}{\natexlab{b}})}\BibitemShut {NoStop}%
\bibitem [{\citenamefont {Leisgang}\ \emph {et~al.}(2020)\citenamefont
  {Leisgang}, \citenamefont {Shree}, \citenamefont {Paradisanos}, \citenamefont
  {Sponfeldner}, \citenamefont {Robert}, \citenamefont {Lagarde}, \citenamefont
  {Balocchi}, \citenamefont {Watanabe}, \citenamefont {Taniguchi},
  \citenamefont {Marie}, \citenamefont {Warburton}, \citenamefont {Gerber},\
  and\ \citenamefont {Urbaszek}}]{Leisgang2020}%
  \BibitemOpen
  \bibfield  {author} {\bibinfo {author} {\bibfnamefont {N.}~\bibnamefont
  {Leisgang}}, \bibinfo {author} {\bibfnamefont {S.}~\bibnamefont {Shree}},
  \bibinfo {author} {\bibfnamefont {I.}~\bibnamefont {Paradisanos}}, \bibinfo
  {author} {\bibfnamefont {L.}~\bibnamefont {Sponfeldner}}, \bibinfo {author}
  {\bibfnamefont {C.}~\bibnamefont {Robert}}, \bibinfo {author} {\bibfnamefont
  {D.}~\bibnamefont {Lagarde}}, \bibinfo {author} {\bibfnamefont
  {A.}~\bibnamefont {Balocchi}}, \bibinfo {author} {\bibfnamefont
  {K.}~\bibnamefont {Watanabe}}, \bibinfo {author} {\bibfnamefont
  {T.}~\bibnamefont {Taniguchi}}, \bibinfo {author} {\bibfnamefont
  {X.}~\bibnamefont {Marie}}, \bibinfo {author} {\bibfnamefont {R.~J.}\
  \bibnamefont {Warburton}}, \bibinfo {author} {\bibfnamefont {I.~C.}\
  \bibnamefont {Gerber}},\ and\ \bibinfo {author} {\bibfnamefont
  {B.}~\bibnamefont {Urbaszek}},\ }\bibfield  {title} {\bibinfo {title} {{Giant
  Stark splitting of an exciton in bilayer MoS$_2$}},\ }\href
  {https://doi.org/10.1038/s41565-020-0750-1} {\bibfield  {journal} {\bibinfo
  {journal} {Nat. Nanotechnol.}\ }\textbf {\bibinfo {volume} {15}},\ \bibinfo
  {pages} {901} (\bibinfo {year} {2020})}\BibitemShut {NoStop}%
\bibitem [{\citenamefont {Peimyoo}\ \emph {et~al.}(2021)\citenamefont
  {Peimyoo}, \citenamefont {Deilmann}, \citenamefont {Withers}, \citenamefont
  {Escolar}, \citenamefont {Nutting}, \citenamefont {Taniguchi}, \citenamefont
  {Watanabe}, \citenamefont {Taghizadeh}, \citenamefont {Craciun},
  \citenamefont {Thygesen},\ and\ \citenamefont {Russo}}]{Peimyoo2021}%
  \BibitemOpen
  \bibfield  {author} {\bibinfo {author} {\bibfnamefont {N.}~\bibnamefont
  {Peimyoo}}, \bibinfo {author} {\bibfnamefont {T.}~\bibnamefont {Deilmann}},
  \bibinfo {author} {\bibfnamefont {F.}~\bibnamefont {Withers}}, \bibinfo
  {author} {\bibfnamefont {J.}~\bibnamefont {Escolar}}, \bibinfo {author}
  {\bibfnamefont {D.}~\bibnamefont {Nutting}}, \bibinfo {author} {\bibfnamefont
  {T.}~\bibnamefont {Taniguchi}}, \bibinfo {author} {\bibfnamefont
  {K.}~\bibnamefont {Watanabe}}, \bibinfo {author} {\bibfnamefont
  {A.}~\bibnamefont {Taghizadeh}}, \bibinfo {author} {\bibfnamefont {M.~F.}\
  \bibnamefont {Craciun}}, \bibinfo {author} {\bibfnamefont {K.~S.}\
  \bibnamefont {Thygesen}},\ and\ \bibinfo {author} {\bibfnamefont
  {S.}~\bibnamefont {Russo}},\ }\bibfield  {title} {\bibinfo {title}
  {{Electrical tuning of optically active interlayer excitons in bilayer
  {MoS}$_2$}},\ }\href {https://doi.org/10.1038/s41565-021-00916-1} {\bibfield
  {journal} {\bibinfo  {journal} {Nat. Nanotechnol.}\ }\textbf {\bibinfo
  {volume} {16}},\ \bibinfo {pages} {888} (\bibinfo {year} {2021})}\BibitemShut
  {NoStop}%
\bibitem [{\citenamefont {Lorchat}\ \emph {et~al.}(2021)\citenamefont
  {Lorchat}, \citenamefont {Selig}, \citenamefont {Katsch}, \citenamefont
  {Yumigeta}, \citenamefont {Tongay}, \citenamefont {Knorr}, \citenamefont
  {Schneider},\ and\ \citenamefont {H\"ofling}}]{Lorchat2021}%
  \BibitemOpen
  \bibfield  {author} {\bibinfo {author} {\bibfnamefont {E.}~\bibnamefont
  {Lorchat}}, \bibinfo {author} {\bibfnamefont {M.}~\bibnamefont {Selig}},
  \bibinfo {author} {\bibfnamefont {F.}~\bibnamefont {Katsch}}, \bibinfo
  {author} {\bibfnamefont {K.}~\bibnamefont {Yumigeta}}, \bibinfo {author}
  {\bibfnamefont {S.}~\bibnamefont {Tongay}}, \bibinfo {author} {\bibfnamefont
  {A.}~\bibnamefont {Knorr}}, \bibinfo {author} {\bibfnamefont
  {C.}~\bibnamefont {Schneider}},\ and\ \bibinfo {author} {\bibfnamefont
  {S.}~\bibnamefont {H\"ofling}},\ }\bibfield  {title} {\bibinfo {title}
  {Excitons in bilayer {MoS}$_{2}$ displaying a colossal electric field
  splitting and tunable magnetic response},\ }\href
  {https://doi.org/10.1103/PhysRevLett.126.037401} {\bibfield  {journal}
  {\bibinfo  {journal} {Phys. Rev. Lett.}\ }\textbf {\bibinfo {volume} {126}},\
  \bibinfo {pages} {037401} (\bibinfo {year} {2021})}\BibitemShut {NoStop}%
\bibitem [{\citenamefont {Altaiary}\ \emph {et~al.}(2021)\citenamefont
  {Altaiary}, \citenamefont {Liu}, \citenamefont {Liang}, \citenamefont
  {Hsiao}, \citenamefont {van Baren}, \citenamefont {Taniguchi}, \citenamefont
  {Watanabe}, \citenamefont {Gabor}, \citenamefont {Chang},\ and\ \citenamefont
  {Lui}}]{Altaiary2021arxiv}%
  \BibitemOpen
  \bibfield  {author} {\bibinfo {author} {\bibfnamefont {M.~M.}\ \bibnamefont
  {Altaiary}}, \bibinfo {author} {\bibfnamefont {E.}~\bibnamefont {Liu}},
  \bibinfo {author} {\bibfnamefont {C.-T.}\ \bibnamefont {Liang}}, \bibinfo
  {author} {\bibfnamefont {F.-C.}\ \bibnamefont {Hsiao}}, \bibinfo {author}
  {\bibfnamefont {J.}~\bibnamefont {van Baren}}, \bibinfo {author}
  {\bibfnamefont {T.}~\bibnamefont {Taniguchi}}, \bibinfo {author}
  {\bibfnamefont {K.}~\bibnamefont {Watanabe}}, \bibinfo {author}
  {\bibfnamefont {N.~M.}\ \bibnamefont {Gabor}}, \bibinfo {author}
  {\bibfnamefont {Y.-C.}\ \bibnamefont {Chang}},\ and\ \bibinfo {author}
  {\bibfnamefont {C.~H.}\ \bibnamefont {Lui}},\ }\bibfield  {title} {\bibinfo
  {title} {{Electric-field-tunable intervalley excitons and phonon replicas in
  bilayer WSe$_2$}},\ }\href {https://arxiv.org/abs/2101.11161v1} {\bibfield
  {journal} {\bibinfo  {journal} {arXiv}\ } (\bibinfo {year} {2021})},\ \Eprint
  {https://arxiv.org/abs/2101.11161} {2101.11161} \BibitemShut {NoStop}%
\bibitem [{\citenamefont {Huang}\ \emph {et~al.}(2021)\citenamefont {Huang},
  \citenamefont {Zhao}, \citenamefont {Bo}, \citenamefont {Chu}, \citenamefont
  {Tian}, \citenamefont {Liu}, \citenamefont {Yuan}, \citenamefont {Wu},
  \citenamefont {Zhao}, \citenamefont {Xian}, \citenamefont {Watanabe},
  \citenamefont {Taniguchi}, \citenamefont {Yang}, \citenamefont {Shi},
  \citenamefont {Du}, \citenamefont {Sun}, \citenamefont {Meng}, \citenamefont
  {Yang},\ and\ \citenamefont {Zhang}}]{Huang2021arxiv}%
  \BibitemOpen
  \bibfield  {author} {\bibinfo {author} {\bibfnamefont {Z.}~\bibnamefont
  {Huang}}, \bibinfo {author} {\bibfnamefont {Y.}~\bibnamefont {Zhao}},
  \bibinfo {author} {\bibfnamefont {T.}~\bibnamefont {Bo}}, \bibinfo {author}
  {\bibfnamefont {Y.}~\bibnamefont {Chu}}, \bibinfo {author} {\bibfnamefont
  {J.}~\bibnamefont {Tian}}, \bibinfo {author} {\bibfnamefont {L.}~\bibnamefont
  {Liu}}, \bibinfo {author} {\bibfnamefont {Y.}~\bibnamefont {Yuan}}, \bibinfo
  {author} {\bibfnamefont {F.}~\bibnamefont {Wu}}, \bibinfo {author}
  {\bibfnamefont {J.}~\bibnamefont {Zhao}}, \bibinfo {author} {\bibfnamefont
  {L.}~\bibnamefont {Xian}}, \bibinfo {author} {\bibfnamefont {K.}~\bibnamefont
  {Watanabe}}, \bibinfo {author} {\bibfnamefont {T.}~\bibnamefont {Taniguchi}},
  \bibinfo {author} {\bibfnamefont {R.}~\bibnamefont {Yang}}, \bibinfo {author}
  {\bibfnamefont {D.}~\bibnamefont {Shi}}, \bibinfo {author} {\bibfnamefont
  {L.}~\bibnamefont {Du}}, \bibinfo {author} {\bibfnamefont {Z.}~\bibnamefont
  {Sun}}, \bibinfo {author} {\bibfnamefont {S.}~\bibnamefont {Meng}}, \bibinfo
  {author} {\bibfnamefont {W.}~\bibnamefont {Yang}},\ and\ \bibinfo {author}
  {\bibfnamefont {G.}~\bibnamefont {Zhang}},\ }\bibfield  {title} {\bibinfo
  {title} {{Spatially indirect intervalley excitons in bilayer WSe$_2$}},\
  }\href {https://arxiv.org/abs/2108.09129v1} {\bibfield  {journal} {\bibinfo
  {journal} {arXiv}\ } (\bibinfo {year} {2021})},\ \Eprint
  {https://arxiv.org/abs/2108.09129} {2108.09129} \BibitemShut {NoStop}%
\bibitem [{\citenamefont {Yuanda}\ \emph {et~al.}(2021)\citenamefont {Yuanda},
  \citenamefont {Kevin}, \citenamefont {Qinghai}, \citenamefont {Timothy},
  \citenamefont {Novoselov~Kostya},\ and\ \citenamefont {Weibo}}]{Liu2021}%
  \BibitemOpen
  \bibfield  {author} {\bibinfo {author} {\bibfnamefont {L.}~\bibnamefont
  {Yuanda}}, \bibinfo {author} {\bibfnamefont {D.}~\bibnamefont {Kevin}},
  \bibinfo {author} {\bibfnamefont {T.}~\bibnamefont {Qinghai}}, \bibinfo
  {author} {\bibfnamefont {L.}~\bibnamefont {Timothy}}, \bibinfo {author}
  {\bibfnamefont {S.}~\bibnamefont {Novoselov~Kostya}},\ and\ \bibinfo {author}
  {\bibfnamefont {G.}~\bibnamefont {Weibo}},\ }\bibfield  {title} {\bibinfo
  {title} {Electrically controllable router of interlayer excitons},\ }\href
  {https://doi.org/10.1126/sciadv.aba1830} {\bibfield  {journal} {\bibinfo
  {journal} {Sci. Adv.}\ }\textbf {\bibinfo {volume} {6}},\ \bibinfo {pages}
  {eaba1830} (\bibinfo {year} {2021})}\BibitemShut {NoStop}%
\bibitem [{\citenamefont {Hsu}\ \emph {et~al.}(2019)\citenamefont {Hsu},
  \citenamefont {Lin}, \citenamefont {Lu}, \citenamefont {Lee}, \citenamefont
  {Chu}, \citenamefont {Li}, \citenamefont {Yao}, \citenamefont {Chang},\ and\
  \citenamefont {Shih}}]{Hsu2019}%
  \BibitemOpen
  \bibfield  {author} {\bibinfo {author} {\bibfnamefont {W.-T.}\ \bibnamefont
  {Hsu}}, \bibinfo {author} {\bibfnamefont {B.-H.}\ \bibnamefont {Lin}},
  \bibinfo {author} {\bibfnamefont {L.-S.}\ \bibnamefont {Lu}}, \bibinfo
  {author} {\bibfnamefont {M.-H.}\ \bibnamefont {Lee}}, \bibinfo {author}
  {\bibfnamefont {M.-W.}\ \bibnamefont {Chu}}, \bibinfo {author} {\bibfnamefont
  {L.-J.}\ \bibnamefont {Li}}, \bibinfo {author} {\bibfnamefont
  {W.}~\bibnamefont {Yao}}, \bibinfo {author} {\bibfnamefont {W.-H.}\
  \bibnamefont {Chang}},\ and\ \bibinfo {author} {\bibfnamefont {C.-K.}\
  \bibnamefont {Shih}},\ }\bibfield  {title} {\bibinfo {title} {Tailoring
  excitonic states of van der waals bilayers through stacking configuration,
  band alignment, and valley spin},\ }\href
  {https://www.science.org/doi/10.1126/sciadv.aax7407} {\bibfield  {journal}
  {\bibinfo  {journal} {Sci. Adv.}\ }\textbf {\bibinfo {volume} {5}},\ \bibinfo
  {pages} {eaax7407} (\bibinfo {year} {2019})}\BibitemShut {NoStop}%
\bibitem [{\citenamefont {Paradisanos}\ \emph {et~al.}(2020)\citenamefont
  {Paradisanos}, \citenamefont {Shree}, \citenamefont {George}, \citenamefont
  {Leisgang}, \citenamefont {Robert}, \citenamefont {Watanabe}, \citenamefont
  {Taniguchi}, \citenamefont {Warburton}, \citenamefont {Turchanin},
  \citenamefont {Marie}, \citenamefont {Gerber},\ and\ \citenamefont
  {Urbaszek}}]{Paradisanos2020}%
  \BibitemOpen
  \bibfield  {author} {\bibinfo {author} {\bibfnamefont {I.}~\bibnamefont
  {Paradisanos}}, \bibinfo {author} {\bibfnamefont {S.}~\bibnamefont {Shree}},
  \bibinfo {author} {\bibfnamefont {A.}~\bibnamefont {George}}, \bibinfo
  {author} {\bibfnamefont {N.}~\bibnamefont {Leisgang}}, \bibinfo {author}
  {\bibfnamefont {C.}~\bibnamefont {Robert}}, \bibinfo {author} {\bibfnamefont
  {K.}~\bibnamefont {Watanabe}}, \bibinfo {author} {\bibfnamefont
  {T.}~\bibnamefont {Taniguchi}}, \bibinfo {author} {\bibfnamefont {R.~J.}\
  \bibnamefont {Warburton}}, \bibinfo {author} {\bibfnamefont {A.}~\bibnamefont
  {Turchanin}}, \bibinfo {author} {\bibfnamefont {X.}~\bibnamefont {Marie}},
  \bibinfo {author} {\bibfnamefont {I.~C.}\ \bibnamefont {Gerber}},\ and\
  \bibinfo {author} {\bibfnamefont {B.}~\bibnamefont {Urbaszek}},\ }\bibfield
  {title} {\bibinfo {title} {Controlling interlayer excitons in {MoS}$_2$
  layers grown by chemical vapor deposition},\ }\href
  {https://doi.org/10.1038/s41467-020-16023-z} {\bibfield  {journal} {\bibinfo
  {journal} {Nat. Commun.}\ }\textbf {\bibinfo {volume} {11}},\ \bibinfo
  {pages} {2391} (\bibinfo {year} {2020})}\BibitemShut {NoStop}%
\bibitem [{\citenamefont {Mak}\ \emph {et~al.}(2010)\citenamefont {Mak},
  \citenamefont {Lee}, \citenamefont {Hone}, \citenamefont {Shan},\ and\
  \citenamefont {Heinz}}]{Mak2010}%
  \BibitemOpen
  \bibfield  {author} {\bibinfo {author} {\bibfnamefont {K.~F.}\ \bibnamefont
  {Mak}}, \bibinfo {author} {\bibfnamefont {C.}~\bibnamefont {Lee}}, \bibinfo
  {author} {\bibfnamefont {J.}~\bibnamefont {Hone}}, \bibinfo {author}
  {\bibfnamefont {J.}~\bibnamefont {Shan}},\ and\ \bibinfo {author}
  {\bibfnamefont {T.~F.}\ \bibnamefont {Heinz}},\ }\bibfield  {title} {\bibinfo
  {title} {Atomically thin {MoS$_{2}$}: A new direct-gap semiconductor},\
  }\href {https://doi.org/10.1103/PhysRevLett.105.136805} {\bibfield  {journal}
  {\bibinfo  {journal} {Phys. Rev. Lett.}\ }\textbf {\bibinfo {volume} {105}},\
  \bibinfo {pages} {136805} (\bibinfo {year} {2010})}\BibitemShut {NoStop}%
\bibitem [{\citenamefont {Splendiani}\ \emph {et~al.}(2010)\citenamefont
  {Splendiani}, \citenamefont {Sun}, \citenamefont {Zhang}, \citenamefont {Li},
  \citenamefont {Kim}, \citenamefont {Chim}, \citenamefont {Galli},\ and\
  \citenamefont {Wang}}]{Splendiani2010}%
  \BibitemOpen
  \bibfield  {author} {\bibinfo {author} {\bibfnamefont {A.}~\bibnamefont
  {Splendiani}}, \bibinfo {author} {\bibfnamefont {L.}~\bibnamefont {Sun}},
  \bibinfo {author} {\bibfnamefont {Y.}~\bibnamefont {Zhang}}, \bibinfo
  {author} {\bibfnamefont {T.}~\bibnamefont {Li}}, \bibinfo {author}
  {\bibfnamefont {J.}~\bibnamefont {Kim}}, \bibinfo {author} {\bibfnamefont
  {C.-Y.}\ \bibnamefont {Chim}}, \bibinfo {author} {\bibfnamefont
  {G.}~\bibnamefont {Galli}},\ and\ \bibinfo {author} {\bibfnamefont
  {F.}~\bibnamefont {Wang}},\ }\bibfield  {title} {\bibinfo {title} {Emerging
  photoluminescence in monolayer {MoS$_{2}$}},\ }\href
  {https://doi.org/10.1021/nl903868w} {\bibfield  {journal} {\bibinfo
  {journal} {Nano Lett.}\ }\textbf {\bibinfo {volume} {10}},\ \bibinfo {pages}
  {1271} (\bibinfo {year} {2010})}\BibitemShut {NoStop}%
\bibitem [{\citenamefont {Lindlau}\ \emph
  {et~al.}(2018{\natexlab{a}})\citenamefont {Lindlau}, \citenamefont {Selig},
  \citenamefont {Neumann}, \citenamefont {Colombier}, \citenamefont
  {F{\ifmmode\ddot{o}\else\"{o}\fi}rste}, \citenamefont {Funk}, \citenamefont
  {F{\ifmmode\ddot{o}\else\"{o}\fi}rg}, \citenamefont {Kim}, \citenamefont
  {Bergh{\ifmmode\ddot{a}\else\"{a}\fi}user}, \citenamefont {Taniguchi},
  \citenamefont {Watanabe}, \citenamefont {Wang}, \citenamefont {Malic},\ and\
  \citenamefont {H{\ifmmode\ddot{o}\else\"{o}\fi}gele}}]{Lindlau2018}%
  \BibitemOpen
  \bibfield  {author} {\bibinfo {author} {\bibfnamefont {J.}~\bibnamefont
  {Lindlau}}, \bibinfo {author} {\bibfnamefont {M.}~\bibnamefont {Selig}},
  \bibinfo {author} {\bibfnamefont {A.}~\bibnamefont {Neumann}}, \bibinfo
  {author} {\bibfnamefont {L.}~\bibnamefont {Colombier}}, \bibinfo {author}
  {\bibfnamefont {J.}~\bibnamefont {F{\ifmmode\ddot{o}\else\"{o}\fi}rste}},
  \bibinfo {author} {\bibfnamefont {V.}~\bibnamefont {Funk}}, \bibinfo {author}
  {\bibfnamefont {M.}~\bibnamefont {F{\ifmmode\ddot{o}\else\"{o}\fi}rg}},
  \bibinfo {author} {\bibfnamefont {J.}~\bibnamefont {Kim}}, \bibinfo {author}
  {\bibfnamefont {G.}~\bibnamefont {Bergh{\ifmmode\ddot{a}\else\"{a}\fi}user}},
  \bibinfo {author} {\bibfnamefont {T.}~\bibnamefont {Taniguchi}}, \bibinfo
  {author} {\bibfnamefont {K.}~\bibnamefont {Watanabe}}, \bibinfo {author}
  {\bibfnamefont {F.}~\bibnamefont {Wang}}, \bibinfo {author} {\bibfnamefont
  {E.}~\bibnamefont {Malic}},\ and\ \bibinfo {author} {\bibfnamefont
  {A.}~\bibnamefont {H{\ifmmode\ddot{o}\else\"{o}\fi}gele}},\ }\bibfield
  {title} {\bibinfo {title} {{The role of momentum-dark excitons in the
  elementary optical response of bilayer WSe$_2$}},\ }\href
  {https://doi.org/10.1038/s41467-018-04877-3} {\bibfield  {journal} {\bibinfo
  {journal} {Nat. Commun.}\ }\textbf {\bibinfo {volume} {9}},\ \bibinfo {pages}
  {1} (\bibinfo {year} {2018}{\natexlab{a}})}\BibitemShut {NoStop}%
\bibitem [{\citenamefont {F\"orste}\ \emph {et~al.}(2020)\citenamefont
  {F\"orste}, \citenamefont {Tepliakov}, \citenamefont {Kruchinin},
  \citenamefont {Lindlau}, \citenamefont {Funk}, \citenamefont
  {F{\ifmmode\ddot{o}\else\"{o}\fi}rg}, \citenamefont {Watanabe}, \citenamefont
  {Taniguchi}, \citenamefont {Baimuratov},\ and\ \citenamefont
  {H\"ogele}}]{Forste2020}%
  \BibitemOpen
  \bibfield  {author} {\bibinfo {author} {\bibfnamefont {J.}~\bibnamefont
  {F\"orste}}, \bibinfo {author} {\bibfnamefont {N.~V.}\ \bibnamefont
  {Tepliakov}}, \bibinfo {author} {\bibfnamefont {S.~{\relax Yu}.}\
  \bibnamefont {Kruchinin}}, \bibinfo {author} {\bibfnamefont {J.}~\bibnamefont
  {Lindlau}}, \bibinfo {author} {\bibfnamefont {V.}~\bibnamefont {Funk}},
  \bibinfo {author} {\bibfnamefont {M.}~\bibnamefont
  {F{\ifmmode\ddot{o}\else\"{o}\fi}rg}}, \bibinfo {author} {\bibfnamefont
  {K.}~\bibnamefont {Watanabe}}, \bibinfo {author} {\bibfnamefont
  {T.}~\bibnamefont {Taniguchi}}, \bibinfo {author} {\bibfnamefont {A.~S.}\
  \bibnamefont {Baimuratov}},\ and\ \bibinfo {author} {\bibfnamefont
  {A.}~\bibnamefont {H\"ogele}},\ }\bibfield  {title} {\bibinfo {title}
  {{Exciton g-factors in monolayer and bilayer WSe$_2$ from experiment and
  theory}},\ }\href {https://doi.org/10.1038/s41467-020-18019-1} {\bibfield
  {journal} {\bibinfo  {journal} {Nat. Commun.}\ }\textbf {\bibinfo {volume}
  {11}},\ \bibinfo {pages} {1} (\bibinfo {year} {2020})}\BibitemShut {NoStop}%
\bibitem [{\citenamefont {Funk}\ \emph {et~al.}(2021)\citenamefont {Funk},
  \citenamefont {Wagner}, \citenamefont {Wietek}, \citenamefont {Ziegler},
  \citenamefont {F\"orste}, \citenamefont {Lindlau}, \citenamefont {F\"org},
  \citenamefont {Watanabe}, \citenamefont {Taniguchi}, \citenamefont
  {Chernikov},\ and\ \citenamefont {H\"ogele}}]{Funk2021}%
  \BibitemOpen
  \bibfield  {author} {\bibinfo {author} {\bibfnamefont {V.}~\bibnamefont
  {Funk}}, \bibinfo {author} {\bibfnamefont {K.}~\bibnamefont {Wagner}},
  \bibinfo {author} {\bibfnamefont {E.}~\bibnamefont {Wietek}}, \bibinfo
  {author} {\bibfnamefont {J.~D.}\ \bibnamefont {Ziegler}}, \bibinfo {author}
  {\bibfnamefont {J.}~\bibnamefont {F\"orste}}, \bibinfo {author}
  {\bibfnamefont {J.}~\bibnamefont {Lindlau}}, \bibinfo {author} {\bibfnamefont
  {M.}~\bibnamefont {F\"org}}, \bibinfo {author} {\bibfnamefont
  {K.}~\bibnamefont {Watanabe}}, \bibinfo {author} {\bibfnamefont
  {T.}~\bibnamefont {Taniguchi}}, \bibinfo {author} {\bibfnamefont
  {A.}~\bibnamefont {Chernikov}},\ and\ \bibinfo {author} {\bibfnamefont
  {A.}~\bibnamefont {H\"ogele}},\ }\bibfield  {title} {\bibinfo {title}
  {Spectral asymmetry of phonon sideband luminescence in monolayer and bilayer
  {WSe}$_{2}$},\ }\href {https://doi.org/10.1103/PhysRevResearch.3.L042019}
  {\bibfield  {journal} {\bibinfo  {journal} {Phys. Rev. Research}\ }\textbf
  {\bibinfo {volume} {3}},\ \bibinfo {pages} {L042019} (\bibinfo {year}
  {2021})}\BibitemShut {NoStop}%
\bibitem [{\citenamefont {Aslan}\ \emph {et~al.}(2020)\citenamefont {Aslan},
  \citenamefont {Deng}, \citenamefont {Brongersma},\ and\ \citenamefont
  {Heinz}}]{Aslan2020}%
  \BibitemOpen
  \bibfield  {author} {\bibinfo {author} {\bibfnamefont {O.~B.}\ \bibnamefont
  {Aslan}}, \bibinfo {author} {\bibfnamefont {M.}~\bibnamefont {Deng}},
  \bibinfo {author} {\bibfnamefont {M.~L.}\ \bibnamefont {Brongersma}},\ and\
  \bibinfo {author} {\bibfnamefont {T.~F.}\ \bibnamefont {Heinz}},\ }\bibfield
  {title} {\bibinfo {title} {{Strained bilayer WSe$_2$ with reduced
  exciton-phonon coupling}},\ }\href
  {https://doi.org/10.1103/PhysRevB.101.115305} {\bibfield  {journal} {\bibinfo
   {journal} {Phys. Rev. B}\ }\textbf {\bibinfo {volume} {101}},\ \bibinfo
  {pages} {115305} (\bibinfo {year} {2020})}\BibitemShut {NoStop}%
\bibitem [{\citenamefont {Liu}\ \emph {et~al.}(2014)\citenamefont {Liu},
  \citenamefont {Zhang}, \citenamefont {Cao}, \citenamefont {Jin},
  \citenamefont {Qiu}, \citenamefont {Zhou}, \citenamefont {Zettl},
  \citenamefont {Yang}, \citenamefont {Louie},\ and\ \citenamefont
  {Wang}}]{Liu2014}%
  \BibitemOpen
  \bibfield  {author} {\bibinfo {author} {\bibfnamefont {K.}~\bibnamefont
  {Liu}}, \bibinfo {author} {\bibfnamefont {L.}~\bibnamefont {Zhang}}, \bibinfo
  {author} {\bibfnamefont {T.}~\bibnamefont {Cao}}, \bibinfo {author}
  {\bibfnamefont {C.}~\bibnamefont {Jin}}, \bibinfo {author} {\bibfnamefont
  {D.}~\bibnamefont {Qiu}}, \bibinfo {author} {\bibfnamefont {Q.}~\bibnamefont
  {Zhou}}, \bibinfo {author} {\bibfnamefont {A.}~\bibnamefont {Zettl}},
  \bibinfo {author} {\bibfnamefont {P.}~\bibnamefont {Yang}}, \bibinfo {author}
  {\bibfnamefont {S.~G.}\ \bibnamefont {Louie}},\ and\ \bibinfo {author}
  {\bibfnamefont {F.}~\bibnamefont {Wang}},\ }\bibfield  {title} {\bibinfo
  {title} {{Evolution of interlayer coupling in twisted molybdenum disulfide
  bilayers}},\ }\href {https://doi.org/10.1038/ncomms5966} {\bibfield
  {journal} {\bibinfo  {journal} {Nat. Commun.}\ }\textbf {\bibinfo {volume}
  {5}},\ \bibinfo {pages} {1} (\bibinfo {year} {2014})}\BibitemShut {NoStop}%
\bibitem [{\citenamefont {Schneider}\ \emph {et~al.}(2019)\citenamefont
  {Schneider}, \citenamefont {Kuhnert}, \citenamefont {Schmitt}, \citenamefont
  {Heimbrodt}, \citenamefont {Huttner}, \citenamefont {Meckbach}, \citenamefont
  {Stroucken}, \citenamefont {Koch}, \citenamefont {Fu}, \citenamefont {Wang},
  \citenamefont {Kang}, \citenamefont {Yang},\ and\ \citenamefont
  {Rahimi-Iman}}]{Schneider2019}%
  \BibitemOpen
  \bibfield  {author} {\bibinfo {author} {\bibfnamefont {L.~M.}\ \bibnamefont
  {Schneider}}, \bibinfo {author} {\bibfnamefont {J.}~\bibnamefont {Kuhnert}},
  \bibinfo {author} {\bibfnamefont {S.}~\bibnamefont {Schmitt}}, \bibinfo
  {author} {\bibfnamefont {W.}~\bibnamefont {Heimbrodt}}, \bibinfo {author}
  {\bibfnamefont {U.}~\bibnamefont {Huttner}}, \bibinfo {author} {\bibfnamefont
  {L.}~\bibnamefont {Meckbach}}, \bibinfo {author} {\bibfnamefont
  {T.}~\bibnamefont {Stroucken}}, \bibinfo {author} {\bibfnamefont {S.~W.}\
  \bibnamefont {Koch}}, \bibinfo {author} {\bibfnamefont {S.}~\bibnamefont
  {Fu}}, \bibinfo {author} {\bibfnamefont {X.}~\bibnamefont {Wang}}, \bibinfo
  {author} {\bibfnamefont {K.}~\bibnamefont {Kang}}, \bibinfo {author}
  {\bibfnamefont {E.-H.}\ \bibnamefont {Yang}},\ and\ \bibinfo {author}
  {\bibfnamefont {A.}~\bibnamefont {Rahimi-Iman}},\ }\bibfield  {title}
  {\bibinfo {title} {{Spin-Layer and Spin-Valley Locking in CVD-Grown AA'- and
  AB-Stacked Tungsten-Disulfide Bilayers}},\ }\href
  {https://doi.org/10.1021/acs.jpcc.9b07213} {\bibfield  {journal} {\bibinfo
  {journal} {J. Phys. Chem. C}\ }\textbf {\bibinfo {volume} {123}},\ \bibinfo
  {pages} {21813} (\bibinfo {year} {2019})}\BibitemShut {NoStop}%
\bibitem [{\citenamefont {Du}\ \emph {et~al.}(2019)\citenamefont {Du},
  \citenamefont {Zhang}, \citenamefont {Zhang}, \citenamefont {Jia},
  \citenamefont {Liang}, \citenamefont {Liu}, \citenamefont {Yang},
  \citenamefont {Shi}, \citenamefont {Xiang}, \citenamefont {Liu},
  \citenamefont {Sun}, \citenamefont {Yao}, \citenamefont {Zhang},\ and\
  \citenamefont {Zhang}}]{Du2019}%
  \BibitemOpen
  \bibfield  {author} {\bibinfo {author} {\bibfnamefont {L.}~\bibnamefont
  {Du}}, \bibinfo {author} {\bibfnamefont {Q.}~\bibnamefont {Zhang}}, \bibinfo
  {author} {\bibfnamefont {T.}~\bibnamefont {Zhang}}, \bibinfo {author}
  {\bibfnamefont {Z.}~\bibnamefont {Jia}}, \bibinfo {author} {\bibfnamefont
  {J.}~\bibnamefont {Liang}}, \bibinfo {author} {\bibfnamefont {G.-B.}\
  \bibnamefont {Liu}}, \bibinfo {author} {\bibfnamefont {R.}~\bibnamefont
  {Yang}}, \bibinfo {author} {\bibfnamefont {D.}~\bibnamefont {Shi}}, \bibinfo
  {author} {\bibfnamefont {J.}~\bibnamefont {Xiang}}, \bibinfo {author}
  {\bibfnamefont {K.}~\bibnamefont {Liu}}, \bibinfo {author} {\bibfnamefont
  {Z.}~\bibnamefont {Sun}}, \bibinfo {author} {\bibfnamefont {Y.}~\bibnamefont
  {Yao}}, \bibinfo {author} {\bibfnamefont {Q.}~\bibnamefont {Zhang}},\ and\
  \bibinfo {author} {\bibfnamefont {G.}~\bibnamefont {Zhang}},\ }\bibfield
  {title} {\bibinfo {title} {{Robust circular polarization of indirect Q-K
  transitions in bilayer $3R\ensuremath{-}\mathrm{W}{\mathrm{S}}_{2}$}},\
  }\href {https://doi.org/10.1103/PhysRevB.100.161404} {\bibfield  {journal}
  {\bibinfo  {journal} {Phys. Rev. B}\ }\textbf {\bibinfo {volume} {100}},\
  \bibinfo {pages} {161404} (\bibinfo {year} {2019})}\BibitemShut {NoStop}%
\bibitem [{\citenamefont {Carr}\ \emph {et~al.}(2018)\citenamefont {Carr},
  \citenamefont {Massatt}, \citenamefont {Torrisi}, \citenamefont {Cazeaux},
  \citenamefont {Luskin},\ and\ \citenamefont {Kaxiras}}]{Carr2018}%
  \BibitemOpen
  \bibfield  {author} {\bibinfo {author} {\bibfnamefont {S.}~\bibnamefont
  {Carr}}, \bibinfo {author} {\bibfnamefont {D.}~\bibnamefont {Massatt}},
  \bibinfo {author} {\bibfnamefont {S.~B.}\ \bibnamefont {Torrisi}}, \bibinfo
  {author} {\bibfnamefont {P.}~\bibnamefont {Cazeaux}}, \bibinfo {author}
  {\bibfnamefont {M.}~\bibnamefont {Luskin}},\ and\ \bibinfo {author}
  {\bibfnamefont {E.}~\bibnamefont {Kaxiras}},\ }\bibfield  {title} {\bibinfo
  {title} {Relaxation and domain formation in incommensurate two-dimensional
  heterostructures},\ }\href {https://doi.org/10.1103/PhysRevB.98.224102}
  {\bibfield  {journal} {\bibinfo  {journal} {Phys. Rev. B}\ }\textbf {\bibinfo
  {volume} {98}},\ \bibinfo {pages} {224102} (\bibinfo {year}
  {2018})}\BibitemShut {NoStop}%
\bibitem [{\citenamefont {Weston}\ \emph {et~al.}(2020)\citenamefont {Weston},
  \citenamefont {Zou}, \citenamefont {Enaldiev}, \citenamefont {Summerfield},
  \citenamefont {Clark}, \citenamefont {Z\'olyomi}, \citenamefont {Graham},
  \citenamefont {Yelgel}, \citenamefont {Magorrian}, \citenamefont {Zhou},
  \citenamefont {Zultak}, \citenamefont {Hopkinson}, \citenamefont {Barinov},
  \citenamefont {Bointon}, \citenamefont {Kretinin}, \citenamefont {Wilson},
  \citenamefont {Beton}, \citenamefont {Fal'ko}, \citenamefont {Haigh},\ and\
  \citenamefont {Gorbachev}}]{Weston2020}%
  \BibitemOpen
  \bibfield  {author} {\bibinfo {author} {\bibfnamefont {A.}~\bibnamefont
  {Weston}}, \bibinfo {author} {\bibfnamefont {Y.}~\bibnamefont {Zou}},
  \bibinfo {author} {\bibfnamefont {V.}~\bibnamefont {Enaldiev}}, \bibinfo
  {author} {\bibfnamefont {A.}~\bibnamefont {Summerfield}}, \bibinfo {author}
  {\bibfnamefont {N.}~\bibnamefont {Clark}}, \bibinfo {author} {\bibfnamefont
  {V.}~\bibnamefont {Z\'olyomi}}, \bibinfo {author} {\bibfnamefont
  {A.}~\bibnamefont {Graham}}, \bibinfo {author} {\bibfnamefont
  {C.}~\bibnamefont {Yelgel}}, \bibinfo {author} {\bibfnamefont
  {S.}~\bibnamefont {Magorrian}}, \bibinfo {author} {\bibfnamefont
  {M.}~\bibnamefont {Zhou}}, \bibinfo {author} {\bibfnamefont {J.}~\bibnamefont
  {Zultak}}, \bibinfo {author} {\bibfnamefont {D.}~\bibnamefont {Hopkinson}},
  \bibinfo {author} {\bibfnamefont {A.}~\bibnamefont {Barinov}}, \bibinfo
  {author} {\bibfnamefont {T.~H.}\ \bibnamefont {Bointon}}, \bibinfo {author}
  {\bibfnamefont {A.}~\bibnamefont {Kretinin}}, \bibinfo {author}
  {\bibfnamefont {N.~R.}\ \bibnamefont {Wilson}}, \bibinfo {author}
  {\bibfnamefont {P.~H.}\ \bibnamefont {Beton}}, \bibinfo {author}
  {\bibfnamefont {V.~I.}\ \bibnamefont {Fal'ko}}, \bibinfo {author}
  {\bibfnamefont {S.~J.}\ \bibnamefont {Haigh}},\ and\ \bibinfo {author}
  {\bibfnamefont {R.}~\bibnamefont {Gorbachev}},\ }\bibfield  {title} {\bibinfo
  {title} {Atomic reconstruction in twisted bilayers of transition metal
  dichalcogenides},\ }\href {https://doi.org/10.1038/s41565-020-0682-9}
  {\bibfield  {journal} {\bibinfo  {journal} {Nat. Nanotechnol.}\ }\textbf
  {\bibinfo {volume} {15}},\ \bibinfo {pages} {592} (\bibinfo {year}
  {2020})}\BibitemShut {NoStop}%
\bibitem [{\citenamefont {Enaldiev}\ \emph {et~al.}(2020)\citenamefont
  {Enaldiev}, \citenamefont {Z\'olyomi}, \citenamefont {Yelgel}, \citenamefont
  {Magorrian},\ and\ \citenamefont {Fal'ko}}]{Enaldiev2020}%
  \BibitemOpen
  \bibfield  {author} {\bibinfo {author} {\bibfnamefont {V.~V.}\ \bibnamefont
  {Enaldiev}}, \bibinfo {author} {\bibfnamefont {V.}~\bibnamefont {Z\'olyomi}},
  \bibinfo {author} {\bibfnamefont {C.}~\bibnamefont {Yelgel}}, \bibinfo
  {author} {\bibfnamefont {S.~J.}\ \bibnamefont {Magorrian}},\ and\ \bibinfo
  {author} {\bibfnamefont {V.~I.}\ \bibnamefont {Fal'ko}},\ }\bibfield
  {title} {\bibinfo {title} {Stacking domains and dislocation networks in
  marginally twisted bilayers of transition metal dichalcogenides},\ }\href
  {https://doi.org/10.1103/PhysRevLett.124.206101} {\bibfield  {journal}
  {\bibinfo  {journal} {Phys. Rev. Lett.}\ }\textbf {\bibinfo {volume} {124}},\
  \bibinfo {pages} {206101} (\bibinfo {year} {2020})}\BibitemShut {NoStop}%
\bibitem [{\citenamefont {Magorrian}\ \emph {et~al.}(2021)\citenamefont
  {Magorrian}, \citenamefont {Enaldiev}, \citenamefont {Z\'olyomi},
  \citenamefont {Ferreira}, \citenamefont {Fal'ko},\ and\ \citenamefont
  {Ruiz-Tijerina}}]{Magorrian2021}%
  \BibitemOpen
  \bibfield  {author} {\bibinfo {author} {\bibfnamefont {S.~J.}\ \bibnamefont
  {Magorrian}}, \bibinfo {author} {\bibfnamefont {V.~V.}\ \bibnamefont
  {Enaldiev}}, \bibinfo {author} {\bibfnamefont {V.}~\bibnamefont {Z\'olyomi}},
  \bibinfo {author} {\bibfnamefont {F.}~\bibnamefont {Ferreira}}, \bibinfo
  {author} {\bibfnamefont {V.~I.}\ \bibnamefont {Fal'ko}},\ and\ \bibinfo
  {author} {\bibfnamefont {D.~A.}\ \bibnamefont {Ruiz-Tijerina}},\ }\bibfield
  {title} {\bibinfo {title} {Multifaceted moir\'e superlattice physics in
  twisted {WSe}$_{2}$ bilayers},\ }\href
  {https://doi.org/10.1103/PhysRevB.104.125440} {\bibfield  {journal} {\bibinfo
   {journal} {Phys. Rev. B}\ }\textbf {\bibinfo {volume} {104}},\ \bibinfo
  {pages} {125440} (\bibinfo {year} {2021})}\BibitemShut {NoStop}%
\bibitem [{\citenamefont {Kenji}\ \emph {et~al.}(2021)\citenamefont {Kenji},
  \citenamefont {Xirui}, \citenamefont {Kenji}, \citenamefont {Takashi},\ and\
  \citenamefont {Pablo}}]{Yasuda2021}%
  \BibitemOpen
  \bibfield  {author} {\bibinfo {author} {\bibfnamefont {Y.}~\bibnamefont
  {Kenji}}, \bibinfo {author} {\bibfnamefont {W.}~\bibnamefont {Xirui}},
  \bibinfo {author} {\bibfnamefont {W.}~\bibnamefont {Kenji}}, \bibinfo
  {author} {\bibfnamefont {T.}~\bibnamefont {Takashi}},\ and\ \bibinfo {author}
  {\bibfnamefont {J.-H.}\ \bibnamefont {Pablo}},\ }\bibfield  {title} {\bibinfo
  {title} {Stacking-engineered ferroelectricity in bilayer boron nitride},\
  }\href {https://doi.org/10.1126/science.abd3230} {\bibfield  {journal}
  {\bibinfo  {journal} {Science}\ }\textbf {\bibinfo {volume} {372}},\ \bibinfo
  {pages} {1458} (\bibinfo {year} {2021})}\BibitemShut {NoStop}%
\bibitem [{\citenamefont {Vizner~Stern}\ \emph {et~al.}(2021)\citenamefont
  {Vizner~Stern}, \citenamefont {Waschitz}, \citenamefont {Cao}, \citenamefont
  {Nevo}, \citenamefont {Watanabe}, \citenamefont {Taniguchi}, \citenamefont
  {Sela}, \citenamefont {Urbakh}, \citenamefont {Hod},\ and\ \citenamefont
  {Ben~Shalom}}]{Vizner2021}%
  \BibitemOpen
  \bibfield  {author} {\bibinfo {author} {\bibfnamefont {M.}~\bibnamefont
  {Vizner~Stern}}, \bibinfo {author} {\bibfnamefont {Y.}~\bibnamefont
  {Waschitz}}, \bibinfo {author} {\bibfnamefont {W.}~\bibnamefont {Cao}},
  \bibinfo {author} {\bibfnamefont {I.}~\bibnamefont {Nevo}}, \bibinfo {author}
  {\bibfnamefont {K.}~\bibnamefont {Watanabe}}, \bibinfo {author}
  {\bibfnamefont {T.}~\bibnamefont {Taniguchi}}, \bibinfo {author}
  {\bibfnamefont {E.}~\bibnamefont {Sela}}, \bibinfo {author} {\bibfnamefont
  {M.}~\bibnamefont {Urbakh}}, \bibinfo {author} {\bibfnamefont
  {O.}~\bibnamefont {Hod}},\ and\ \bibinfo {author} {\bibfnamefont
  {M.}~\bibnamefont {Ben~Shalom}},\ }\bibfield  {title} {\bibinfo {title}
  {Interfacial ferroelectricity by van der {Waals} sliding},\ }\href
  {https://www.science.org/doi/10.1126/science.abe8177} {\bibfield  {journal}
  {\bibinfo  {journal} {Science}\ }\textbf {\bibinfo {volume} {372}},\ \bibinfo
  {pages} {1462} (\bibinfo {year} {2021})}\BibitemShut {NoStop}%
\bibitem [{\citenamefont {Woods}\ \emph {et~al.}(2021)\citenamefont {Woods},
  \citenamefont {Ares}, \citenamefont {Nevison-Andrews}, \citenamefont
  {Holwill}, \citenamefont {Fabregas}, \citenamefont {Guinea}, \citenamefont
  {Geim}, \citenamefont {Novoselov}, \citenamefont {Walet},\ and\ \citenamefont
  {Fumagalli}}]{Woods2021}%
  \BibitemOpen
  \bibfield  {author} {\bibinfo {author} {\bibfnamefont {C.}~\bibnamefont
  {Woods}}, \bibinfo {author} {\bibfnamefont {P.}~\bibnamefont {Ares}},
  \bibinfo {author} {\bibfnamefont {H.}~\bibnamefont {Nevison-Andrews}},
  \bibinfo {author} {\bibfnamefont {M.}~\bibnamefont {Holwill}}, \bibinfo
  {author} {\bibfnamefont {R.}~\bibnamefont {Fabregas}}, \bibinfo {author}
  {\bibfnamefont {F.}~\bibnamefont {Guinea}}, \bibinfo {author} {\bibfnamefont
  {A.}~\bibnamefont {Geim}}, \bibinfo {author} {\bibfnamefont {K.}~\bibnamefont
  {Novoselov}}, \bibinfo {author} {\bibfnamefont {N.}~\bibnamefont {Walet}},\
  and\ \bibinfo {author} {\bibfnamefont {L.}~\bibnamefont {Fumagalli}},\
  }\bibfield  {title} {\bibinfo {title} {Charge-polarized interfacial
  superlattices in marginally twisted hexagonal boron nitride},\ }\href
  {https://www.nature.com/articles/s41467-020-20667-2?proof=t%C2%A0} {\bibfield
   {journal} {\bibinfo  {journal} {Nat. Commun.}\ }\textbf {\bibinfo {volume}
  {12}},\ \bibinfo {pages} {1} (\bibinfo {year} {2021})}\BibitemShut {NoStop}%
\bibitem [{\citenamefont {Ferreira}\ \emph {et~al.}(2021)\citenamefont
  {Ferreira}, \citenamefont {Enaldiev}, \citenamefont {Fal'ko},\ and\
  \citenamefont {Magorrian}}]{Ferreira2021}%
  \BibitemOpen
  \bibfield  {author} {\bibinfo {author} {\bibfnamefont {F.}~\bibnamefont
  {Ferreira}}, \bibinfo {author} {\bibfnamefont {V.~V.}\ \bibnamefont
  {Enaldiev}}, \bibinfo {author} {\bibfnamefont {V.~I.}\ \bibnamefont
  {Fal'ko}},\ and\ \bibinfo {author} {\bibfnamefont {S.~J.}\ \bibnamefont
  {Magorrian}},\ }\bibfield  {title} {\bibinfo {title} {{Weak ferroelectric
  charge transfer in layer-asymmetric bilayers of 2D semiconductors}},\ }\href
  {https://doi.org/10.1038/s41598-021-92710-1} {\bibfield  {journal} {\bibinfo
  {journal} {Sci. Rep.}\ }\textbf {\bibinfo {volume} {11}},\ \bibinfo {pages}
  {13422} (\bibinfo {year} {2021})}\BibitemShut {NoStop}%
\bibitem [{\citenamefont {Weston}\ \emph {et~al.}(2021)\citenamefont {Weston},
  \citenamefont {Castanon}, \citenamefont {Enaldiev}, \citenamefont {Ferreira},
  \citenamefont {Bhattacharjee}, \citenamefont {Xu}, \citenamefont
  {Corte-Leon}, \citenamefont {Wu}, \citenamefont {Clark}, \citenamefont
  {Summerfield}, \citenamefont {Hashimoto}, \citenamefont {Gao}, \citenamefont
  {Wang}, \citenamefont {Hamer}, \citenamefont {Read}, \citenamefont
  {Fumagalli}, \citenamefont {Kretinin}, \citenamefont {Haigh}, \citenamefont
  {Kazakova}, \citenamefont {Geim}, \citenamefont {Fal'ko},\ and\ \citenamefont
  {Gorbachev}}]{Weston2021}%
  \BibitemOpen
  \bibfield  {author} {\bibinfo {author} {\bibfnamefont {A.}~\bibnamefont
  {Weston}}, \bibinfo {author} {\bibfnamefont {E.~G.}\ \bibnamefont
  {Castanon}}, \bibinfo {author} {\bibfnamefont {V.}~\bibnamefont {Enaldiev}},
  \bibinfo {author} {\bibfnamefont {F.}~\bibnamefont {Ferreira}}, \bibinfo
  {author} {\bibfnamefont {S.}~\bibnamefont {Bhattacharjee}}, \bibinfo {author}
  {\bibfnamefont {S.}~\bibnamefont {Xu}}, \bibinfo {author} {\bibfnamefont
  {H.}~\bibnamefont {Corte-Leon}}, \bibinfo {author} {\bibfnamefont
  {Z.}~\bibnamefont {Wu}}, \bibinfo {author} {\bibfnamefont {N.}~\bibnamefont
  {Clark}}, \bibinfo {author} {\bibfnamefont {A.}~\bibnamefont {Summerfield}},
  \bibinfo {author} {\bibfnamefont {T.}~\bibnamefont {Hashimoto}}, \bibinfo
  {author} {\bibfnamefont {Y.}~\bibnamefont {Gao}}, \bibinfo {author}
  {\bibfnamefont {W.}~\bibnamefont {Wang}}, \bibinfo {author} {\bibfnamefont
  {M.}~\bibnamefont {Hamer}}, \bibinfo {author} {\bibfnamefont
  {H.}~\bibnamefont {Read}}, \bibinfo {author} {\bibfnamefont {L.}~\bibnamefont
  {Fumagalli}}, \bibinfo {author} {\bibfnamefont {A.~V.}\ \bibnamefont
  {Kretinin}}, \bibinfo {author} {\bibfnamefont {S.~J.}\ \bibnamefont {Haigh}},
  \bibinfo {author} {\bibfnamefont {O.}~\bibnamefont {Kazakova}}, \bibinfo
  {author} {\bibfnamefont {A.~K.}\ \bibnamefont {Geim}}, \bibinfo {author}
  {\bibfnamefont {V.~I.}\ \bibnamefont {Fal'ko}},\ and\ \bibinfo {author}
  {\bibfnamefont {R.}~\bibnamefont {Gorbachev}},\ }\bibfield  {title} {\bibinfo
  {title} {Interfacial ferroelectricity in marginally twisted {2D}
  semiconductors},\ }\href@noop {} {\  (\bibinfo {year} {2021})},\ \Eprint
  {https://arxiv.org/abs/2108.06489} {arXiv:2108.06489 [cond-mat.mes-hall]}
  \BibitemShut {NoStop}%
\bibitem [{\citenamefont {Li}\ \emph {et~al.}(2013)\citenamefont {Li},
  \citenamefont {Lu}, \citenamefont {Wang}, \citenamefont {Yin}, \citenamefont
  {Cong}, \citenamefont {He}, \citenamefont {Wang}, \citenamefont {Ding},
  \citenamefont {Yu},\ and\ \citenamefont {Zhang}}]{Li2013}%
  \BibitemOpen
  \bibfield  {author} {\bibinfo {author} {\bibfnamefont {H.}~\bibnamefont
  {Li}}, \bibinfo {author} {\bibfnamefont {G.}~\bibnamefont {Lu}}, \bibinfo
  {author} {\bibfnamefont {Y.}~\bibnamefont {Wang}}, \bibinfo {author}
  {\bibfnamefont {Z.}~\bibnamefont {Yin}}, \bibinfo {author} {\bibfnamefont
  {C.}~\bibnamefont {Cong}}, \bibinfo {author} {\bibfnamefont {Q.}~\bibnamefont
  {He}}, \bibinfo {author} {\bibfnamefont {L.}~\bibnamefont {Wang}}, \bibinfo
  {author} {\bibfnamefont {F.}~\bibnamefont {Ding}}, \bibinfo {author}
  {\bibfnamefont {T.}~\bibnamefont {Yu}},\ and\ \bibinfo {author}
  {\bibfnamefont {H.}~\bibnamefont {Zhang}},\ }\bibfield  {title} {\bibinfo
  {title} {{Mechanical Exfoliation and Characterization of Single- and
  Few-Layer Nanosheets of WSe$_2$, TaS$_2$, and TaSe$_2$}},\ }\href
  {https://doi.org/10.1002/smll.201202919} {\bibfield  {journal} {\bibinfo
  {journal} {Small}\ }\textbf {\bibinfo {volume} {9}},\ \bibinfo {pages} {1974}
  (\bibinfo {year} {2013})}\BibitemShut {NoStop}%
\bibitem [{\citenamefont {Zeng}\ \emph {et~al.}(2013)\citenamefont {Zeng},
  \citenamefont {Liu}, \citenamefont {Dai}, \citenamefont {Yan}, \citenamefont
  {Zhu}, \citenamefont {He}, \citenamefont {Xie}, \citenamefont {Xu},
  \citenamefont {Chen}, \citenamefont {Yao},\ and\ \citenamefont
  {Cui}}]{Zeng2013}%
  \BibitemOpen
  \bibfield  {author} {\bibinfo {author} {\bibfnamefont {H.}~\bibnamefont
  {Zeng}}, \bibinfo {author} {\bibfnamefont {G.-B.}\ \bibnamefont {Liu}},
  \bibinfo {author} {\bibfnamefont {J.}~\bibnamefont {Dai}}, \bibinfo {author}
  {\bibfnamefont {Y.}~\bibnamefont {Yan}}, \bibinfo {author} {\bibfnamefont
  {B.}~\bibnamefont {Zhu}}, \bibinfo {author} {\bibfnamefont {R.}~\bibnamefont
  {He}}, \bibinfo {author} {\bibfnamefont {L.}~\bibnamefont {Xie}}, \bibinfo
  {author} {\bibfnamefont {S.}~\bibnamefont {Xu}}, \bibinfo {author}
  {\bibfnamefont {X.}~\bibnamefont {Chen}}, \bibinfo {author} {\bibfnamefont
  {W.}~\bibnamefont {Yao}},\ and\ \bibinfo {author} {\bibfnamefont
  {X.}~\bibnamefont {Cui}},\ }\bibfield  {title} {\bibinfo {title} {{Optical
  signature of symmetry variations and spin-valley coupling in atomically thin
  tungsten dichalcogenides}},\ }\href {https://doi.org/10.1038/srep01608}
  {\bibfield  {journal} {\bibinfo  {journal} {Sci. Rep.}\ }\textbf {\bibinfo
  {volume} {3}},\ \bibinfo {pages} {1} (\bibinfo {year} {2013})}\BibitemShut
  {NoStop}%
\bibitem [{\citenamefont {Sahin}\ \emph {et~al.}(2013)\citenamefont {Sahin},
  \citenamefont {Tongay}, \citenamefont {Horzum}, \citenamefont {Fan},
  \citenamefont {Zhou}, \citenamefont {Li}, \citenamefont {Wu},\ and\
  \citenamefont {Peeters}}]{Sahin2013}%
  \BibitemOpen
  \bibfield  {author} {\bibinfo {author} {\bibfnamefont {H.}~\bibnamefont
  {Sahin}}, \bibinfo {author} {\bibfnamefont {S.}~\bibnamefont {Tongay}},
  \bibinfo {author} {\bibfnamefont {S.}~\bibnamefont {Horzum}}, \bibinfo
  {author} {\bibfnamefont {W.}~\bibnamefont {Fan}}, \bibinfo {author}
  {\bibfnamefont {J.}~\bibnamefont {Zhou}}, \bibinfo {author} {\bibfnamefont
  {J.}~\bibnamefont {Li}}, \bibinfo {author} {\bibfnamefont {J.}~\bibnamefont
  {Wu}},\ and\ \bibinfo {author} {\bibfnamefont {F.~M.}\ \bibnamefont
  {Peeters}},\ }\bibfield  {title} {\bibinfo {title} {{Anomalous Raman spectra
  and thickness-dependent electronic properties of WSe${}_{2}$}},\ }\href
  {https://doi.org/10.1103/PhysRevB.87.165409} {\bibfield  {journal} {\bibinfo
  {journal} {Phys. Rev. B}\ }\textbf {\bibinfo {volume} {87}},\ \bibinfo
  {pages} {165409} (\bibinfo {year} {2013})}\BibitemShut {NoStop}%
\bibitem [{\citenamefont {Zhao}\ \emph
  {et~al.}(2013{\natexlab{b}})\citenamefont {Zhao}, \citenamefont
  {Ghorannevis}, \citenamefont {Amara}, \citenamefont {Pang}, \citenamefont
  {Toh}, \citenamefont {Zhang}, \citenamefont {Kloc}, \citenamefont {Tan},\
  and\ \citenamefont {Eda}}]{Zhao2013}%
  \BibitemOpen
  \bibfield  {author} {\bibinfo {author} {\bibfnamefont {W.}~\bibnamefont
  {Zhao}}, \bibinfo {author} {\bibfnamefont {Z.}~\bibnamefont {Ghorannevis}},
  \bibinfo {author} {\bibfnamefont {K.~K.}\ \bibnamefont {Amara}}, \bibinfo
  {author} {\bibfnamefont {J.~R.}\ \bibnamefont {Pang}}, \bibinfo {author}
  {\bibfnamefont {M.}~\bibnamefont {Toh}}, \bibinfo {author} {\bibfnamefont
  {X.}~\bibnamefont {Zhang}}, \bibinfo {author} {\bibfnamefont
  {C.}~\bibnamefont {Kloc}}, \bibinfo {author} {\bibfnamefont {P.~H.}\
  \bibnamefont {Tan}},\ and\ \bibinfo {author} {\bibfnamefont {G.}~\bibnamefont
  {Eda}},\ }\bibfield  {title} {\bibinfo {title} {{Lattice dynamics in mono-
  and few-layer sheets of WS$_2$ and WSe$_2$}},\ }\href
  {https://doi.org/10.1039/C3NR03052K} {\bibfield  {journal} {\bibinfo
  {journal} {Nanoscale}\ }\textbf {\bibinfo {volume} {5}},\ \bibinfo {pages}
  {9677} (\bibinfo {year} {2013}{\natexlab{b}})}\BibitemShut {NoStop}%
\bibitem [{\citenamefont {Luo}\ \emph {et~al.}(2013)\citenamefont {Luo},
  \citenamefont {Zhao}, \citenamefont {Zhang}, \citenamefont {Toh},
  \citenamefont {Kloc}, \citenamefont {Xiong},\ and\ \citenamefont
  {Quek}}]{Luo2013}%
  \BibitemOpen
  \bibfield  {author} {\bibinfo {author} {\bibfnamefont {X.}~\bibnamefont
  {Luo}}, \bibinfo {author} {\bibfnamefont {Y.}~\bibnamefont {Zhao}}, \bibinfo
  {author} {\bibfnamefont {J.}~\bibnamefont {Zhang}}, \bibinfo {author}
  {\bibfnamefont {M.}~\bibnamefont {Toh}}, \bibinfo {author} {\bibfnamefont
  {C.}~\bibnamefont {Kloc}}, \bibinfo {author} {\bibfnamefont {Q.}~\bibnamefont
  {Xiong}},\ and\ \bibinfo {author} {\bibfnamefont {S.~Y.}\ \bibnamefont
  {Quek}},\ }\bibfield  {title} {\bibinfo {title} {{Effects of lower symmetry
  and dimensionality on Raman spectra in two-dimensional WSe${}_{2}$}},\ }\href
  {https://doi.org/10.1103/PhysRevB.88.195313} {\bibfield  {journal} {\bibinfo
  {journal} {Phys. Rev. B}\ }\textbf {\bibinfo {volume} {88}},\ \bibinfo
  {pages} {195313} (\bibinfo {year} {2013})}\BibitemShut {NoStop}%
\bibitem [{\citenamefont {Terrones}\ \emph {et~al.}(2014)\citenamefont
  {Terrones}, \citenamefont {Corro}, \citenamefont {Feng}, \citenamefont
  {Poumirol}, \citenamefont {Rhodes}, \citenamefont {Smirnov}, \citenamefont
  {Pradhan}, \citenamefont {Lin}, \citenamefont {Nguyen}, \citenamefont
  {Elias}, \citenamefont {Mallouk}, \citenamefont {Balicas}, \citenamefont
  {Pimenta},\ and\ \citenamefont {Terrones}}]{Terrones2014}%
  \BibitemOpen
  \bibfield  {author} {\bibinfo {author} {\bibfnamefont {H.}~\bibnamefont
  {Terrones}}, \bibinfo {author} {\bibfnamefont {E.~D.}\ \bibnamefont {Corro}},
  \bibinfo {author} {\bibfnamefont {S.}~\bibnamefont {Feng}}, \bibinfo {author}
  {\bibfnamefont {J.~M.}\ \bibnamefont {Poumirol}}, \bibinfo {author}
  {\bibfnamefont {D.}~\bibnamefont {Rhodes}}, \bibinfo {author} {\bibfnamefont
  {D.}~\bibnamefont {Smirnov}}, \bibinfo {author} {\bibfnamefont {N.~R.}\
  \bibnamefont {Pradhan}}, \bibinfo {author} {\bibfnamefont {Z.}~\bibnamefont
  {Lin}}, \bibinfo {author} {\bibfnamefont {M.~A.~T.}\ \bibnamefont {Nguyen}},
  \bibinfo {author} {\bibfnamefont {A.~L.}\ \bibnamefont {Elias}}, \bibinfo
  {author} {\bibfnamefont {T.~E.}\ \bibnamefont {Mallouk}}, \bibinfo {author}
  {\bibfnamefont {L.}~\bibnamefont {Balicas}}, \bibinfo {author} {\bibfnamefont
  {M.~A.}\ \bibnamefont {Pimenta}},\ and\ \bibinfo {author} {\bibfnamefont
  {M.}~\bibnamefont {Terrones}},\ }\bibfield  {title} {\bibinfo {title} {{New
  First Order Raman-active Modes in Few Layered Transition Metal
  Dichalcogenides}},\ }\href {https://doi.org/10.1038/srep04215} {\bibfield
  {journal} {\bibinfo  {journal} {Sci. Rep.}\ }\textbf {\bibinfo {volume}
  {4}},\ \bibinfo {pages} {1} (\bibinfo {year} {2014})}\BibitemShut {NoStop}%
\bibitem [{\citenamefont {Puretzky}\ \emph {et~al.}(2015)\citenamefont
  {Puretzky}, \citenamefont {Liang}, \citenamefont {Li}, \citenamefont {Xiao},
  \citenamefont {Wang}, \citenamefont {Mahjouri-Samani}, \citenamefont
  {Basile}, \citenamefont {Idrobo}, \citenamefont {Sumpter}, \citenamefont
  {Meunier},\ and\ \citenamefont {Geohegan}}]{Puretzky2015}%
  \BibitemOpen
  \bibfield  {author} {\bibinfo {author} {\bibfnamefont {A.~A.}\ \bibnamefont
  {Puretzky}}, \bibinfo {author} {\bibfnamefont {L.}~\bibnamefont {Liang}},
  \bibinfo {author} {\bibfnamefont {X.}~\bibnamefont {Li}}, \bibinfo {author}
  {\bibfnamefont {K.}~\bibnamefont {Xiao}}, \bibinfo {author} {\bibfnamefont
  {K.}~\bibnamefont {Wang}}, \bibinfo {author} {\bibfnamefont {M.}~\bibnamefont
  {Mahjouri-Samani}}, \bibinfo {author} {\bibfnamefont {L.}~\bibnamefont
  {Basile}}, \bibinfo {author} {\bibfnamefont {J.~C.}\ \bibnamefont {Idrobo}},
  \bibinfo {author} {\bibfnamefont {B.~G.}\ \bibnamefont {Sumpter}}, \bibinfo
  {author} {\bibfnamefont {V.}~\bibnamefont {Meunier}},\ and\ \bibinfo {author}
  {\bibfnamefont {D.~B.}\ \bibnamefont {Geohegan}},\ }\bibfield  {title}
  {\bibinfo {title} {{Low-Frequency Raman Fingerprints of Two-Dimensional Metal
  Dichalcogenide Layer Stacking Configurations}},\ }\href
  {https://doi.org/10.1021/acsnano.5b01884} {\bibfield  {journal} {\bibinfo
  {journal} {ACS Nano}\ }\textbf {\bibinfo {volume} {9}},\ \bibinfo {pages}
  {6333} (\bibinfo {year} {2015})}\BibitemShut {NoStop}%
\bibitem [{\citenamefont {Cadiz}\ \emph {et~al.}(2017)\citenamefont {Cadiz},
  \citenamefont {Courtade}, \citenamefont {Robert}, \citenamefont {Wang},
  \citenamefont {Shen}, \citenamefont {Cai}, \citenamefont {Taniguchi},
  \citenamefont {Watanabe}, \citenamefont {Carrere}, \citenamefont {Lagarde},
  \citenamefont {Manca}, \citenamefont {Amand}, \citenamefont {Renucci},
  \citenamefont {Tongay}, \citenamefont {Marie},\ and\ \citenamefont
  {Urbaszek}}]{Cadiz2017}%
  \BibitemOpen
  \bibfield  {author} {\bibinfo {author} {\bibfnamefont {F.}~\bibnamefont
  {Cadiz}}, \bibinfo {author} {\bibfnamefont {E.}~\bibnamefont {Courtade}},
  \bibinfo {author} {\bibfnamefont {C.}~\bibnamefont {Robert}}, \bibinfo
  {author} {\bibfnamefont {G.}~\bibnamefont {Wang}}, \bibinfo {author}
  {\bibfnamefont {Y.}~\bibnamefont {Shen}}, \bibinfo {author} {\bibfnamefont
  {H.}~\bibnamefont {Cai}}, \bibinfo {author} {\bibfnamefont {T.}~\bibnamefont
  {Taniguchi}}, \bibinfo {author} {\bibfnamefont {K.}~\bibnamefont {Watanabe}},
  \bibinfo {author} {\bibfnamefont {H.}~\bibnamefont {Carrere}}, \bibinfo
  {author} {\bibfnamefont {D.}~\bibnamefont {Lagarde}}, \bibinfo {author}
  {\bibfnamefont {M.}~\bibnamefont {Manca}}, \bibinfo {author} {\bibfnamefont
  {T.}~\bibnamefont {Amand}}, \bibinfo {author} {\bibfnamefont
  {P.}~\bibnamefont {Renucci}}, \bibinfo {author} {\bibfnamefont
  {S.}~\bibnamefont {Tongay}}, \bibinfo {author} {\bibfnamefont
  {X.}~\bibnamefont {Marie}},\ and\ \bibinfo {author} {\bibfnamefont
  {B.}~\bibnamefont {Urbaszek}},\ }\bibfield  {title} {\bibinfo {title}
  {{Excitonic linewidth approaching the homogeneous limit in MoS$_2$-based van
  der Waals heterostructures}},\ }\href
  {https://doi.org/10.1103/PhysRevX.7.021026} {\bibfield  {journal} {\bibinfo
  {journal} {Phys. Rev. X}\ }\textbf {\bibinfo {volume} {7}},\ \bibinfo {pages}
  {021026} (\bibinfo {year} {2017})}\BibitemShut {NoStop}%
\bibitem [{\citenamefont {Ajayi}\ \emph {et~al.}(2017)\citenamefont {Ajayi},
  \citenamefont {Ardelean}, \citenamefont {Shepard}, \citenamefont {Wang},
  \citenamefont {Antony}, \citenamefont {Taniguchi}, \citenamefont {Watanabe},
  \citenamefont {Heinz}, \citenamefont {Strauf}, \citenamefont {Zhu},\ and\
  \citenamefont {Hone}}]{Ajayi2017}%
  \BibitemOpen
  \bibfield  {author} {\bibinfo {author} {\bibfnamefont {O.~A.}\ \bibnamefont
  {Ajayi}}, \bibinfo {author} {\bibfnamefont {J.~V.}\ \bibnamefont {Ardelean}},
  \bibinfo {author} {\bibfnamefont {G.~D.}\ \bibnamefont {Shepard}}, \bibinfo
  {author} {\bibfnamefont {J.}~\bibnamefont {Wang}}, \bibinfo {author}
  {\bibfnamefont {A.}~\bibnamefont {Antony}}, \bibinfo {author} {\bibfnamefont
  {T.}~\bibnamefont {Taniguchi}}, \bibinfo {author} {\bibfnamefont
  {K.}~\bibnamefont {Watanabe}}, \bibinfo {author} {\bibfnamefont {T.~F.}\
  \bibnamefont {Heinz}}, \bibinfo {author} {\bibfnamefont {S.}~\bibnamefont
  {Strauf}}, \bibinfo {author} {\bibfnamefont {X.-Y.}\ \bibnamefont {Zhu}},\
  and\ \bibinfo {author} {\bibfnamefont {J.~C.}\ \bibnamefont {Hone}},\
  }\bibfield  {title} {\bibinfo {title} {{Approaching the intrinsic
  photoluminescence linewidth in transition metal dichalcogenide monolayers}},\
  }\href {https://doi.org/10.1088/2053-1583/aa6aa1} {\bibfield  {journal}
  {\bibinfo  {journal} {2D Mater.}\ }\textbf {\bibinfo {volume} {4}},\ \bibinfo
  {pages} {031011} (\bibinfo {year} {2017})}\BibitemShut {NoStop}%
\bibitem [{\citenamefont {Wierzbowski}\ \emph {et~al.}(2017)\citenamefont
  {Wierzbowski}, \citenamefont {Klein}, \citenamefont {Sigger}, \citenamefont
  {Straubinger}, \citenamefont {Kremser}, \citenamefont {Taniguchi},
  \citenamefont {Watanabe}, \citenamefont {Wurstbauer}, \citenamefont
  {Holleitner}, \citenamefont {Kaniber}, \citenamefont {M\"uller},\ and\
  \citenamefont {Finley}}]{Wierzbowski2017}%
  \BibitemOpen
  \bibfield  {author} {\bibinfo {author} {\bibfnamefont {J.}~\bibnamefont
  {Wierzbowski}}, \bibinfo {author} {\bibfnamefont {J.}~\bibnamefont {Klein}},
  \bibinfo {author} {\bibfnamefont {F.}~\bibnamefont {Sigger}}, \bibinfo
  {author} {\bibfnamefont {C.}~\bibnamefont {Straubinger}}, \bibinfo {author}
  {\bibfnamefont {M.}~\bibnamefont {Kremser}}, \bibinfo {author} {\bibfnamefont
  {T.}~\bibnamefont {Taniguchi}}, \bibinfo {author} {\bibfnamefont
  {K.}~\bibnamefont {Watanabe}}, \bibinfo {author} {\bibfnamefont
  {U.}~\bibnamefont {Wurstbauer}}, \bibinfo {author} {\bibfnamefont {A.~W.}\
  \bibnamefont {Holleitner}}, \bibinfo {author} {\bibfnamefont
  {M.}~\bibnamefont {Kaniber}}, \bibinfo {author} {\bibfnamefont
  {K.}~\bibnamefont {M\"uller}},\ and\ \bibinfo {author} {\bibfnamefont
  {J.~J.}\ \bibnamefont {Finley}},\ }\bibfield  {title} {\bibinfo {title}
  {Direct exciton emission from atomically thin transition metal dichalcogenide
  heterostructures near the lifetime limit},\ }\href
  {https://www.nature.com/articles/s41598-017-09739-4} {\bibfield  {journal}
  {\bibinfo  {journal} {Sci. Rep.}\ }\textbf {\bibinfo {volume} {7}},\ \bibinfo
  {pages} {12383} (\bibinfo {year} {2017})}\BibitemShut {NoStop}%
\bibitem [{\citenamefont {Shree}\ \emph {et~al.}(2019)\citenamefont {Shree},
  \citenamefont {George}, \citenamefont {Lehnert}, \citenamefont {Neumann},
  \citenamefont {Benelajla}, \citenamefont {Robert}, \citenamefont {Marie},
  \citenamefont {Watanabe}, \citenamefont {Taniguchi}, \citenamefont {Kaiser},
  \citenamefont {Urbaszek},\ and\ \citenamefont {Turchanin}}]{Shree2019}%
  \BibitemOpen
  \bibfield  {author} {\bibinfo {author} {\bibfnamefont {S.}~\bibnamefont
  {Shree}}, \bibinfo {author} {\bibfnamefont {A.}~\bibnamefont {George}},
  \bibinfo {author} {\bibfnamefont {T.}~\bibnamefont {Lehnert}}, \bibinfo
  {author} {\bibfnamefont {C.}~\bibnamefont {Neumann}}, \bibinfo {author}
  {\bibfnamefont {M.}~\bibnamefont {Benelajla}}, \bibinfo {author}
  {\bibfnamefont {C.}~\bibnamefont {Robert}}, \bibinfo {author} {\bibfnamefont
  {X.}~\bibnamefont {Marie}}, \bibinfo {author} {\bibfnamefont
  {K.}~\bibnamefont {Watanabe}}, \bibinfo {author} {\bibfnamefont
  {T.}~\bibnamefont {Taniguchi}}, \bibinfo {author} {\bibfnamefont
  {U.}~\bibnamefont {Kaiser}}, \bibinfo {author} {\bibfnamefont
  {B.}~\bibnamefont {Urbaszek}},\ and\ \bibinfo {author} {\bibfnamefont
  {A.}~\bibnamefont {Turchanin}},\ }\bibfield  {title} {\bibinfo {title} {High
  optical quality of {MoS}$_2$ monolayers grown by chemical vapor deposition},\
  }\href {https://doi.org/10.1088/2053-1583/ab4f1f} {\bibfield  {journal}
  {\bibinfo  {journal} {2D Mater.}\ }\textbf {\bibinfo {volume} {7}},\ \bibinfo
  {pages} {015011} (\bibinfo {year} {2019})}\BibitemShut {NoStop}%
\bibitem [{\citenamefont {You}\ \emph {et~al.}(2015)\citenamefont {You},
  \citenamefont {Zhang}, \citenamefont {Berkelbach}, \citenamefont {Hybertsen},
  \citenamefont {Reichman},\ and\ \citenamefont {Heinz}}]{You2015}%
  \BibitemOpen
  \bibfield  {author} {\bibinfo {author} {\bibfnamefont {Y.}~\bibnamefont
  {You}}, \bibinfo {author} {\bibfnamefont {X.-X.}\ \bibnamefont {Zhang}},
  \bibinfo {author} {\bibfnamefont {T.~C.}\ \bibnamefont {Berkelbach}},
  \bibinfo {author} {\bibfnamefont {M.~S.}\ \bibnamefont {Hybertsen}}, \bibinfo
  {author} {\bibfnamefont {D.~R.}\ \bibnamefont {Reichman}},\ and\ \bibinfo
  {author} {\bibfnamefont {T.~F.}\ \bibnamefont {Heinz}},\ }\bibfield  {title}
  {\bibinfo {title} {Observation of biexcitons in monolayer {WSe}$_2$},\ }\href
  {https://doi.org/10.1038/nphys3324} {\bibfield  {journal} {\bibinfo
  {journal} {Nat. Phys.}\ }\textbf {\bibinfo {volume} {11}},\ \bibinfo {pages}
  {477} (\bibinfo {year} {2015})}\BibitemShut {NoStop}%
\bibitem [{\citenamefont {Barbone}\ \emph {et~al.}(2018)\citenamefont
  {Barbone}, \citenamefont {Montblanch}, \citenamefont {Kara}, \citenamefont
  {Palacios-Berraquero}, \citenamefont {Cadore}, \citenamefont {De~Fazio},
  \citenamefont {Pingault}, \citenamefont {Mostaani}, \citenamefont {Li},
  \citenamefont {Chen}, \citenamefont {Watanabe}, \citenamefont {Taniguchi},
  \citenamefont {Tongay}, \citenamefont {Wang}, \citenamefont {Ferrari},\ and\
  \citenamefont {Atat{\ifmmode\ddot{u}\else\"{u}\fi}re}}]{Barbone2018}%
  \BibitemOpen
  \bibfield  {author} {\bibinfo {author} {\bibfnamefont {M.}~\bibnamefont
  {Barbone}}, \bibinfo {author} {\bibfnamefont {A.~R.-P.}\ \bibnamefont
  {Montblanch}}, \bibinfo {author} {\bibfnamefont {D.~M.}\ \bibnamefont
  {Kara}}, \bibinfo {author} {\bibfnamefont {C.}~\bibnamefont
  {Palacios-Berraquero}}, \bibinfo {author} {\bibfnamefont {A.~R.}\
  \bibnamefont {Cadore}}, \bibinfo {author} {\bibfnamefont {D.}~\bibnamefont
  {De~Fazio}}, \bibinfo {author} {\bibfnamefont {B.}~\bibnamefont {Pingault}},
  \bibinfo {author} {\bibfnamefont {E.}~\bibnamefont {Mostaani}}, \bibinfo
  {author} {\bibfnamefont {H.}~\bibnamefont {Li}}, \bibinfo {author}
  {\bibfnamefont {B.}~\bibnamefont {Chen}}, \bibinfo {author} {\bibfnamefont
  {K.}~\bibnamefont {Watanabe}}, \bibinfo {author} {\bibfnamefont
  {T.}~\bibnamefont {Taniguchi}}, \bibinfo {author} {\bibfnamefont
  {S.}~\bibnamefont {Tongay}}, \bibinfo {author} {\bibfnamefont
  {G.}~\bibnamefont {Wang}}, \bibinfo {author} {\bibfnamefont {A.~C.}\
  \bibnamefont {Ferrari}},\ and\ \bibinfo {author} {\bibfnamefont
  {M.}~\bibnamefont {Atat{\ifmmode\ddot{u}\else\"{u}\fi}re}},\ }\bibfield
  {title} {\bibinfo {title} {{Charge-tuneable biexciton complexes in monolayer
  WSe$_2$}},\ }\href {https://doi.org/10.1038/s41467-018-05632-4} {\bibfield
  {journal} {\bibinfo  {journal} {Nat. Commun.}\ }\textbf {\bibinfo {volume}
  {9}},\ \bibinfo {pages} {1} (\bibinfo {year} {2018})}\BibitemShut {NoStop}%
\bibitem [{\citenamefont {Steinhoff}\ \emph {et~al.}(2018)\citenamefont
  {Steinhoff}, \citenamefont {Florian}, \citenamefont {Singh}, \citenamefont
  {Tran}, \citenamefont {Kolarczik}, \citenamefont {Helmrich}, \citenamefont
  {Achtstein}, \citenamefont {Woggon}, \citenamefont {Owschimikow},
  \citenamefont {Jahnke},\ and\ \citenamefont {Li}}]{Steinhoff2018}%
  \BibitemOpen
  \bibfield  {author} {\bibinfo {author} {\bibfnamefont {A.}~\bibnamefont
  {Steinhoff}}, \bibinfo {author} {\bibfnamefont {M.}~\bibnamefont {Florian}},
  \bibinfo {author} {\bibfnamefont {A.}~\bibnamefont {Singh}}, \bibinfo
  {author} {\bibfnamefont {K.}~\bibnamefont {Tran}}, \bibinfo {author}
  {\bibfnamefont {M.}~\bibnamefont {Kolarczik}}, \bibinfo {author}
  {\bibfnamefont {S.}~\bibnamefont {Helmrich}}, \bibinfo {author}
  {\bibfnamefont {A.~W.}\ \bibnamefont {Achtstein}}, \bibinfo {author}
  {\bibfnamefont {U.}~\bibnamefont {Woggon}}, \bibinfo {author} {\bibfnamefont
  {N.}~\bibnamefont {Owschimikow}}, \bibinfo {author} {\bibfnamefont
  {F.}~\bibnamefont {Jahnke}},\ and\ \bibinfo {author} {\bibfnamefont
  {X.}~\bibnamefont {Li}},\ }\bibfield  {title} {\bibinfo {title} {Biexciton
  fine structure in monolayer transition metal dichalcogenides},\ }\href
  {https://doi.org/10.1038/s41567-018-0282-x} {\bibfield  {journal} {\bibinfo
  {journal} {Nat. Phys.}\ }\textbf {\bibinfo {volume} {14}},\ \bibinfo {pages}
  {1199} (\bibinfo {year} {2018})}\BibitemShut {NoStop}%
\bibitem [{\citenamefont {Li}\ \emph {et~al.}(2018)\citenamefont {Li},
  \citenamefont {Wang}, \citenamefont {Lu}, \citenamefont {Jin}, \citenamefont
  {Chen}, \citenamefont {Meng}, \citenamefont {Lian}, \citenamefont
  {Taniguchi}, \citenamefont {Watanabe}, \citenamefont {Zhang}, \citenamefont
  {Smirnov},\ and\ \citenamefont {Shi}}]{Li2018}%
  \BibitemOpen
  \bibfield  {author} {\bibinfo {author} {\bibfnamefont {Z.}~\bibnamefont
  {Li}}, \bibinfo {author} {\bibfnamefont {T.}~\bibnamefont {Wang}}, \bibinfo
  {author} {\bibfnamefont {Z.}~\bibnamefont {Lu}}, \bibinfo {author}
  {\bibfnamefont {C.}~\bibnamefont {Jin}}, \bibinfo {author} {\bibfnamefont
  {Y.}~\bibnamefont {Chen}}, \bibinfo {author} {\bibfnamefont {Y.}~\bibnamefont
  {Meng}}, \bibinfo {author} {\bibfnamefont {Z.}~\bibnamefont {Lian}}, \bibinfo
  {author} {\bibfnamefont {T.}~\bibnamefont {Taniguchi}}, \bibinfo {author}
  {\bibfnamefont {K.}~\bibnamefont {Watanabe}}, \bibinfo {author}
  {\bibfnamefont {S.}~\bibnamefont {Zhang}}, \bibinfo {author} {\bibfnamefont
  {D.}~\bibnamefont {Smirnov}},\ and\ \bibinfo {author} {\bibfnamefont {S.-F.}\
  \bibnamefont {Shi}},\ }\bibfield  {title} {\bibinfo {title} {Revealing the
  biexciton and trion-exciton complexes in {BN} encapsulated {WSe}$_2$},\
  }\href {https://doi.org/10.1038/s41467-018-05863-5} {\bibfield  {journal}
  {\bibinfo  {journal} {Nat. Commun.}\ }\textbf {\bibinfo {volume} {9}},\
  \bibinfo {pages} {3719} (\bibinfo {year} {2018})}\BibitemShut {NoStop}%
\bibitem [{\citenamefont {Ye}\ \emph {et~al.}(2018)\citenamefont {Ye},
  \citenamefont {Waldecker}, \citenamefont {Ma}, \citenamefont {Rhodes},
  \citenamefont {Antony}, \citenamefont {Kim}, \citenamefont {Zhang},
  \citenamefont {Deng}, \citenamefont {Jiang}, \citenamefont {Lu},
  \citenamefont {Smirnov}, \citenamefont {Watanabe}, \citenamefont {Taniguchi},
  \citenamefont {Hone},\ and\ \citenamefont {Heinz}}]{Ye2018}%
  \BibitemOpen
  \bibfield  {author} {\bibinfo {author} {\bibfnamefont {Z.}~\bibnamefont
  {Ye}}, \bibinfo {author} {\bibfnamefont {L.}~\bibnamefont {Waldecker}},
  \bibinfo {author} {\bibfnamefont {E.~Y.}\ \bibnamefont {Ma}}, \bibinfo
  {author} {\bibfnamefont {D.}~\bibnamefont {Rhodes}}, \bibinfo {author}
  {\bibfnamefont {A.}~\bibnamefont {Antony}}, \bibinfo {author} {\bibfnamefont
  {B.}~\bibnamefont {Kim}}, \bibinfo {author} {\bibfnamefont {X.-X.}\
  \bibnamefont {Zhang}}, \bibinfo {author} {\bibfnamefont {M.}~\bibnamefont
  {Deng}}, \bibinfo {author} {\bibfnamefont {Y.}~\bibnamefont {Jiang}},
  \bibinfo {author} {\bibfnamefont {Z.}~\bibnamefont {Lu}}, \bibinfo {author}
  {\bibfnamefont {D.}~\bibnamefont {Smirnov}}, \bibinfo {author} {\bibfnamefont
  {K.}~\bibnamefont {Watanabe}}, \bibinfo {author} {\bibfnamefont
  {T.}~\bibnamefont {Taniguchi}}, \bibinfo {author} {\bibfnamefont
  {J.}~\bibnamefont {Hone}},\ and\ \bibinfo {author} {\bibfnamefont {T.~F.}\
  \bibnamefont {Heinz}},\ }\bibfield  {title} {\bibinfo {title} {{Efficient
  generation of neutral and charged biexcitons in encapsulated WSe$_2$
  monolayers}},\ }\href {https://doi.org/10.1038/s41467-018-05917-8} {\bibfield
   {journal} {\bibinfo  {journal} {Nat. Commun.}\ }\textbf {\bibinfo {volume}
  {9}},\ \bibinfo {pages} {3718} (\bibinfo {year} {2018})}\BibitemShut
  {NoStop}%
\bibitem [{\citenamefont {Arora}\ \emph {et~al.}(2015)\citenamefont {Arora},
  \citenamefont {Koperski}, \citenamefont {Nogajewski}, \citenamefont {Marcus},
  \citenamefont {Faugeras},\ and\ \citenamefont {Potemski}}]{Arora2015}%
  \BibitemOpen
  \bibfield  {author} {\bibinfo {author} {\bibfnamefont {A.}~\bibnamefont
  {Arora}}, \bibinfo {author} {\bibfnamefont {M.}~\bibnamefont {Koperski}},
  \bibinfo {author} {\bibfnamefont {K.}~\bibnamefont {Nogajewski}}, \bibinfo
  {author} {\bibfnamefont {J.}~\bibnamefont {Marcus}}, \bibinfo {author}
  {\bibfnamefont {C.}~\bibnamefont {Faugeras}},\ and\ \bibinfo {author}
  {\bibfnamefont {M.}~\bibnamefont {Potemski}},\ }\bibfield  {title} {\bibinfo
  {title} {{Excitonic resonances in thin films of WSe$_2$: from monolayer to
  bulk material}},\ }\href {https://doi.org/10.1039/c5nr01536g} {\bibfield
  {journal} {\bibinfo  {journal} {Nanoscale}\ }\textbf {\bibinfo {volume}
  {7}},\ \bibinfo {pages} {10421} (\bibinfo {year} {2015})}\BibitemShut
  {NoStop}%
\bibitem [{\citenamefont {Lindlau}\ \emph
  {et~al.}(2018{\natexlab{b}})\citenamefont {Lindlau}, \citenamefont {Selig},
  \citenamefont {Neumann}, \citenamefont {Colombier}, \citenamefont
  {F{\"o}rste}, \citenamefont {Funk}, \citenamefont {F{\"o}rg}, \citenamefont
  {Kim}, \citenamefont {Bergh{\"a}user}, \citenamefont {Taniguchi},
  \citenamefont {Watanabe}, \citenamefont {Wang}, \citenamefont {Malic},\ and\
  \citenamefont {H{\"o}gele}}]{Lindlau2017BL}%
  \BibitemOpen
  \bibfield  {author} {\bibinfo {author} {\bibfnamefont {J.}~\bibnamefont
  {Lindlau}}, \bibinfo {author} {\bibfnamefont {M.}~\bibnamefont {Selig}},
  \bibinfo {author} {\bibfnamefont {A.}~\bibnamefont {Neumann}}, \bibinfo
  {author} {\bibfnamefont {L.}~\bibnamefont {Colombier}}, \bibinfo {author}
  {\bibfnamefont {J.}~\bibnamefont {F{\"o}rste}}, \bibinfo {author}
  {\bibfnamefont {V.}~\bibnamefont {Funk}}, \bibinfo {author} {\bibfnamefont
  {M.}~\bibnamefont {F{\"o}rg}}, \bibinfo {author} {\bibfnamefont
  {J.}~\bibnamefont {Kim}}, \bibinfo {author} {\bibfnamefont {G.}~\bibnamefont
  {Bergh{\"a}user}}, \bibinfo {author} {\bibfnamefont {T.}~\bibnamefont
  {Taniguchi}}, \bibinfo {author} {\bibfnamefont {K.}~\bibnamefont {Watanabe}},
  \bibinfo {author} {\bibfnamefont {F.}~\bibnamefont {Wang}}, \bibinfo {author}
  {\bibfnamefont {E.}~\bibnamefont {Malic}},\ and\ \bibinfo {author}
  {\bibfnamefont {A.}~\bibnamefont {H{\"o}gele}},\ }\bibfield  {title}
  {\bibinfo {title} {The role of momentum-dark excitons in the elementary
  optical response of bilayer {WSe}$_2$},\ }\href
  {https://doi.org/10.1038/s41467-018-04877-3} {\bibfield  {journal} {\bibinfo
  {journal} {Nat. Commun.}\ }\textbf {\bibinfo {volume} {9}},\ \bibinfo {pages}
  {2586} (\bibinfo {year} {2018}{\natexlab{b}})}\BibitemShut {NoStop}%
\bibitem [{\citenamefont {Wickramaratne}\ \emph {et~al.}(2014)\citenamefont
  {Wickramaratne}, \citenamefont {Zahid},\ and\ \citenamefont
  {Lake}}]{Wickramaratne2014}%
  \BibitemOpen
  \bibfield  {author} {\bibinfo {author} {\bibfnamefont {D.}~\bibnamefont
  {Wickramaratne}}, \bibinfo {author} {\bibfnamefont {F.}~\bibnamefont
  {Zahid}},\ and\ \bibinfo {author} {\bibfnamefont {R.~K.}\ \bibnamefont
  {Lake}},\ }\bibfield  {title} {\bibinfo {title} {Electronic and
  thermoelectric properties of few-layer transition metal dichalcogenides},\
  }\href {https://doi.org/10.1063/1.4869142} {\bibfield  {journal} {\bibinfo
  {journal} {J. Chem. Phys.}\ }\textbf {\bibinfo {volume} {140}},\ \bibinfo
  {pages} {124710} (\bibinfo {year} {2014})}\BibitemShut {NoStop}%
\bibitem [{\citenamefont {Slobodeniuk}\ \emph {et~al.}(2019)\citenamefont
  {Slobodeniuk}, \citenamefont {Koperski}, \citenamefont {Molas}, \citenamefont
  {Kossacki}, \citenamefont {Nogajewski}, \citenamefont {Bartos}, \citenamefont
  {Watanabe}, \citenamefont {Taniguchi}, \citenamefont {Faugeras},\ and\
  \citenamefont {Potemski}}]{slobodeniuk2019fine}%
  \BibitemOpen
  \bibfield  {author} {\bibinfo {author} {\bibfnamefont {A.~O.}\ \bibnamefont
  {Slobodeniuk}}, \bibinfo {author} {\bibfnamefont {M.}~\bibnamefont
  {Koperski}}, \bibinfo {author} {\bibfnamefont {M.~R.}\ \bibnamefont {Molas}},
  \bibinfo {author} {\bibfnamefont {P.}~\bibnamefont {Kossacki}}, \bibinfo
  {author} {\bibfnamefont {K.}~\bibnamefont {Nogajewski}}, \bibinfo {author}
  {\bibfnamefont {M.}~\bibnamefont {Bartos}}, \bibinfo {author} {\bibfnamefont
  {K.}~\bibnamefont {Watanabe}}, \bibinfo {author} {\bibfnamefont
  {T.}~\bibnamefont {Taniguchi}}, \bibinfo {author} {\bibfnamefont
  {C.}~\bibnamefont {Faugeras}},\ and\ \bibinfo {author} {\bibfnamefont
  {M.}~\bibnamefont {Potemski}},\ }\bibfield  {title} {\bibinfo {title} {Fine
  structure of k-excitons in multilayers of transition metal dichalcogenides},\
  }\href {https://iopscience.iop.org/article/10.1088/2053-1583/ab0776}
  {\bibfield  {journal} {\bibinfo  {journal} {2D Mater.}\ }\textbf {\bibinfo
  {volume} {6}},\ \bibinfo {pages} {025026} (\bibinfo {year}
  {2019})}\BibitemShut {NoStop}%
\bibitem [{\citenamefont {Horng}\ \emph {et~al.}(2018)\citenamefont {Horng},
  \citenamefont {Stroucken}, \citenamefont {Zhang}, \citenamefont {Paik},
  \citenamefont {Deng},\ and\ \citenamefont {Koch}}]{horng2018observation}%
  \BibitemOpen
  \bibfield  {author} {\bibinfo {author} {\bibfnamefont {J.}~\bibnamefont
  {Horng}}, \bibinfo {author} {\bibfnamefont {T.}~\bibnamefont {Stroucken}},
  \bibinfo {author} {\bibfnamefont {L.}~\bibnamefont {Zhang}}, \bibinfo
  {author} {\bibfnamefont {E.~Y.}\ \bibnamefont {Paik}}, \bibinfo {author}
  {\bibfnamefont {H.}~\bibnamefont {Deng}},\ and\ \bibinfo {author}
  {\bibfnamefont {S.~W.}\ \bibnamefont {Koch}},\ }\bibfield  {title} {\bibinfo
  {title} {Observation of interlayer excitons in {MoSe}$_2$ single crystals},\
  }\href {https://journals.aps.org/prb/abstract/10.1103/PhysRevB.97.241404}
  {\bibfield  {journal} {\bibinfo  {journal} {Phys. Rev. B}\ }\textbf {\bibinfo
  {volume} {97}},\ \bibinfo {pages} {241404} (\bibinfo {year}
  {2018})}\BibitemShut {NoStop}%
\bibitem [{\citenamefont {Shree}\ \emph {et~al.}(2021)\citenamefont {Shree},
  \citenamefont {Paradisanos}, \citenamefont {Marie}, \citenamefont {Robert},\
  and\ \citenamefont {Urbaszek}}]{shree2021guide}%
  \BibitemOpen
  \bibfield  {author} {\bibinfo {author} {\bibfnamefont {S.}~\bibnamefont
  {Shree}}, \bibinfo {author} {\bibfnamefont {I.}~\bibnamefont {Paradisanos}},
  \bibinfo {author} {\bibfnamefont {X.}~\bibnamefont {Marie}}, \bibinfo
  {author} {\bibfnamefont {C.}~\bibnamefont {Robert}},\ and\ \bibinfo {author}
  {\bibfnamefont {B.}~\bibnamefont {Urbaszek}},\ }\bibfield  {title} {\bibinfo
  {title} {Guide to optical spectroscopy of layered semiconductors},\ }\href
  {https://www.nature.com/articles/s42254-020-00259-1} {\bibfield  {journal}
  {\bibinfo  {journal} {Nat. Rev. Phys.}\ }\textbf {\bibinfo {volume} {3}},\
  \bibinfo {pages} {39} (\bibinfo {year} {2021})}\BibitemShut {NoStop}%
\bibitem [{\citenamefont {F\"org}\ \emph {et~al.}(2021)\citenamefont {F\"org},
  \citenamefont {Baimuratov}, \citenamefont {Kruchinin}, \citenamefont {Vovk},
  \citenamefont {Scherzer},\ and\ \citenamefont {H\"ogele}}]{Forg2021}%
  \BibitemOpen
  \bibfield  {author} {\bibinfo {author} {\bibfnamefont {M.}~\bibnamefont
  {F\"org}}, \bibinfo {author} {\bibfnamefont {A.~S.}\ \bibnamefont
  {Baimuratov}}, \bibinfo {author} {\bibfnamefont {S.~{\relax Yu}.}\
  \bibnamefont {Kruchinin}}, \bibinfo {author} {\bibfnamefont {I.~A.}\
  \bibnamefont {Vovk}}, \bibinfo {author} {\bibfnamefont {J.}~\bibnamefont
  {Scherzer}},\ and\ \bibinfo {author} {\bibfnamefont {A.}~\bibnamefont
  {H\"ogele}},\ }\bibfield  {title} {\bibinfo {title}
  {{Moir{\ifmmode\acute{e}\else\'{e}\fi} excitons in MoSe$_2$-WSe$_2$
  heterobilayers and heterotrilayers}},\ }\href
  {https://doi.org/10.1038/s41467-021-21822-z} {\bibfield  {journal} {\bibinfo
  {journal} {Nat. Commun.}\ }\textbf {\bibinfo {volume} {12}},\ \bibinfo
  {pages} {1} (\bibinfo {year} {2021})}\BibitemShut {NoStop}%
\bibitem [{\citenamefont {Aivazian}\ \emph {et~al.}(2015)\citenamefont
  {Aivazian}, \citenamefont {Gong}, \citenamefont {Jones}, \citenamefont {Chu},
  \citenamefont {Yan}, \citenamefont {Mandrus}, \citenamefont {Zhang},
  \citenamefont {Cobden}, \citenamefont {Yao},\ and\ \citenamefont
  {Xu}}]{Aivazian2015}%
  \BibitemOpen
  \bibfield  {author} {\bibinfo {author} {\bibfnamefont {G.}~\bibnamefont
  {Aivazian}}, \bibinfo {author} {\bibfnamefont {Z.}~\bibnamefont {Gong}},
  \bibinfo {author} {\bibfnamefont {A.~M.}\ \bibnamefont {Jones}}, \bibinfo
  {author} {\bibfnamefont {R.-L.}\ \bibnamefont {Chu}}, \bibinfo {author}
  {\bibfnamefont {J.}~\bibnamefont {Yan}}, \bibinfo {author} {\bibfnamefont
  {D.~G.}\ \bibnamefont {Mandrus}}, \bibinfo {author} {\bibfnamefont
  {C.}~\bibnamefont {Zhang}}, \bibinfo {author} {\bibfnamefont
  {D.}~\bibnamefont {Cobden}}, \bibinfo {author} {\bibfnamefont
  {W.}~\bibnamefont {Yao}},\ and\ \bibinfo {author} {\bibfnamefont
  {X.}~\bibnamefont {Xu}},\ }\bibfield  {title} {\bibinfo {title} {{Magnetic
  control of valley pseudospin in monolayer WSe$_2$}},\ }\href
  {https://doi.org/10.1038/nphys3201} {\bibfield  {journal} {\bibinfo
  {journal} {Nat. Phys.}\ }\textbf {\bibinfo {volume} {11}},\ \bibinfo {pages}
  {148} (\bibinfo {year} {2015})}\BibitemShut {NoStop}%
\bibitem [{\citenamefont {Srivastava}\ \emph {et~al.}(2015)\citenamefont
  {Srivastava}, \citenamefont {Sidler}, \citenamefont {Allain}, \citenamefont
  {Lembke}, \citenamefont {Kis},\ and\ \citenamefont
  {Imamo{\ifmmode\breve{g}\else\u{g}\fi}lu}}]{Srivastava2015}%
  \BibitemOpen
  \bibfield  {author} {\bibinfo {author} {\bibfnamefont {A.}~\bibnamefont
  {Srivastava}}, \bibinfo {author} {\bibfnamefont {M.}~\bibnamefont {Sidler}},
  \bibinfo {author} {\bibfnamefont {A.~V.}\ \bibnamefont {Allain}}, \bibinfo
  {author} {\bibfnamefont {D.~S.}\ \bibnamefont {Lembke}}, \bibinfo {author}
  {\bibfnamefont {A.}~\bibnamefont {Kis}},\ and\ \bibinfo {author}
  {\bibfnamefont {A.}~\bibnamefont {Imamo{\ifmmode\breve{g}\else\u{g}\fi}lu}},\
  }\bibfield  {title} {\bibinfo {title} {{Valley Zeeman effect in elementary
  optical excitations of monolayer WSe$_2$}},\ }\href
  {https://doi.org/10.1038/nphys3203} {\bibfield  {journal} {\bibinfo
  {journal} {Nat. Phys.}\ }\textbf {\bibinfo {volume} {11}},\ \bibinfo {pages}
  {141} (\bibinfo {year} {2015})}\BibitemShut {NoStop}%
\bibitem [{\citenamefont {Wang}\ \emph {et~al.}(2015)\citenamefont {Wang},
  \citenamefont {Bouet}, \citenamefont {Glazov}, \citenamefont {Amand},
  \citenamefont {Ivchenko}, \citenamefont {Palleau}, \citenamefont {Marie},\
  and\ \citenamefont {Urbaszek}}]{Wang2015}%
  \BibitemOpen
  \bibfield  {author} {\bibinfo {author} {\bibfnamefont {G.}~\bibnamefont
  {Wang}}, \bibinfo {author} {\bibfnamefont {L.}~\bibnamefont {Bouet}},
  \bibinfo {author} {\bibfnamefont {M.~M.}\ \bibnamefont {Glazov}}, \bibinfo
  {author} {\bibfnamefont {T.}~\bibnamefont {Amand}}, \bibinfo {author}
  {\bibfnamefont {E.~L.}\ \bibnamefont {Ivchenko}}, \bibinfo {author}
  {\bibfnamefont {E.}~\bibnamefont {Palleau}}, \bibinfo {author} {\bibfnamefont
  {X.}~\bibnamefont {Marie}},\ and\ \bibinfo {author} {\bibfnamefont
  {B.}~\bibnamefont {Urbaszek}},\ }\bibfield  {title} {\bibinfo {title}
  {{Magneto-optics in transition metal diselenide monolayers}},\ }\href
  {https://doi.org/10.1088/2053-1583/2/3/034002} {\bibfield  {journal}
  {\bibinfo  {journal} {2D Mater.}\ }\textbf {\bibinfo {volume} {2}},\ \bibinfo
  {pages} {034002} (\bibinfo {year} {2015})}\BibitemShut {NoStop}%
\bibitem [{\citenamefont {Koperski}\ \emph {et~al.}(2018)\citenamefont
  {Koperski}, \citenamefont {Molas}, \citenamefont {Arora}, \citenamefont
  {Nogajewski}, \citenamefont {Bartos}, \citenamefont {Wyzula}, \citenamefont
  {Vaclavkova}, \citenamefont {Kossacki},\ and\ \citenamefont
  {Potemski}}]{Koperski2018}%
  \BibitemOpen
  \bibfield  {author} {\bibinfo {author} {\bibfnamefont {M.}~\bibnamefont
  {Koperski}}, \bibinfo {author} {\bibfnamefont {M.~R.}\ \bibnamefont {Molas}},
  \bibinfo {author} {\bibfnamefont {A.}~\bibnamefont {Arora}}, \bibinfo
  {author} {\bibfnamefont {K.}~\bibnamefont {Nogajewski}}, \bibinfo {author}
  {\bibfnamefont {M.}~\bibnamefont {Bartos}}, \bibinfo {author} {\bibfnamefont
  {J.}~\bibnamefont {Wyzula}}, \bibinfo {author} {\bibfnamefont
  {D.}~\bibnamefont {Vaclavkova}}, \bibinfo {author} {\bibfnamefont
  {P.}~\bibnamefont {Kossacki}},\ and\ \bibinfo {author} {\bibfnamefont
  {M.}~\bibnamefont {Potemski}},\ }\bibfield  {title} {\bibinfo {title}
  {{Orbital, spin and valley contributions to Zeeman splitting of excitonic
  resonances in MoSe$_2$, WSe$_2$ and WS$_2$ Monolayers}},\ }\href
  {https://doi.org/10.1088/2053-1583/aae14b} {\bibfield  {journal} {\bibinfo
  {journal} {2D Mater.}\ }\textbf {\bibinfo {volume} {6}},\ \bibinfo {pages}
  {015001} (\bibinfo {year} {2018})}\BibitemShut {NoStop}%
\bibitem [{\citenamefont {Wozniak}\ \emph {et~al.}(2020)\citenamefont
  {Wozniak}, \citenamefont {Faria~Junior}, \citenamefont {Seifert},
  \citenamefont {Chaves},\ and\ \citenamefont {Kunstmann}}]{Wozniak2020}%
  \BibitemOpen
  \bibfield  {author} {\bibinfo {author} {\bibfnamefont {T.}~\bibnamefont
  {Wozniak}}, \bibinfo {author} {\bibfnamefont {P.~E.}\ \bibnamefont
  {Faria~Junior}}, \bibinfo {author} {\bibfnamefont {G.}~\bibnamefont
  {Seifert}}, \bibinfo {author} {\bibfnamefont {A.}~\bibnamefont {Chaves}},\
  and\ \bibinfo {author} {\bibfnamefont {J.}~\bibnamefont {Kunstmann}},\
  }\bibfield  {title} {\bibinfo {title} {Exciton $g$ factors of van der waals
  heterostructures from first-principles calculations},\ }\href
  {https://link.aps.org/doi/10.1103/PhysRevB.101.235408} {\bibfield  {journal}
  {\bibinfo  {journal} {Phys. Rev. B}\ }\textbf {\bibinfo {volume} {101}},\
  \bibinfo {pages} {235408} (\bibinfo {year} {2020})}\BibitemShut {NoStop}%
\bibitem [{\citenamefont {Deilmann}\ \emph {et~al.}(2020)\citenamefont
  {Deilmann}, \citenamefont {Kr\"uger},\ and\ \citenamefont
  {Rohlfing}}]{Deilmann2020}%
  \BibitemOpen
  \bibfield  {author} {\bibinfo {author} {\bibfnamefont {T.}~\bibnamefont
  {Deilmann}}, \bibinfo {author} {\bibfnamefont {P.}~\bibnamefont {Kr\"uger}},\
  and\ \bibinfo {author} {\bibfnamefont {M.}~\bibnamefont {Rohlfing}},\
  }\bibfield  {title} {\bibinfo {title} {Ab initio studies of exciton $g$
  factors: Monolayer transition metal dichalcogenides in magnetic fields},\
  }\href {https://doi.org/10.1103/PhysRevLett.124.226402} {\bibfield  {journal}
  {\bibinfo  {journal} {Phys. Rev. Lett.}\ }\textbf {\bibinfo {volume} {124}},\
  \bibinfo {pages} {226402} (\bibinfo {year} {2020})}\BibitemShut {NoStop}%
\bibitem [{\citenamefont {Xuan}\ and\ \citenamefont {Quek}(2020)}]{Xuan2020}%
  \BibitemOpen
  \bibfield  {author} {\bibinfo {author} {\bibfnamefont {F.}~\bibnamefont
  {Xuan}}\ and\ \bibinfo {author} {\bibfnamefont {S.~Y.}\ \bibnamefont
  {Quek}},\ }\bibfield  {title} {\bibinfo {title} {Valley zeeman effect and
  landau levels in two-dimensional transition metal dichalcogenides},\ }\href
  {https://doi.org/10.1103/PhysRevResearch.2.033256} {\bibfield  {journal}
  {\bibinfo  {journal} {Phys. Rev. Research}\ }\textbf {\bibinfo {volume}
  {2}},\ \bibinfo {pages} {033256} (\bibinfo {year} {2020})}\BibitemShut
  {NoStop}%
\bibitem [{\citenamefont {Bilgin}\ \emph {et~al.}(2015)\citenamefont {Bilgin},
  \citenamefont {Liu}, \citenamefont {Vargas}, \citenamefont {Winchester},
  \citenamefont {Man}, \citenamefont {Upmanyu}, \citenamefont {Dani},
  \citenamefont {Gupta}, \citenamefont {Talapatra}, \citenamefont {Mohite},\
  and\ \citenamefont {Kar}}]{Bilgin2015}%
  \BibitemOpen
  \bibfield  {author} {\bibinfo {author} {\bibfnamefont {I.}~\bibnamefont
  {Bilgin}}, \bibinfo {author} {\bibfnamefont {F.}~\bibnamefont {Liu}},
  \bibinfo {author} {\bibfnamefont {A.}~\bibnamefont {Vargas}}, \bibinfo
  {author} {\bibfnamefont {A.}~\bibnamefont {Winchester}}, \bibinfo {author}
  {\bibfnamefont {M.~K.~L.}\ \bibnamefont {Man}}, \bibinfo {author}
  {\bibfnamefont {M.}~\bibnamefont {Upmanyu}}, \bibinfo {author} {\bibfnamefont
  {K.~M.}\ \bibnamefont {Dani}}, \bibinfo {author} {\bibfnamefont
  {G.}~\bibnamefont {Gupta}}, \bibinfo {author} {\bibfnamefont
  {S.}~\bibnamefont {Talapatra}}, \bibinfo {author} {\bibfnamefont {A.~D.}\
  \bibnamefont {Mohite}},\ and\ \bibinfo {author} {\bibfnamefont
  {S.}~\bibnamefont {Kar}},\ }\bibfield  {title} {\bibinfo {title} {{Chemical
  Vapor Deposition Synthesized Atomically Thin Molybdenum Disulfide with
  Optoelectronic-Grade Crystalline Quality}},\ }\href
  {https://doi.org/10.1021/acsnano.5b02019} {\bibfield  {journal} {\bibinfo
  {journal} {ACS Nano}\ }\textbf {\bibinfo {volume} {9}},\ \bibinfo {pages}
  {8822} (\bibinfo {year} {2015})}\BibitemShut {NoStop}%
\bibitem [{\citenamefont {Bilgin}\ \emph {et~al.}(2018)\citenamefont {Bilgin},
  \citenamefont {Raeliarijaona}, \citenamefont {Lucking}, \citenamefont
  {Hodge}, \citenamefont {Mohite}, \citenamefont {de~Luna~Bugallo},
  \citenamefont {Terrones},\ and\ \citenamefont {Kar}}]{Bilgin2018}%
  \BibitemOpen
  \bibfield  {author} {\bibinfo {author} {\bibfnamefont {I.}~\bibnamefont
  {Bilgin}}, \bibinfo {author} {\bibfnamefont {A.~S.}\ \bibnamefont
  {Raeliarijaona}}, \bibinfo {author} {\bibfnamefont {M.~C.}\ \bibnamefont
  {Lucking}}, \bibinfo {author} {\bibfnamefont {S.~C.}\ \bibnamefont {Hodge}},
  \bibinfo {author} {\bibfnamefont {A.~D.}\ \bibnamefont {Mohite}}, \bibinfo
  {author} {\bibfnamefont {A.}~\bibnamefont {de~Luna~Bugallo}}, \bibinfo
  {author} {\bibfnamefont {H.}~\bibnamefont {Terrones}},\ and\ \bibinfo
  {author} {\bibfnamefont {S.}~\bibnamefont {Kar}},\ }\bibfield  {title}
  {\bibinfo {title} {{Resonant Raman and Exciton Coupling in High-Quality
  Single Crystals of Atomically Thin Molybdenum Diselenide Grown by Vapor-Phase
  Chalcogenization}},\ }\href {https://doi.org/10.1021/acsnano.7b07933}
  {\bibfield  {journal} {\bibinfo  {journal} {ACS Nano}\ }\textbf {\bibinfo
  {volume} {12}},\ \bibinfo {pages} {740} (\bibinfo {year} {2018})}\BibitemShut
  {NoStop}%
\bibitem [{\citenamefont {Li}\ \emph {et~al.}(2015)\citenamefont {Li},
  \citenamefont {Wang}, \citenamefont {Tang}, \citenamefont {Zhao},
  \citenamefont {Xu}, \citenamefont {Chu}, \citenamefont {Bando}, \citenamefont
  {Golberg},\ and\ \citenamefont {Eda}}]{Li2015}%
  \BibitemOpen
  \bibfield  {author} {\bibinfo {author} {\bibfnamefont {S.}~\bibnamefont
  {Li}}, \bibinfo {author} {\bibfnamefont {S.}~\bibnamefont {Wang}}, \bibinfo
  {author} {\bibfnamefont {D.-M.}\ \bibnamefont {Tang}}, \bibinfo {author}
  {\bibfnamefont {W.}~\bibnamefont {Zhao}}, \bibinfo {author} {\bibfnamefont
  {H.}~\bibnamefont {Xu}}, \bibinfo {author} {\bibfnamefont {L.}~\bibnamefont
  {Chu}}, \bibinfo {author} {\bibfnamefont {Y.}~\bibnamefont {Bando}}, \bibinfo
  {author} {\bibfnamefont {D.}~\bibnamefont {Golberg}},\ and\ \bibinfo {author}
  {\bibfnamefont {G.}~\bibnamefont {Eda}},\ }\bibfield  {title} {\bibinfo
  {title} {{Halide-assisted atmospheric pressure growth of large WSe$_2$ and
  WS$_2$ monolayer crystals}},\ }\href
  {https://doi.org/10.1016/j.apmt.2015.09.001} {\bibfield  {journal} {\bibinfo
  {journal} {Appl. Mater. Today}\ }\textbf {\bibinfo {volume} {1}},\ \bibinfo
  {pages} {60} (\bibinfo {year} {2015})}\BibitemShut {NoStop}%
\bibitem [{\citenamefont {Pizzocchero}\ \emph {et~al.}(2016)\citenamefont
  {Pizzocchero}, \citenamefont {Gammelgaard}, \citenamefont {Jessen},
  \citenamefont {Caridad}, \citenamefont {Wang}, \citenamefont {Hone},
  \citenamefont {B{\o}ggild},\ and\ \citenamefont {Booth}}]{Pizzocchero2016}%
  \BibitemOpen
  \bibfield  {author} {\bibinfo {author} {\bibfnamefont {F.}~\bibnamefont
  {Pizzocchero}}, \bibinfo {author} {\bibfnamefont {L.}~\bibnamefont
  {Gammelgaard}}, \bibinfo {author} {\bibfnamefont {B.~S.}\ \bibnamefont
  {Jessen}}, \bibinfo {author} {\bibfnamefont {J.~M.}\ \bibnamefont {Caridad}},
  \bibinfo {author} {\bibfnamefont {L.}~\bibnamefont {Wang}}, \bibinfo {author}
  {\bibfnamefont {J.}~\bibnamefont {Hone}}, \bibinfo {author} {\bibfnamefont
  {P.}~\bibnamefont {B{\o}ggild}},\ and\ \bibinfo {author} {\bibfnamefont
  {T.~J.}\ \bibnamefont {Booth}},\ }\bibfield  {title} {\bibinfo {title} {{The
  hot pick-up technique for batch assembly of van der Waals
  heterostructures}},\ }\href {https://doi.org/10.1038/ncomms11894} {\bibfield
  {journal} {\bibinfo  {journal} {Nat. Commun.}\ }\textbf {\bibinfo {volume}
  {7}},\ \bibinfo {pages} {1} (\bibinfo {year} {2016})}\BibitemShut {NoStop}%
\bibitem [{\citenamefont {Purdie}\ \emph {et~al.}(2018)\citenamefont {Purdie},
  \citenamefont {Pugno}, \citenamefont {Taniguchi}, \citenamefont {Watanabe},
  \citenamefont {Ferrari},\ and\ \citenamefont {Lombardo}}]{Purdie2018}%
  \BibitemOpen
  \bibfield  {author} {\bibinfo {author} {\bibfnamefont {D.~G.}\ \bibnamefont
  {Purdie}}, \bibinfo {author} {\bibfnamefont {N.~M.}\ \bibnamefont {Pugno}},
  \bibinfo {author} {\bibfnamefont {T.}~\bibnamefont {Taniguchi}}, \bibinfo
  {author} {\bibfnamefont {K.}~\bibnamefont {Watanabe}}, \bibinfo {author}
  {\bibfnamefont {A.~C.}\ \bibnamefont {Ferrari}},\ and\ \bibinfo {author}
  {\bibfnamefont {A.}~\bibnamefont {Lombardo}},\ }\bibfield  {title} {\bibinfo
  {title} {{Cleaning interfaces in layered materials heterostructures}},\
  }\href {https://doi.org/10.1038/s41467-018-07558-3} {\bibfield  {journal}
  {\bibinfo  {journal} {Nat. Commun.}\ }\textbf {\bibinfo {volume} {9}},\
  \bibinfo {pages} {1} (\bibinfo {year} {2018})}\BibitemShut {NoStop}%
\bibitem [{\citenamefont {Kresse}\ and\ \citenamefont
  {Hafner}(1994)}]{Kresse:1993a}%
  \BibitemOpen
  \bibfield  {author} {\bibinfo {author} {\bibfnamefont {G.}~\bibnamefont
  {Kresse}}\ and\ \bibinfo {author} {\bibfnamefont {J.}~\bibnamefont
  {Hafner}},\ }\bibfield  {title} {\bibinfo {title} {Ab initio
  molecular-dynamics simulation of the liquid-metal--amorphous-semiconductor
  transition in germanium},\ }\href {https://doi.org/10.1103/PhysRevB.49.14251}
  {\bibfield  {journal} {\bibinfo  {journal} {Phys. Rev. B}\ }\textbf {\bibinfo
  {volume} {49}},\ \bibinfo {pages} {14251} (\bibinfo {year}
  {1994})}\BibitemShut {NoStop}%
\bibitem [{\citenamefont {Kresse}\ and\ \citenamefont
  {Furthm\"uller}(1996)}]{Kresse:1996a}%
  \BibitemOpen
  \bibfield  {author} {\bibinfo {author} {\bibfnamefont {G.}~\bibnamefont
  {Kresse}}\ and\ \bibinfo {author} {\bibfnamefont {J.}~\bibnamefont
  {Furthm\"uller}},\ }\bibfield  {title} {\bibinfo {title} {Efficient iterative
  schemes for \textit{ab initio} total-energy calculations using a plane-wave
  basis set},\ }\href {https://doi.org/10.1103/PhysRevB.54.11169} {\bibfield
  {journal} {\bibinfo  {journal} {Phys. Rev. B}\ }\textbf {\bibinfo {volume}
  {54}},\ \bibinfo {pages} {11169} (\bibinfo {year} {1996})}\BibitemShut
  {NoStop}%
\bibitem [{\citenamefont {Bl\"{o}chl}(1994)}]{Blochl1994}%
  \BibitemOpen
  \bibfield  {author} {\bibinfo {author} {\bibfnamefont {P.~E.}\ \bibnamefont
  {Bl\"{o}chl}},\ }\bibfield  {title} {\bibinfo {title} {{Projector
  augmented-wave method}},\ }\href {https://doi.org/10.1103/PhysRevB.50.17953}
  {\bibfield  {journal} {\bibinfo  {journal} {Phys. Rev. B}\ }\textbf {\bibinfo
  {volume} {50}},\ \bibinfo {pages} {17953} (\bibinfo {year}
  {1994})}\BibitemShut {NoStop}%
\bibitem [{\citenamefont {Mostofi}\ \emph {et~al.}(2008)\citenamefont
  {Mostofi}, \citenamefont {Yates}, \citenamefont {Lee}, \citenamefont {Souza},
  \citenamefont {Vanderbilt},\ and\ \citenamefont {Marzari}}]{Mostofi2008}%
  \BibitemOpen
  \bibfield  {author} {\bibinfo {author} {\bibfnamefont {A.~A.}\ \bibnamefont
  {Mostofi}}, \bibinfo {author} {\bibfnamefont {J.~R.}\ \bibnamefont {Yates}},
  \bibinfo {author} {\bibfnamefont {Y.-S.}\ \bibnamefont {Lee}}, \bibinfo
  {author} {\bibfnamefont {I.}~\bibnamefont {Souza}}, \bibinfo {author}
  {\bibfnamefont {D.}~\bibnamefont {Vanderbilt}},\ and\ \bibinfo {author}
  {\bibfnamefont {N.}~\bibnamefont {Marzari}},\ }\bibfield  {title} {\bibinfo
  {title} {{wannier90: A tool for obtaining maximally-localised Wannier
  functions}},\ }\href {https://doi.org/10.1016/j.cpc.2007.11.016} {\bibfield
  {journal} {\bibinfo  {journal} {Comput. Phys. Commun.}\ }\textbf {\bibinfo
  {volume} {178}},\ \bibinfo {pages} {685} (\bibinfo {year}
  {2008})}\BibitemShut {NoStop}%
\bibitem [{\citenamefont {Grimme}\ \emph {et~al.}(2010)\citenamefont {Grimme},
  \citenamefont {Antony}, \citenamefont {Ehrlich},\ and\ \citenamefont
  {Krieg}}]{Grimme2010}%
  \BibitemOpen
  \bibfield  {author} {\bibinfo {author} {\bibfnamefont {S.}~\bibnamefont
  {Grimme}}, \bibinfo {author} {\bibfnamefont {J.}~\bibnamefont {Antony}},
  \bibinfo {author} {\bibfnamefont {S.}~\bibnamefont {Ehrlich}},\ and\ \bibinfo
  {author} {\bibfnamefont {H.}~\bibnamefont {Krieg}},\ }\bibfield  {title}
  {\bibinfo {title} {{A consistent and accurate ab initio parametrization of
  density functional dispersion correction (DFT-D) for the 94 elements H-Pu}},\
  }\href {https://doi.org/10.1063/1.3382344} {\bibfield  {journal} {\bibinfo
  {journal} {J. Chem. Phys.}\ }\textbf {\bibinfo {volume} {132}},\ \bibinfo
  {pages} {154104} (\bibinfo {year} {2010})}\BibitemShut {NoStop}%
\bibitem [{\citenamefont {Heyd}\ and\ \citenamefont
  {Scuseria}(2004)}]{Heyd2004}%
  \BibitemOpen
  \bibfield  {author} {\bibinfo {author} {\bibfnamefont {J.}~\bibnamefont
  {Heyd}}\ and\ \bibinfo {author} {\bibfnamefont {G.~E.}\ \bibnamefont
  {Scuseria}},\ }\bibfield  {title} {\bibinfo {title} {{Assessment and
  validation of a screened Coulomb hybrid density functional}},\ }\href
  {https://doi.org/10.1063/1.1668634} {\bibfield  {journal} {\bibinfo
  {journal} {J. Chem. Phys.}\ }\textbf {\bibinfo {volume} {120}},\ \bibinfo
  {pages} {7274} (\bibinfo {year} {2004})}\BibitemShut {NoStop}%
\bibitem [{\citenamefont {Heyd}\ \emph {et~al.}(2005)\citenamefont {Heyd},
  \citenamefont {Peralta}, \citenamefont {Scuseria},\ and\ \citenamefont
  {Martin}}]{Heyd2005}%
  \BibitemOpen
  \bibfield  {author} {\bibinfo {author} {\bibfnamefont {J.}~\bibnamefont
  {Heyd}}, \bibinfo {author} {\bibfnamefont {J.~E.}\ \bibnamefont {Peralta}},
  \bibinfo {author} {\bibfnamefont {G.~E.}\ \bibnamefont {Scuseria}},\ and\
  \bibinfo {author} {\bibfnamefont {R.~L.}\ \bibnamefont {Martin}},\ }\bibfield
   {title} {\bibinfo {title} {{Energy band gaps and lattice parameters
  evaluated with the Heyd-Scuseria-Ernzerhof screened hybrid functional}},\
  }\href {https://doi.org/10.1063/1.2085170} {\bibfield  {journal} {\bibinfo
  {journal} {J. Chem. Phys.}\ }\textbf {\bibinfo {volume} {123}},\ \bibinfo
  {pages} {174101} (\bibinfo {year} {2005})}\BibitemShut {NoStop}%
\bibitem [{\citenamefont {Paier}\ \emph {et~al.}(2006)\citenamefont {Paier},
  \citenamefont {Marsman}, \citenamefont {Hummer}, \citenamefont {Kresse},
  \citenamefont {Gerber},\ and\ \citenamefont
  {{\ifmmode\acute{A}\else\'{A}\fi}ngy{\ifmmode\acute{a}\else\'{a}\fi}n}}]{Paier2006}%
  \BibitemOpen
  \bibfield  {author} {\bibinfo {author} {\bibfnamefont {J.}~\bibnamefont
  {Paier}}, \bibinfo {author} {\bibfnamefont {M.}~\bibnamefont {Marsman}},
  \bibinfo {author} {\bibfnamefont {K.}~\bibnamefont {Hummer}}, \bibinfo
  {author} {\bibfnamefont {G.}~\bibnamefont {Kresse}}, \bibinfo {author}
  {\bibfnamefont {I.~C.}\ \bibnamefont {Gerber}},\ and\ \bibinfo {author}
  {\bibfnamefont {J.~G.}\ \bibnamefont
  {{\ifmmode\acute{A}\else\'{A}\fi}ngy{\ifmmode\acute{a}\else\'{a}\fi}n}},\
  }\bibfield  {title} {\bibinfo {title} {{Screened hybrid density functionals
  applied to solids}},\ }\href {https://doi.org/10.1063/1.2187006} {\bibfield
  {journal} {\bibinfo  {journal} {J. Chem. Phys.}\ }\textbf {\bibinfo {volume}
  {124}},\ \bibinfo {pages} {154709} (\bibinfo {year} {2006})}\BibitemShut
  {NoStop}%
\bibitem [{\citenamefont {Shishkin}\ and\ \citenamefont
  {Kresse}(2006)}]{Shishkin2006}%
  \BibitemOpen
  \bibfield  {author} {\bibinfo {author} {\bibfnamefont {M.}~\bibnamefont
  {Shishkin}}\ and\ \bibinfo {author} {\bibfnamefont {G.}~\bibnamefont
  {Kresse}},\ }\bibfield  {title} {\bibinfo {title} {{Implementation and
  performance of the frequency-dependent $GW$ method within the PAW
  framework}},\ }\href {https://doi.org/10.1103/PhysRevB.74.035101} {\bibfield
  {journal} {\bibinfo  {journal} {Phys. Rev. B}\ }\textbf {\bibinfo {volume}
  {74}},\ \bibinfo {pages} {035101} (\bibinfo {year} {2006})}\BibitemShut
  {NoStop}%
\bibitem [{\citenamefont {Hanke}\ and\ \citenamefont {Sham}(1979)}]{Hanke1979}%
  \BibitemOpen
  \bibfield  {author} {\bibinfo {author} {\bibfnamefont {W.}~\bibnamefont
  {Hanke}}\ and\ \bibinfo {author} {\bibfnamefont {L.~J.}\ \bibnamefont
  {Sham}},\ }\bibfield  {title} {\bibinfo {title} {{Many-Particle Effects in
  the Optical Excitations of a Semiconductor}},\ }\href
  {https://doi.org/10.1103/PhysRevLett.43.387} {\bibfield  {journal} {\bibinfo
  {journal} {Phys. Rev. Lett.}\ }\textbf {\bibinfo {volume} {43}},\ \bibinfo
  {pages} {387} (\bibinfo {year} {1979})}\BibitemShut {NoStop}%
\bibitem [{\citenamefont {Rohlfing}\ and\ \citenamefont
  {Louie}(1998)}]{Rohlfing1998}%
  \BibitemOpen
  \bibfield  {author} {\bibinfo {author} {\bibfnamefont {M.}~\bibnamefont
  {Rohlfing}}\ and\ \bibinfo {author} {\bibfnamefont {S.~G.}\ \bibnamefont
  {Louie}},\ }\bibfield  {title} {\bibinfo {title} {{Electron-Hole Excitations
  in Semiconductors and Insulators}},\ }\href
  {https://doi.org/10.1103/PhysRevLett.81.2312} {\bibfield  {journal} {\bibinfo
   {journal} {Phys. Rev. Lett.}\ }\textbf {\bibinfo {volume} {81}},\ \bibinfo
  {pages} {2312} (\bibinfo {year} {1998})}\BibitemShut {NoStop}%
\bibitem [{\citenamefont {McCreary}\ \emph {et~al.}(2021)\citenamefont
  {McCreary}, \citenamefont {Phillips}, \citenamefont {Chuang}, \citenamefont
  {Wickramaratne}, \citenamefont {Rosenberger}, \citenamefont {Hellberg},\ and\
  \citenamefont {Jonker}}]{Mccreary2021}%
  \BibitemOpen
  \bibfield  {author} {\bibinfo {author} {\bibfnamefont {K.~M.}\ \bibnamefont
  {McCreary}}, \bibinfo {author} {\bibfnamefont {M.}~\bibnamefont {Phillips}},
  \bibinfo {author} {\bibfnamefont {H.-J.}\ \bibnamefont {Chuang}}, \bibinfo
  {author} {\bibfnamefont {D.}~\bibnamefont {Wickramaratne}}, \bibinfo {author}
  {\bibfnamefont {M.}~\bibnamefont {Rosenberger}}, \bibinfo {author}
  {\bibfnamefont {C.~S.}\ \bibnamefont {Hellberg}},\ and\ \bibinfo {author}
  {\bibfnamefont {B.~T.}\ \bibnamefont {Jonker}},\ }\bibfield  {title}
  {\bibinfo {title} {{Stacking-dependent optical properties in bilayer
  WSe$_2$}},\ }\href {https://arxiv.org/abs/2111.05704} {\  (\bibinfo {year}
  {2021})},\ \Eprint {https://arxiv.org/abs/2111.05704} {arXiv:2111.05704
  [cond-mat.mtrl-sci]} \BibitemShut {NoStop}%
\end{thebibliography}
\end{document}